\documentclass[%
 aip,
 pof,%
 amsmath,amssymb,
 reprint,%
]{revtex4-1}

\usepackage{graphicx}
\usepackage{dcolumn}
\usepackage{bm}

\begin{document}


\title[arXiv version]{Lift Evaluation of a 2D Flapping Flat Plate}

\author{X. Xia}
\affiliation{Department of Mechanical and Aerospace Engineering, University of Florida, Gainesville, FL, 32611-6250, USA.}%

\author{K. Mohseni}%
\thanks{ Email address for correspondence: mohseni@ufl.edu \hfill \mbox{}}
\affiliation{Department of Mechanical and Aerospace Engineering, University of Florida, Gainesville, FL, 32611-6250, USA.}
\affiliation{Department of Electrical and Computer Engineering, University of Florida, Gainesville, FL, 32611-6250, USA.}
\affiliation{Institution for Networked Autonomous Systems, University of Florida, Gainesville, FL, 32611-6250, USA.}%

\date{\today}

\begin{abstract}
Several previous experimental and theoretical studies have shown that a leading edge vortex (LEV) on an airfoil or wing can provide lift enhancement. In this paper, unsteady 2D potential flow theory is employed to model the flow field of a flapping flat plate wing. A multi-vortices model is developed to model both the leading edge and trailing edge vortices (TEVs), which offers improved accuracy compared with using only single vortex at each separation location. The lift is obtained by integrating the unsteady Blasius equation. It is found that the motion of vortices contributes significantly to the overall aerodynamic force on the flat plate. The shedding of TEVs and the stabilization of LEVs explicitly contributes to lift enhancement. A Kutta-like condition is used to determine the vortex intensity and location at the leading edge for large angle of attack cases; however, it is proposed to relax this condition for small angle of attack cases and apply a 2D shear layer model to calculate the circulation of the new added vortex. The results of the simulation are then compared with classical numerical, modeled and experimental data for canonical unsteady flat plat problems. Good agreement with these data is observed. Moreover, these results suggested that the leading edge vortex shedding for small angles of attack should be modeled differently than that for large angles of attack. Finally, the results of vortex motion vs. lift indicate that both a motion against the streamwise direction of the LEV and a streamwise motion of the TEV contributes positive lift. This also provides the insights for future active flow control of MAVs that the formation and shedding process of LEVs and TEVs can be manipulated to provide lift enhancement.

\end{abstract}

\maketitle

\section{\label{sec:level1}Introduction}

Over the last several decades, researchers have been trying to unveil the aerodynamic secrets of natural flyers that have demonstrated the capacity for high performance hovering flight with unrivaled maneuverability and speed. To understand the basic physical flight mechanisms, early investigations have been focused on experimentations and have attributed this high lift performance to, among other things, an attached leading edge vortex (LEV). This flow phenomenon has been observed to induce a dynamic stall condition at high angles of attack and provide lift enhancement for flapping wings \cite{EllingtonCP:84a, Dickinson:93a}. Although many full numerical studies have been carried out recently to understand the nature of the LEV \cite{Dickinson:04a, EllingtonCP:98a, SunM:02a, Peskin:04a}, simplified models and lower computational cost are equally desirable due to the requirements in potential MAV realtime control applications. To theoretically model the flapping wing and identify the effect of the leading edge vortex on lift, early researchers employed steady potential flow models with a single point vortex representing the LEV and a flat plate representing the wing. The first theoretical evidence for lift enhancement in an attached free vortex over a flat plate was offered by Saffman and Sheffield \cite{Saffman:77a}. They considered a steady two-dimensional irrotational flow over a thin wing with an attached free line vortex. In their model, the presence of a vortex results in increased lift by inducing a stronger bound circulation around the wing. Following this research, Huang and Chow \cite{ChowCY:81a} extended Saffman and Sheffield's work to include the effects of airfoil thickness and camber. Similar to Saffman and Sheffield's approach, Rossow \cite{RossowVJ:78a} modeled an airfoil and added a nose flap to trap a vortex. He also added a sink at the vortex core to represent the spanwise flow which was observed to bleed vorticity from the LEV; this inspired further investigations by Mourtos and Brooks \cite{MourtosNJ:96a}.  This preliminary research produced the same conclusion: attaching and stabilizing a vortex on the upper side of an airfoil will increase the lift coefficient. However, these investigations inherently assumed that the free trapped vortex could somehow be stabilized over the lifting surface and that the vortex should be located at its equilibrium point (velocity of the vortex center equals zero at this point) in order to perform the Kutta-Joukowski lift calculation. Furthermore, these early models are basically steady flows which are incapable of capturing the unsteady features caused by flapping motion or vortex formation and shedding. Nevertheless, the validity of using the steady flow model and the stabilized vortex was supported by many studies of aerodynamic forces in insects which demonstrated that a significant part of lift production is generated during the translational phases of the stroke (down-stroke); see Willmott et al. \cite{EllingtonCP:97c}. Moreover, for flapping flyers the LEV appears to be attached to the wing throughout the translation phase of flapping and seems to be stabilized in the wake region above the wing \cite{Dickinson:01a}.   

Therefore, in order to accurately evaluate the lift and understand the physics of the flapping wing, later modeling attempts have concentrated on resolving the unsteady dynamics of the flat plate and incorporating vortex dynamics into the wake evolution. Minotti \cite{MinottiFO:02a} was the first to consider simple 2D unsteady potential flow for a flapping flat plate with a single point vortex to model the LEV, which is still assumed to be stabilized during flapping motion. Later, Yu et al. \cite{YuY:03a} implemented a perturbation potential flow model with discretized point vortices shed from both the leading and trailing edges to account for the dynamic effect of the wake. Ansari et al. \cite{AnsariSA:06b, AnsariSA:06a} carried out similar investigations and compared their results with Dickinson's flapping plate experiments \cite{Dickinson:93a,Dickinson:99a}, which showed a good agreement in force calculation and wake evolution behaviors. Pullin and Wang \cite{Pullin:04b} established an evolution model for the vortex sheets at the shedding edges and applied it to an accelerating plate. Related work has been done on the unsteady dynamics of a sharp-edged body by Michelin and Smith \cite{LlewellynSmith:09a}, however, they assumed variable circulation of the shed vortices and employed a momentum conservative approach (the Brown-Michael equation \cite{BrownCE:54a}) to solve for the dynamics of the vortices. Other than that, Mason \cite{Mason:03a} and Cochran et al. \cite{Kelly:09a} also made contributions to field by simulating fish locomotion using a deforming airfoil in an unsteady 2D potential flow. 

From these studies, it can be learned that the discretized vortices formulation is an appropriate reduced order model as it saves computational cost while preserving the physics of the wake. Therefore, this work follows the approach of most previous studies (e.g. \cite{YuY:03a, AnsariSA:06b}) by treating each individual vortex as a free vortex, the motion of which is governed by Kirchhoff's law. However, one of the more complicated issues is to determine the intensity and the placement of the shedding vortices near the shedding edges (especially the leading edge) for unsteady flows. Dickinson $\&$ Gotz \cite{Dickinson:93a} proposed that the treatment of the leading edge might depend on the size and configuration of the leading edge vortex or the separation bubble which is related to the angle of attack. Following this thought, the authors suggest that for a fully separated flow at high angles of attack, the classical Kutta condition should be satisfied by enforcing a stagnation point at the leading edge and placing the new vortex in the tangential direction of the edge. At lower angles of attack, a new condition dealing with the placement and circulation of the new vortices is proposed to release the Kutta condition at the leading edge which yields a much better representation of the observed experimental lift. 

Another major focus of this study is to derive the unsteady force equations based on a potential flow model. Several existing models have been used by previous researchers to estimate the lift or drag; Eldredge et al. \cite{Eldredge:09a} and Wang et al. \cite{Dickinson:04a} applied empirical force coefficient models as references of comparison with their computational results; Ansari et al. \cite{AnsariSA:06b} and Michelin and Smith \cite{LlewellynSmith:09a} followed Kelvin's theorem to obtain the unsteady force equations; Yu et al. \cite{YuY:03a} and Miller and Peskin \cite{Peskin:04a} adopted the aerodynamic force equation derived by Wu \cite{Wu:81a} which is based on the first moment of the vorticity field. In this study, the treatment of the unsteadiness in lift calculation is inspired by Minotti \cite{MinottiFO:02a}, who calculated the force from an unsteady form of the Blasius equation. However, several modifications are made here to incorporate the multi-vortices model. This new approach yields additional terms in the lift equation which are found to contribute significantly to the net lift of the wing. The resulting lift expression is then compared with other similar models \cite{LlewellynSmith:09a,Pullin:04b} that have been obtained from different approaches. In summary, the goal of this study is to build a simple model for flapping wing simulations based on earlier studies and to provide an accurate model for lift estimation. Such a low-dimensional model is expected to facilitate future active flow control strategies for flapping wing MAVs.

This paper is organized as follows: Section II provides the unsteady potential flow model of a flat plate with vortices shedding at the leading edge and the trailing edge. Section III provides the derivation of the unsteady force calculations. Section IV presents the unsteady conditions imposed at the edges to determine the locations and intensities of the shedding vortices. Section V provides the validation of the lift equation and compares the simulation results with experimental data as well as other models for canonical cases. Finally, concluding remarks are given in Section VI. 

\section{The 2D Potential Flow Model}

\subsection{Assumptions and Simplifications}

This study employs a potential flow theory in seeking of a simplied model thanks to the inherently explicit representation of the flow field. In order to model a flapping wing problem by a potential flow theory, several assumptions are made here in advance. First of all, the flow is assumed to be 2D; this is justifiable as a first approximation since the aspect ratio of insect wings usually range from 2 to over 10 (see Dudley\cite{Dudley:02a}). Then, under the 2D flow assumption, the flat plate simplification is adopted as the thickness and camber of most insect wings are usually negligible compared to the chord length. This assumption is made to simplify the calculations and it could be extended to Joukowski's airfoil with cambers. Finally, we assume the flow to be inviscid. Combining this with the irrotationality, which should be implicitly satisfied for potential flow, the governing equation reduces to a simple equation in which the Laplacian of the complex potential equals zero. Therefore, the superposition principle can be applied, which implies that the flow field could be built up by combining a background flow with singularities while satisfying proper boundary conditions. Consequently, the flow around a flat plate is obtained by a Joukowski transformation from the flow around a cylinder. This is valid for the flat plate motion without rotation, however, the following section will deal with the case in which the rotational motion needs to be modeled carefully.

\subsection{Potential Flow Model} 

\begin{figure}
\begin{center}
\scalebox{0.6}{\includegraphics{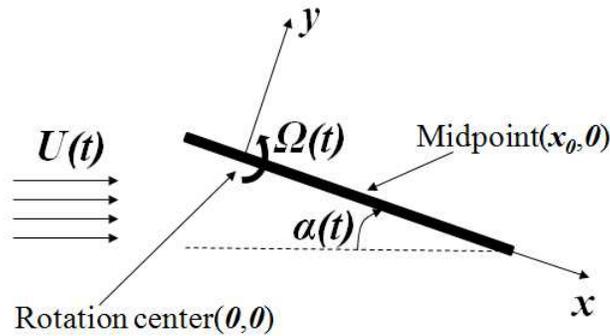}}
\caption{\small Diagram representing the unsteady flow model of a flapping flat plate. The plate pivots around its rotation center. The original translational motion of the flat plate is toward left, while it is substituted by a uniform background flow towards the right represented by $U(t)$.}
\label{fig:physical_flow}
\end{center}
\end{figure}

The flow configuration of a flapping flat plate is shown in Figure. \ref{fig:physical_flow}. Basically, the flapping motion can be decomposed into a translational motion and a rotational motion around the rotation center. A typical way of describing the translational motion is to fix the rotational center of the plate and then to add an opposite velocity to the background flow. The rotational motion is then described by a time dependent angle of attack. Next, we shall take steps to handle these two elemental motions. 

The main difficulty of this unsteady problem is created by the rotational motion. To better understand this we first consider the case with no rotary motion. The first step is to map the complex potential of the physical plane ($z$-plane) to that of the corresponding cylinder plane ($\zeta$-plane). Since the plate is fixed in the background flow, the contour closely encircling the plate is a streamline which indicates that the stream function around the cylinder in $\zeta$-plane is a constant. With this boundary condition satisfied, we can write the complex potential explicitly using Milne-Thomson's circle theorem \cite{MilneThomsonLM:58a}. Now, we consider the case with a plate rotating at an angular velocity $\omega$. Assuming the physical plane is still within a globally fixed inertial coordinate system, then the plate has an angular velocity of $\omega$ in this coordinate; this implies that the contour closely around the plate is no longer a streamline. Consequently, Milne-Thomson's circle theorem can not be applied as the stream function at the cylinder boundary is not a constant in the corresponding $\zeta$-plane. This means that it is difficult to write out the complex potential if the coordinate system is globally irrotational. Therefore, in order to apply the Milne-Thomson's circle theorem, it is natural to consider a non-inertial coordinate system that is fixed on the plate and rotates with it. In this way, we recover the beneficial property that the contour immediately around the plate can be treated as a streamline in this new non-inertial coordinate. However, an observer in this new coordinate would observe the flow field to be rotational with a vorticity magnitude of $-2\omega$. This indicates that the flow in this non-inertial coordinate system can not be represented by any potential flow since it is rotational. Once again, it seems that we can not find a complex potential that is representative of the physical flow.

While various approaches to this problem exist, here we present a new solution to this problem which is inspired by an earlier work by Minotti \cite{MinottiFO:02a}. Again, consider the flow configuration shown in Figure. \ref{fig:physical_flow} with the background flow denoted $\textbf{U}(t)=(U_x,U_y)$. We set up a Cartesian coordinate system $oxy$, with the origin at the plate rotation center and the $x$ axis along the plate's chord. We assume this reference frame, noted as $oxy$, to be an inertial coordinate which is not rotating with the flat plate; therefore the flow in this reference frame can be viewed as the original flow in a global system. Since the flow in this reference frame is irrotational, it is justifiable to assume it to be a potential flow. It then follows that the complex potential $w(z)$ and the velocity field $\textbf{u}=(u_x,u_y)$ of this flow satisfy
\begin{equation}
\label{e:CompPoten1}
w(z) = \phi + i\psi, 
\end{equation}
\begin{subequations}
where
\begin{equation}
\label{e:PotenFunc}
u_x = \frac{\partial \phi}{\partial x}, \;\; u_y = \frac{\partial \phi}{\partial y},
\end{equation}
\begin{equation}
\label{e:FlowFunc}
u_x = \frac{\partial \psi}{\partial y}, \;\; u_y = -\frac{\partial \psi}{\partial x}.
\end{equation}
\end{subequations}

Now we consider a second reference frame, denoted by $ox'y'$, that is attached to and rotates with the plate. Correspondingly, the velocity in this new reference frame is denoted as $\textbf{u}'=(u'_x,u'_y)$ and the relation between $\textbf{u}$ and $\textbf{u}'$ can be expressed as,
\begin{subequations}
\begin{equation}
\label{e:VelRel1}
u'_x = u_x + \Omega(t)y, 
\end{equation}
\begin{equation}
\label{e:VelRel2}
u'_y = u_y - \Omega(t)x,
\end{equation}
\end{subequations}
where $\Omega(t)$ is the angular velocity of the rotational motion that is related to the angle of attack $\alpha(t)$ as
\begin{equation}
\label{e:angle_rel}
\Omega(t) = - \dot{\alpha}(t).
\end{equation}

As indicated above, the flow in the reference frame $ox'y'$ is rotational. So next we will verify the validity of the conditions of continuity and irrotationality in the reference frame $ox'y'$. The continuity equation is
\begin{equation}
\label{e:Contin1}
\frac{\partial u_x'}{\partial x} + \frac{\partial u_y'}{\partial y} = \frac{\partial u_x}{\partial x} + \frac{\partial u_y}{\partial y} =0,
\end{equation}
which indicates that the stream function $\psi'$ exists and is related to $\psi$ by
\begin{equation}
\label{e:StreamRel1}
\psi' = \psi + \frac{1}{2} \Omega (x^2+y^2).
\end{equation}
Next, the vorticity is evaluated
\begin{equation}
\label{e:vorticity}
 \omega'_z = \frac{\partial u'_x}{\partial y} - \frac{\partial u'_y}{\partial x} = 2\Omega.
\end{equation}
This verifies that the flow is rotational. Therefore, $\phi'$ does not exist and the flow in the reference frame $oxy'$ can not be represented by a potential flow. To resolve this problem, Minotti \cite{MinottiFO:02a} proposed a second virtual reference frame, $ox''y''$, in which a potential function will exist. This virtual reference frame is given as
\begin{subequations}
\begin{equation}
\label{e:flowfield_new1}
 \psi'' = \psi' - \Omega y^2 = \psi + \frac{1}{2} \Omega (x^2-y^2),  
\end{equation}
\begin{equation}
\label{e:flowfield_new2}
 \phi'' = \phi - \Omega xy.
\end{equation}
\end{subequations}
Therefore, the velocity in the virtual reference frame $ox''y''$ is related to the velocity of the original flow (in the reference frame $oxy$) by
\begin{subequations}
\begin{equation}
\label{e:Virtual_velocity1}
 u''_x = u'_x - 2\Omega y = u_x - \Omega y,
\end{equation}
\begin{equation}
\label{e:Virtual_velocity2}
 u''_y = u'_y = u_y - \Omega x.
\end{equation}
\end{subequations}

It is apparent that this virtual flow satisfies both the continuity and irrotationality conditions. Finally, the complex potential $w''$ is related to $w$ as
\begin{equation}
\label{e:Potential_func}
 w''(z) = \phi'' + i\psi'' = w(z) + \frac{i}{2} \Omega z^2.
\end{equation}
It should be noted here that an essential point that distinguishes this work from Minotti's lies in the usage of the virtual reference frame. In the following sections, we will show that the virtual reference frame in this study merely serves as a platform to obtain the complex potential of the original flow before the vortex dynamics and force calculation are performed in the original flow. Minotti, however, performed aerodynamic force calculation in the virtual reference frame and treated that as the force in the original flow which does not appear suitable in our case. Moreover, this approach does not have to deal with the complex non-inertial form of the Euler equation which makes it simpler to implement.

\subsection{The Complex Potential} 

In this part the complex potential $w(z)$ representing the potential flow is derived. Here, the Joukowski transformation
\begin{equation}
 \label{e:joukouski-T}
 z = \zeta + \frac{a^2}{\zeta} + x_0,
\end{equation}
is used to link the physical flow ($z$-plane) to a virtual flow ($\zeta$-plane), which maps the flat plate to a cylinder. Here $a=c/4$, with $c$ being the chord length of the flat plate. In the far field, we have the following relations
\begin{subequations}
\begin{equation}
 \label{e:joukouski-T-property1}
 \zeta = z -x_0 \; \text{as} \; \left| z \right| \rightarrow \infty,
\end{equation}
\begin{equation}
 \label{e:joukouski-T-property2}
 \frac{d \zeta}{d z} = 1 \; \text{as} \; \left| z \right| \rightarrow \infty.
 \end{equation}
\end{subequations}

Next, boundary conditions (BCs) in the reference frame $oxy''$ need to be determined to obtain the complex potential.\\
\vspace{1 mm}\\
\textit{BC on the flat plate.} Because the thickness of the flat plate is ignored, $y=0$ represent the flat plate boundary. As $\psi'$ is constant, Equation~\ref{e:flowfield_new1} shows that $\psi''$ is also constant. This is the prerequisite for applying Milne-Thomson's circle theorem in the $\zeta$-plane corresponding to the reference frame $ox''y''$.\\
\vspace{1 mm}\\
\textit{BC at far field.} The original flow velocity is  
\begin{subequations}
\label{e:vel_infty}
\begin{equation}
u_x = U_x(t),
\end{equation}
\begin{equation}
u_y = U_y(t).
\end{equation}
\end{subequations}
Combining Equation~\ref{e:Potential_func} with $U(t) = U_x(t) + iU_y(t)$ yields
\begin{equation}
\label{e:Potential_infty}
 w''_{\infty}(z) = U^*z + i \frac{\Omega z^2}{2},
\end{equation}
where $U^*$ is the complex conjugate of $U$. Now, considering the mapping between the physical plane and the virtual plane via Joukowski transformation, the complex potentials of the two planes satisfy $w_{\zeta}(\zeta) = w[z(\zeta)]$. Here, $w$ and $w_{\zeta}$ denote the complex potentials in the physical plane and the virtual plane respectively. In the reference frame $ox''y''$, Equation~\ref{e:Potential_infty} gives the complex potential at the far field, so the corresponding complex potential at the far field of $\zeta$-plane is given by
\begin{equation}
\label{e:Potential_zeta_infty}
 w''_{\zeta \infty}(\zeta) \approx w''_{\infty}(\zeta+x_0) = U^*(\zeta+x_0) + i \frac{\Omega (\zeta+x_0)^2}{2}.
\end{equation}
Therefore, with the stream function on the flat plate boundary in reference frame $ox''y''$ being zero, the corresponding complex potential of the near field in $\zeta$-plane can be derived using the Milne-Thomson circle theorem as
\begin{equation}
\label{e:Potential_zeta_far}
\begin{split}
 w''_{\zeta}(\zeta) & = w''_{\zeta \infty}(\zeta) + (w''_{\zeta \infty})^*(\frac{a^2}{\bar{\zeta}}) \\
 & = |U|e^{-i\alpha}(\zeta+x_0) + |U|e^{i\alpha}(\frac{a^2}{\zeta}+x_0) + i \frac{\Omega (\zeta+x_0)^2}{2} - i \frac{\Omega (\frac{a^2}{\zeta}+x_0)^2}{2}.
\end{split} 
\end{equation}
This is the complex potential for the background flow. Further incorporating the effects from the singularities (vortex/sink) into Equation~\ref{e:Potential_zeta_far} gives the complete complex potential representing the flow field in reference frame $ox''y''$. Here, we first consider a case with a single free point vortex-sink singularity located at $z_1$ in the $z$-plane, mapped to $\zeta_1$ in the $\zeta$-plane, with vortex and sink intensities of $\Gamma_1$ and $Q$, respectively. The resulting complex potential in $\zeta$-plane and reference frame $ox''y''$ is
\begin{equation}
\label{e:Potential_zeta0}
\begin{split}
 w''_{\zeta}(\zeta) & = |U|e^{-i\alpha}(\zeta+x_0) + |U|e^{i\alpha}(\frac{a^2}{\zeta}+x_0) + i \frac{\Omega (\zeta+x_0)^2}{2} - i \frac{\Omega (\frac{a^2}{\zeta}+x_0)^2}{2}\\
 & -\frac{1}{2\pi}\left[\left(i \Gamma_0 - Q \right) \ln(\zeta) + (Q + i\Gamma_1)\ln(\zeta - \zeta_1) + (Q - i\Gamma_1)\ln \left(\zeta - \frac{a^2}{\bar \zeta_1}\right)\right].
\end{split} 
\end{equation}
Recalling Equation~\ref{e:Potential_func}, the complex potential in $\zeta$-plane and reference frame $oxy$ can be written as
\begin{equation}
\label{e:Potential_zeta}
\begin{split}
 w_{\zeta}(\zeta) & = \underbrace{|U|e^{-i\alpha}(\zeta+x_0) + |U|e^{i\alpha}(\frac{a^2}{\zeta}+x_0)}_{\text{Translational effect}} + \underbrace{i\Omega \frac{ x_0^2-2a^2-2(\frac{a^2}{\zeta} + x_0)^2}{2}}_{\text{Rotational effect}}\\
 & \underbrace{-\frac{1}{2\pi}\left[\left(i \Gamma_0 - Q \right)\ln(\zeta) + (Q + i\Gamma_1)\ln(\zeta - \zeta_1) + (Q - i\Gamma_1)\ln \left(\zeta - \frac{a^2}{\bar \zeta_1}\right)\right]}_{\text{Singularities}}.
\end{split} 
\end{equation}
Here $\Gamma_0$ is a bound circulation at the center of the cylinder that is used to compensate for the circulation deficit determined by flow conditions (i.e. Kutta condition at the trailing edge). Note that in Equation~\ref{e:Potential_zeta} the first two terms represent the translational effect as the third term represents the rotational effect; the last term describes the contribution from the singularities.

\section{Lift Evaluation}

\subsection{Unsteady Blasius Equation}

In order to calculate the totle force on the plate, we first employ the unsteady Bernoulli equation to obtain pressure distribution around the plate. We then calculate the total force by integrating the pressure along the entire body. Considering the physical flow in the inertial reference frame $oxy$, the unsteady Bernoulli equation is written as
\begin{equation}
\label{e:Bernoulli}
-\frac{P}{\rho}  = \frac{\partial \phi}{\partial t} + \frac{1}{2} |\nabla \phi|^2.
\end{equation}

The aerodynamic force over the flat plate is then calculated by integrating the pressure along a closed contour around the plate surface,
\begin{equation}
\label{e:Force}
\begin{split}
\mathbf F & = -\oint P \mathbf n \; dl,
\end{split} 
\end{equation}
where $\mathbf n$ is the normal vector of the 2D integral surface, and $dl$ denotes the integral path along the 2D surface. Therefore, the expression for the unsteady Blasius theorem, written in the complex domain, is
\begin{equation}
\label{e:Force1}
\begin{split}
F_x - iF_y & = - i \oint P dz^*\\
& = i\rho \oint \left( \frac{\partial \phi}{\partial t} + \frac{1}{2} |\nabla \phi|^2 \right) dz^*\\
& = \frac{i\rho}{2} \oint \left(\frac{dw}{dz}\right)^2 dz + i\rho \left(\oint \frac{\partial \phi}{\partial t} dz \right)^*,
\end{split} 
\end{equation}
where $(.)^*$ denotes the complex conjugate in this context. 

The following subsections describe the steps taken to evaluate the integrals in Equation~\ref{e:Force1}. We will treat the steady term and unsteady term separately. First, the force equation will be derived based on the single singularity model specified by Equation~\ref{e:Potential_zeta}. 

\subsection{Steady Blasius Integral}

\begin{figure}
 \center{\includegraphics[width=0.4\textwidth]{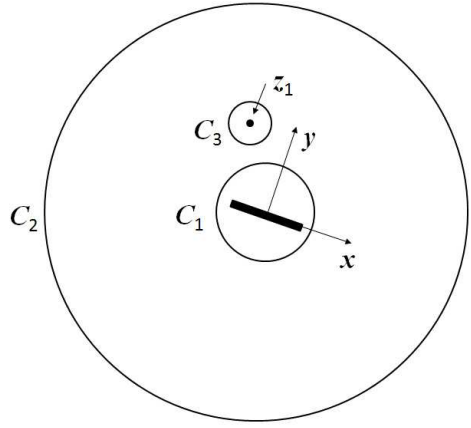}}
 \caption{\small Integration contours $C_1$, $C_2$, $C_3$ for the Blasius theorem. $C_1$ contains the flat plate only, $C_2$ is at the infinity and encloses all, $C_3$ only contains the vortex-sink.}
 \label{fig:Blasius}
\end{figure}

The solution of the integrals in Equation~\ref{e:Force1} requires defining three separate integration contours as shown in Figure~\ref{fig:Blasius}. $C_1$ is the path that encircles the flat plate boundary and is used to compute the steady Blasius integral, $\oint \left(\frac{dw}{dz}\right)^2 dz$. Another contour $C_2$ at the infinity is drawn to include both the flat plate and the vortex-sink point at $z_1$ in the $z$-plane. Then, a smaller $C_3$ contour is drawn to include the vortex-sink point only. As $(\frac{dw}{dz})^2$ is analytic inside the $C_2-C_1-C_3$ region, we obtain  
\begin{equation}
 \label{e:blasius}
 \oint_{C_1}\left(\frac{dw}{dz}\right)^2dz = \oint_{C_2}\left(\frac{dw}{dz}\right)^2dz-\oint_{C_3}\left(\frac{dw}{dz}\right)^2dz.
\end{equation}
In the following, the integrals on the right hand side of Equation~\ref{e:blasius} are computed successively.
\\
\vspace{1 mm}\\
\textit{The integral at infinity.} Since the integral $\oint_{C_2} \left(\frac{dw}{dz}\right)^2 dz$ is calculated in the far field, we have the approximation $z  \rightarrow \zeta+x_0$. Consequently, Equation~\ref{e:Potential_zeta} can be converted to a complex potential in $z$-plane as
\begin{equation}
\label{e:Potential_z_infty}
\begin{split}
 w_{\infty}(z) & = |U|e^{-i\alpha}z + |U|e^{i\alpha}x_0 + i \Omega \frac{-x_0^2 - 2a^2}{2} + \frac{|U| e^{i \alpha} a^2}{z - x_0} - i\frac{2 \Omega a^2 x_0}{z - x_0}\\
 & - \frac{\left(i\Gamma_0 - Q \right) \ln(z-x_0)}{2\pi}  - \frac{(Q + i\Gamma_1)\ln(z - x_0 - \zeta_1)}{2\pi} \\
 & - \frac{(Q - i\Gamma_1)\ln \left(z - x_0 - \frac{a^2}{\bar \zeta_1}\right)}{2\pi}.
\end{split}
\end{equation}

Under the limit $\left| z \right| \rightarrow \infty$, the terms $z - x_0$, $z - x_0 - \zeta_1$, and $z - x_0 - \frac{a^2}{\bar \zeta_1}$ are all approximately equal to $z$; therefore, by taking the derivative of the velocity potential this yields

\begin{equation}
 \label{e:complexvelocity-z-infinity-simple}
  \frac{d w_{\infty}(z)}{d z} = |U| e^{-i \alpha} - \frac{i\Gamma_0 + Q}{2\pi z} + \frac{a^2(2 i\Omega x_0 - |U| e^{i \alpha})}{z^2}.
\end{equation}
Making use of Cauchy's residue theorem, the integral along contour $C_2$ can be computed as
\begin{equation}
 \label{e:C2}
 \oint_{C_2}\left(\frac{dw}{dz}\right)^2dz = 2\pi i \cdot \text{Res} \left[ \left(\frac{d w_{\infty}(z)}{d z} \right)^2,z=0 \right] = -2 i|U| e^{-i \alpha}\left(Q + i\Gamma_0 \right).
\end{equation}
\\
\vspace{1 mm}\\
\textit{The integral around the singularity.} As $\zeta_1$ is a singular point in the $\zeta$-plane, it implies that the corresponding $z_1$ is also a singular point in the transformed $z$-plane. In order to calculate the integral around the contour $C_3$, we divide the velocity potential into two parts
\begin{equation}
\label{e:velocitypotential-singular}
 w(z) = w'(z) - \frac{Q + i\Gamma_1}{2\pi} \ln(z - z_1).
\end{equation}
Here, $w'(z)$ represents the velocity potential that excludes the contribution from the vortex-sink; it is therefore analytic over the region surrounded by the contour $C_3$. Taking the derivative of $w(z)$ yields
\begin{equation}
\label{e:complexvelocity-singular}
 \frac{d w(z)}{d z} = \frac{d w'(z)}{d z} - \frac{Q + i\Gamma_1}{2\pi(z - z_1)}.
\end{equation}
By taking square of both sides, we obtain
\begin{equation}
\label{e:complexvelocity-singular2}
 \left( \frac{dw (z)}{d z} \right)^2 = \left( \frac{d w'(z)}{d z} \right)^2 - \frac{Q + i\Gamma_1}{\pi} \left( \frac{d w'(z)}{d z} \right) \frac{1}{z - z_1} + \frac{(Q + i\Gamma_1)^2}{4\pi^2} \frac{1}{(z - z_1)^2}.
\end{equation}
As discussed previously, $w'(z)$ is analytic inside contour $C_3$. It follows that $\frac{dw'(z)}{dz}$ and $(\frac{dw'(z)}{dz})^2$ are also analytic inside contour $C_3$; therefore, the second integral on the right hand side of Equation~\ref{e:blasius} can be computed with Equation~\ref{e:complexvelocity-singular2} using Cauchy's residue theorem to obtain
\begin{equation}
\label{e:C3}
\begin{split}
 \oint_{C_3}\left(\frac{d w}{d z}\right)^2dz & = 2\pi i\left\{ 0 + \text{Res} \left[-\frac{Q + i\Gamma_1}{\pi} \left(\frac{d w'(z)}{d z}\right) \frac{1}{z - z_1},z=z_1 \right] + 0 \right\}\\
 & = -2 i \left( Q + i\Gamma_1\right) \left(\frac{d w'(z)}{d z}\right)|_{z=z_1}.
 \end{split}
\end{equation}
where the term $(\frac{d w'(z)}{d z})|_{z=z_1}$ equals the conjugate of the velocity of the free vortex-sink that can be derived to be
\begin{equation}
 \label{e:vel-vs}
 \begin{split}
\left(\frac{d w'(z)}{d z}\right)|_{z=z_1} & = \frac{\zeta_1^2}{\zeta_1^2 - a^2}\left[|U| \left(e^{-i\alpha} - \frac{a^2e^{i\alpha}}{\zeta_1^2} \right) + i\Omega \frac{2a^2}{\zeta_1^2}\left(\frac{a^2}{\zeta_1}+x_0\right) \right]\\
& - \frac{\zeta_1^2}{\zeta_1^2 - a^2}\left(\frac{1}{2\pi}\frac{i\Gamma_0-Q}{\zeta_1} + \frac{1}{2\pi}\frac{Q-i\Gamma_1}{\zeta_1-\frac{a^2}{\bar{\zeta_1}}}\right) + \frac{Q + i\Gamma_1}{\pi}\frac{\zeta_1a^2}{(a^2 - \zeta_1^2)^2}.
 \end{split}
\end{equation}
Note here that the last term in $(\frac{d w'(z)}{d z})|_{z=z_1}$ is known as the Routh correction \cite{LinCC:41a, ClementsRR:73a}. This term represents the difference of velocity at the location of the sigularity between the actual flow and a virtual flow without the point sigularity (this virtual flow only considers the effect of the image of the sigularity).

\subsection{Unsteady Blasius Integral}

To compute the integral $\oint \frac{\partial \phi}{\partial t} dz$, the velocity potential $\phi$ is divided into four components based on different methods used to evaluate the integral and the corresponding complex potential $w_{\zeta}(\zeta)$ can be expressed as
\begin{subequations}
\label{e:Potential_zeta_parts}
\begin{equation}
\begin{split}
 w_{\zeta1}(\zeta) & = |U|e^{-i\alpha}(\zeta+x_0) + |U|e^{i\alpha}\left(\frac{a^2}{\zeta}+x_0\right) + i \Omega \frac{ (\zeta+x_0)^2}{2} - i \Omega \frac{(\frac{a^2}{\zeta}+x_0)^2}{2},
\end{split} 
\end{equation}
\begin{equation}
\begin{split}
 w_{\zeta2}(\zeta) & = -i\Omega \frac{z(\zeta)^2}{2} = -i\Omega \frac{(\zeta + \frac{a^2}{\zeta} + x_0)^2}{2},
\end{split} 
\end{equation}
\begin{equation}
\begin{split}
 w_{\zeta3}(\zeta) & = -\frac{1}{2\pi}\left(i \Gamma_0 - Q \right)\ln(\zeta),
\end{split} 
\end{equation}
\begin{equation}
\begin{split}
 w_{\zeta4}(\zeta) & = -\frac{1}{2\pi}\left[(Q + i\Gamma_1)\ln(\zeta - \zeta_1) + (Q - i\Gamma_1)\ln \left(\zeta - \frac{a^2}{\bar \zeta_1}\right)\right].
\end{split} 
\end{equation}
\end{subequations}
In this way, by distinguishing the analytic part from the logarithmic part, the integral terms of $\oint \frac{\partial \phi}{\partial t} dz$ are handled through different approaches as shown below. Basically, the analytic part will be integrated by the residue theorem, while the logarithmic part will be integrated by carefully dealing with different sigularities inside and outside the integration contour.
\\
\vspace{1 mm}\\
\textit{First Unsteady Integral.} Note that the integral is taken around the contour $C_1$. It is not difficult to verify that $\psi_1 = 0$ on $C_1$. Therefore, by expanding the contour $C_1$ to infinity and employing the residue theorem, this integral is evaluated to be
\begin{equation}
\label{e:Unsteady_1}
\begin{split}
\oint_{C_1} \frac{\partial \phi_1}{\partial t} dz & = \oint_{C_1} \frac{\partial w_1(z)}{\partial t} dz\\
& = 2\pi i \; \text{Res} \left[\frac{\partial w_{\zeta1}}{\partial t} \left( 1-\frac{a^2}{\zeta^2} \right),\zeta=0 \right]\\
& = 2\pi i a^2(\underbrace{|\dot{U}| (e^{i \alpha}-e^{-i \alpha}) - 2i\dot{\Omega}x_0}_{\text{Added Mass}} \underbrace{- 2i|U| e^{i \alpha}\Omega}_{\text{Rotational}}).
\end{split}
\end{equation}
where $\dot{U}$ and $\dot{\Omega}$ represent the translational and angular accelerations of the plate, respectively. The results of the calculation give the added mass and rotational force terms as shown above.
\\
\vspace{1 mm}\\
\textit{Second Unsteady Integral.} Since $w_2(z) = -i\Omega \frac{z^2}{2}$, $z = z^*$ on the contour $C_1$ which is near the wall of the flat plate. This yields $w_2(z) + w_2^*(z) = 0$, and the unsteady integral associated with $\phi_2$ becomes
\begin{equation}
\label{e:Unsteady_2}
\oint_{C_1} \frac{\partial \phi_2}{\partial t} dz = \frac{1}{2} \oint_{C_1} \frac{\partial (w_2(z) + w_2^*(z))}{\partial t} dz = 0.
\end{equation}
\\
\vspace{1 mm}\\
\textit{Third Unsteady Integral.} On the contour $C_1$ that is closely around the flat plate, the following relations are satisfied: $\text{Real} (Q \ln(\zeta)) = Q \ln(a)$ and $\text{Imag} (i \Gamma_0 \ln(\zeta)) = \Gamma_0 \ln(a)$ which are constants on $C_{\zeta1}$ at any time. However, their integral over a contour should be equal to zero at any time even if $Q$ and $\Gamma_0$ are time variant. In this case, the unsteady integral associated with $\phi_3$ is computed as
\begin{equation}
\label{e:Unsteady_3}
\begin{split}
\oint_{C_1} \frac{\partial \phi_3}{\partial t} dz & = -\frac{1}{2 \pi} \oint_{C_1} \frac{\partial (\text{Real}(i \Gamma_0 \ln(\zeta) - Q \ln(\zeta)))}{\partial t} dz\\
& = -\frac{1}{2 \pi} \oint_{C_{\zeta1}} \frac{\partial (i \Gamma_0 \ln(\zeta))}{\partial t} \frac{dz}{d\zeta} d\zeta = 2a\dot{\Gamma_0}.
\end{split}
\end{equation}
Here, $\dot{\Gamma_0}$ represents the rate at which $\Gamma_0$ changes with time.
\\
\vspace{1 mm}\\
\textit{Fourth Unsteady Integral.} This integral actually reflects the unsteady force generated by the singularity point in the flow field. Similar to $\psi_1$, $\psi_4$ also equals zero on the contour $C_1$ that is closely around the flat plate. Therefore, this integral is derived as
\begin{equation}
\label{e:Unsteady_4}
\begin{split}
\oint_{C_1} \frac{\partial \phi_4}{\partial t} dz & = \oint_{C_1} \frac{\partial w_4(z)}{\partial t} dz\\
& = -\frac{1}{2 \pi} \oint_{C_{\zeta1}} \frac{\partial}{\partial t} \left[(Q + i\Gamma_1)\ln(\zeta - \zeta_1) + (Q - i\Gamma_1)\ln \left(\zeta - \frac{a^2}{\bar \zeta_1}\right)\right] \frac{dz}{d\zeta} d\zeta.\\
\end{split}
\end{equation}
Note that as the contour $C_1$ is rotating with the plate, the temporal partial derivative can not be taken outside the contour integral. To integrate the above equation, $C_1$ should not be expanded directly to infinity without removing the effect of the singularity points. Noting that the singularity point $\zeta=a^2/\bar{\zeta_1}$ is inside the contour $C_1$ and the resultant integral contains the logarithmic function $\ln(\zeta - a^2/\bar{\zeta_1})$, the integral term with $\ln(\zeta - a^2/\bar{\zeta_1})$ should be multiplied by $2\pi i$. Here, for convenience in integration, $\zeta_1$ is set to be $\frac{a^2}{b}e^{i\theta_0}$ and the result is derived to be
\begin{equation}
\label{e:Unsteady_41}
\begin{split}
\oint_{C_1} \frac{\partial \phi_4}{\partial t} dz & = -i [(\dot{Q} + i\dot{\Gamma_1}) be^{-i\theta_0} + (\dot{Q} - i\dot{\Gamma_1}) ((a-b)e^{i\theta_0} + ae^{-i\theta_0} )\\
& - (Q + i\Gamma_1) \dot{\zeta_1} b^2e^{-2i\theta_0}/a^2  + (Q - i\Gamma_1) \dot{\bar{\zeta_1}} b^2e^{2i\theta_0}/a^2]\\
& =  -i [(\dot{Q} + i\dot{\Gamma_1}) \frac{a^2}{\zeta_1} + (\dot{Q} - i\dot{\Gamma_1}) (a (e^{i\theta_0} + e^{-i\theta_0}) - \frac{a^2}{\bar \zeta_1})\\
& + (Q + i\Gamma_1) (\dot{z_1} - \dot{\zeta_1}) + (Q - i\Gamma_1) (\dot{\bar{\zeta_1}} -  \dot{\bar{z_1}})].
\end{split}
\end{equation}
where $\dot{Q}$, $\dot{\Gamma_1}$, $\dot{z_1}$ and $\dot{\zeta_1}$ represent the time derivatives of $Q$, $\Gamma_1$, $z_1$ and $\zeta_1$, respectively. Combining Equations~\ref{e:blasius}, ~\ref{e:Unsteady_1}, ~\ref{e:Unsteady_2}, ~\ref{e:Unsteady_3}, and ~\ref{e:Unsteady_41} gives the second term in force calculation (Equation~\ref{e:Force1}) of $F_x - iF_y$.

\subsection{The Final Expression of the Unsteady Force}

Combining Equations~\ref{e:Force1}, ~\ref{e:blasius}, ~\ref{e:C2}, ~\ref{e:C3}, ~\ref{e:Unsteady_1}, ~\ref{e:Unsteady_2}, ~\ref{e:Unsteady_3}, and ~\ref{e:Unsteady_41} together with the observation that $\dot{\zeta_1} - \dot{z_1} = \frac{d}{dt}(-\frac{a^2}{\zeta_1})$, one could derive the total force to be
\begin{equation}
\label{e:Lift-Drag-6}
\begin{split}
F_x - iF_y  & = \rho |U| e^{-i \alpha}\left(Q + i\Gamma_0 \right) - \rho (Q + i\Gamma_1) \dot{\bar{z_1}}\\
& + 2\pi \rho a^2 \left( |\dot{U}| (e^{-i \alpha}-e^{i \alpha}) + 2i\dot{\Omega}x_0 + 2i|U| e^{-i \alpha}\Omega \right)\\
& - \rho \left( (Q + i\Gamma_1) \frac{d}{dt}(-\frac{a^2}{\zeta_1}) - (Q - i\Gamma_1) (\dot{\bar{\zeta_1}} -  \dot{\bar{z_1}}) \right)\\
& - \rho \left( (\dot{Q} - i\dot{\Gamma_1}) \frac{a^2}{\bar \zeta_1} + (\dot{Q} + i\dot{\Gamma_1}) (a (e^{i\theta_0} + e^{-i\theta_0}) - \frac{a^2}{\zeta_1}) \right).
\end{split}
\end{equation}

At this point, we have completed the derivation of the unsteady force for a flapping flat plate with a single attached vortex-sink singularity. Further setting  $Q$ and $\Gamma_0$ to be zero (no sink is considered at the location of the vortex and no bound circulation is placed inside the flat plate), the above equation can simplified to be
\begin{equation}
\label{e:Lift-Drag-7}
\begin{split}
F_x - iF_y  & = 2\pi \rho a^2 ( \underbrace{|\dot{U}| (e^{-i \alpha}-e^{i \alpha}) + 2i\dot{\Omega}x_0}_{\text{Added Mass}} + \underbrace{2i|U| e^{-i \alpha}\Omega}_{\text{Rotational}} )\\
& - i \rho \underbrace{ \Gamma_1  \frac{d}{dt}\left(-\frac{a^2}{\zeta_1} + \bar{\zeta_1} \right) }_{\text{Vortex Convection}} - i \rho \underbrace{ \dot{\Gamma_1} \left(  \frac{a^2}{\bar \zeta_1} - \frac{a^2}{\zeta_1} +  a (e^{i\theta_0} + e^{-i\theta_0})  \right) }_{\text{Vortex Variation}}.
\end{split}
\end{equation}
Here, the last two terms show the lift contributions from the vortex. This indicates that both the variation of intensity and the convection in space of a vortex could generate a force on the flat plate. Comparing this equation with the calculations obtained by Pullin $\&$ Wang \cite{Pullin:04b} and Michelin $\&$ Smith \cite{LlewellynSmith:09a}, the difference lies in the vortex variation term. Actually, this is originated from the derivation of Equation~\ref{e:Unsteady_41}. If the singularity point $\zeta=a^2/\bar{\zeta_1}$ were not treated as one inside the integration contour, the vortex variation term would be $-i \rho \dot{\Gamma_1} \left(  \bar \zeta_1 - \frac{a^2}{\zeta_1}   \right)$, which is the same as reported by Pullin $\&$ Wang and Michelin $\&$ Smith. 

\section{The Multi-vortices Model}

In previous sections, the complex potential of the flow field and the force equations are computed based on a single point singularity model that features both a vortex and a sink/source. The physical nature of the flow around a flapping wing, however, is significantly affected by the shedding of vortices from the leading and trailing edges. The formation of these vortices and the evolution of their structure will necessarily impact the flow field as well as the interaction between the flow and the flat plate which is readily indicated from the derived force calculations. This additional consideration of the flow physics is highly relevant to the force model and is not taken into account by the single vortex model. Therefore, the following section introduces a discretized multi-vortices model to represent the vortex structures in the vicinity of the leading and trailing edges. Moreover, new vortices will be added near the vortex shedding edges at each time step to simulate the behavior of the separated shear layer, which in reality serves as the source of vorticity \cite{YuY:03a, AnsariSA:06a, LlewellynSmith:09a}. 

\subsection{Complex Potential and Force Calculation}

In essense, the multi-vortices model extends the theory of the previous sections to a system of many vortices shed from the edges of the plate. The complex potential of this flow can be obtained by summing the same background flow with multiple free vortices instead of the single singularity point. It should be mentioned that due to the existence of spanwise flow, future models might need to consider incorporating $Q$ into the model to account for some aspects of 3D flow effect; however, for the current study, we assume $Q=0$ for all vortices. To implement this model, two new vortices are generated at the leading and trailing edges, denoted by $1n$ and $2n$ respectively, at the $n^{\text{th}}$ time step. The positions of the vortices, $z_{1n}$ and $z_{2n}$, are determined by the vortex shedding conditions that will be discussed below. Typically, the circulations $\Gamma_{1n}$ and $\Gamma_{2n}$ are updated by implementing Kutta conditions both at the leading edge and the trailing edge \cite{YuY:03a, AnsariSA:06a}. However, in this study, we will also present a new method for computing $\Gamma_{1n}$ for the shed vortices at the leading edge as well as introducing a shedding condition for low angles of attack. With these initializations, the complex potential can be written by adapting Equation~\ref{e:Potential_zeta} for the multi-vortices model
\begin{equation}
\label{e:Potential_zetam}
\begin{split}
 w_{\zeta}(\zeta) & = |U|e^{-i\alpha}(\zeta+x_0) + |U|e^{i\alpha}(\frac{a^2}{\zeta}+x_0) + i\Omega \frac{ x_0^2-2a^2-2(\frac{a^2}{\zeta} + x_0)^2}{2}\\
 & -\frac{i}{2\pi} \sum_{n}^{N} \left[ \Gamma_{1n} \ln \left( \frac{\zeta - \zeta_{1n}}{\zeta - \frac{a^2}{\bar \zeta_{1n}}} \right) + \Gamma_{2n} \ln \left( \frac{\zeta - \zeta_{2n}}{\zeta - \frac{a^2}{\bar \zeta_{2n}}} \right) \right],
\end{split} 
\end{equation}
where $N$ is the number of total vortices shed from the leading or trailing edge. Again, it is assumed that the intensities of the vortices are constant once they are generated. This is reasonable because there is no need to resolve vortex generation and diffusion mechanism at the wall, the time scale of which is much smaller than the simulation time step. We further note that the bound circulation, $\Gamma_0$, vanishes in this expression. This is because all shed vortices are represented in this model, thus the effect of the bound circulation is explicitly resolved by the imaginary vortices inside the cylinder. Therefore, the corresponding force calculation is expressed as
\begin{equation}
\label{e:Lift-Drag2}
\begin{split}
F_x - iF_y  & = 2\pi \rho a^2 \left( |\dot{U}| (e^{-i \alpha} - e^{i \alpha}) + 2i\dot{\Omega}x_0 + 2i|U| e^{-i \alpha}\Omega \right)\\
& -  i \rho \sum_{n}^{N}  \left[ \Gamma_{1n} \left( \dot{\zeta_{1n}} - \dot{z_{1n}} + \dot{\bar{\zeta_{1n}}} \right) + \Gamma_{2n} \left( \dot{\zeta_{2n}} - \dot{z_{2n}} + \dot{\bar{\zeta_{2n}}} \right) \right].\\
& = 2\pi \rho a^2 ( \underbrace{|\dot{U}| (e^{-i \alpha} - e^{i \alpha}) + 2i\dot{\Omega}x_0}_{\text{Added Mass}} + \underbrace{2i|U| e^{-i \alpha}\Omega}_{\text{Rotational}} )\\
& - i \rho \sum_{n}^{N} [ \underbrace{\Gamma_{1n}  \frac{d}{dt}(\bar{\zeta_{1n}} -\frac{a^2}{\zeta_{1n}})}_{\text{LEV Effect}} + \underbrace{\Gamma_{2n} \frac{d}{dt}(\bar{\zeta_{2n}} - \frac{a^2}{\zeta_{2n}})}_{\text{TEV Effect}} ].
\end{split}
\end{equation}
Here, the velocities of the vortices can be easily obtained in a similar manner as for the single singularity model in Equation~\ref{e:vel-vs}; the detailed calculations are not presented here for brevity. It should be pointed out that the force contribution from the vortices, as shown in Equation~\ref{e:Lift-Drag2}, is similar to that obtained by Pullin $\&$ Wang \cite{Pullin:04b} and Michelin $\&$ Smith \cite{LlewellynSmith:09a} due to the vortex variation term in Equation~\ref{e:Lift-Drag-7} being zero under the constant circulation assumption.  

\subsection{Kutta Condition and Vortex Placement for Large Angle of Attack}

To simulate the dynamics of the flat plate as well as evaluating the aerodynamic forces, the intensities and locations of the shedding vortices at the shedding edges are important components in Equation~\ref{e:Potential_zeta} and ~\ref{e:Lift-Drag2}. However, in the physical flow, the presence of the shedding vortices is related to the LE or TE shear layer, which in turn is a product of the viscous effects near the plate. Since viscosity is ignored in a potential flow model, a typical way of reconciling that is to apply the Kutta condition. This means that all the viscous effects can be incorporated into a single edge condition \cite{CrightonDG:85a} which allows the Blasius theorem to be applied for computing the aerodynamic forces. A common way of describing the Kutta condition for steady flows is known as the steady state trailing edge Kutta condition which requires a finite velocity at the trailing edge \cite{Saffman:77a,ChowCY:81a,MourtosNJ:96a}. After some algebra, the effect of the steady state Kutta condition is to simply place a stagnation point at the trailing edge in the $\zeta$-plane at all time. By prescribing a value for the circulation of the shed vortex which satisfies the trailing edge Kutta condition, the flow becomes physically accurate and the aerodynamic forces can be estimated.

\begin{figure}
\begin{center}

\begin{minipage}{0.2\linewidth}\begin{center} \textbf{(a) Large angle of attack}  \end{center}
\end{minipage}
\begin{minipage}{0.7\linewidth} \begin{center}
\includegraphics[width=.99\linewidth]{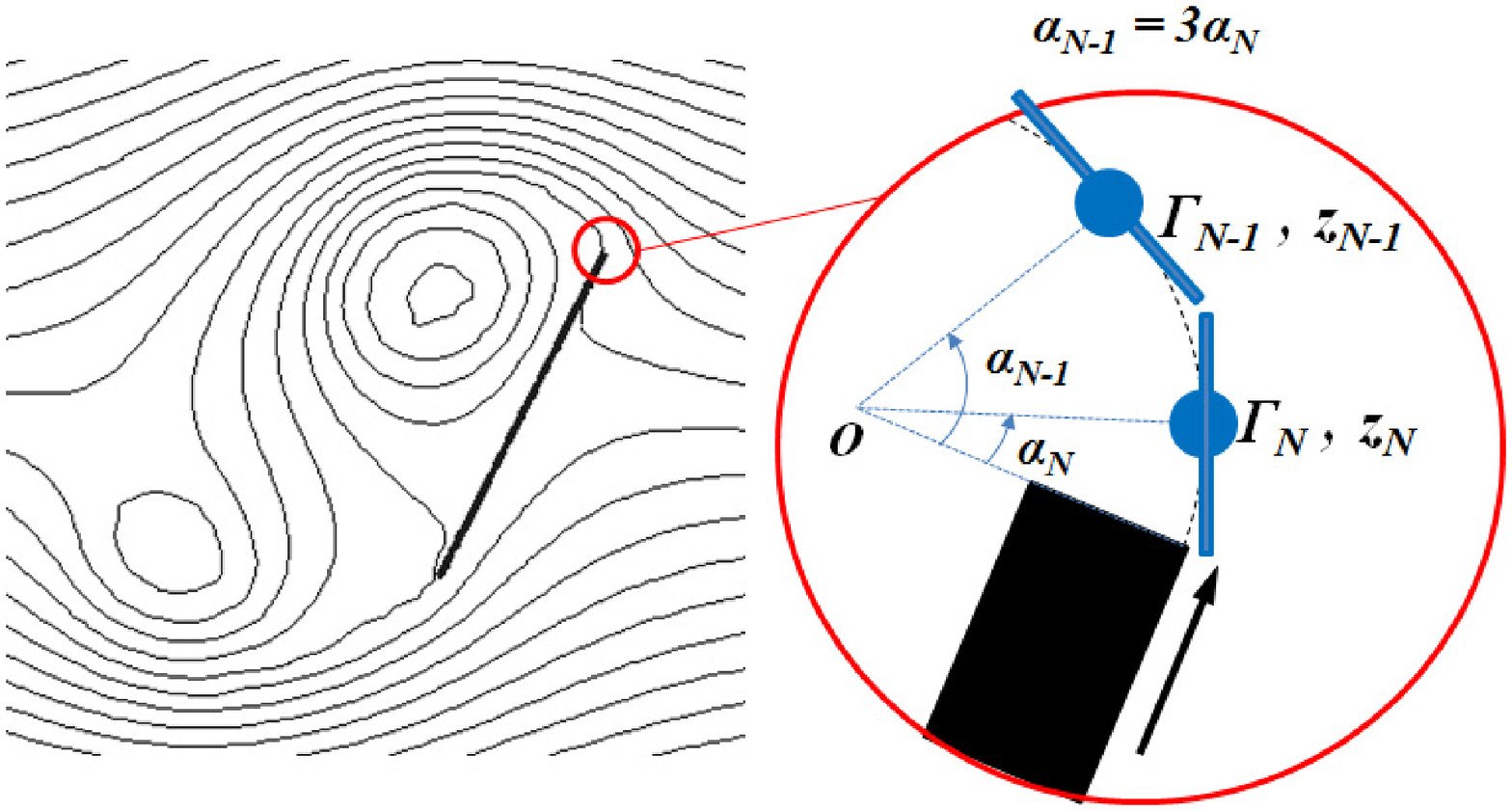}
\end{center} \end{minipage}\\

\begin{minipage}{0.2\linewidth}\begin{center} \textbf{(b) Small angle of attack}  \end{center}
\end{minipage}
\begin{minipage}{0.7\linewidth} \begin{center}
\includegraphics[width=.99\linewidth]{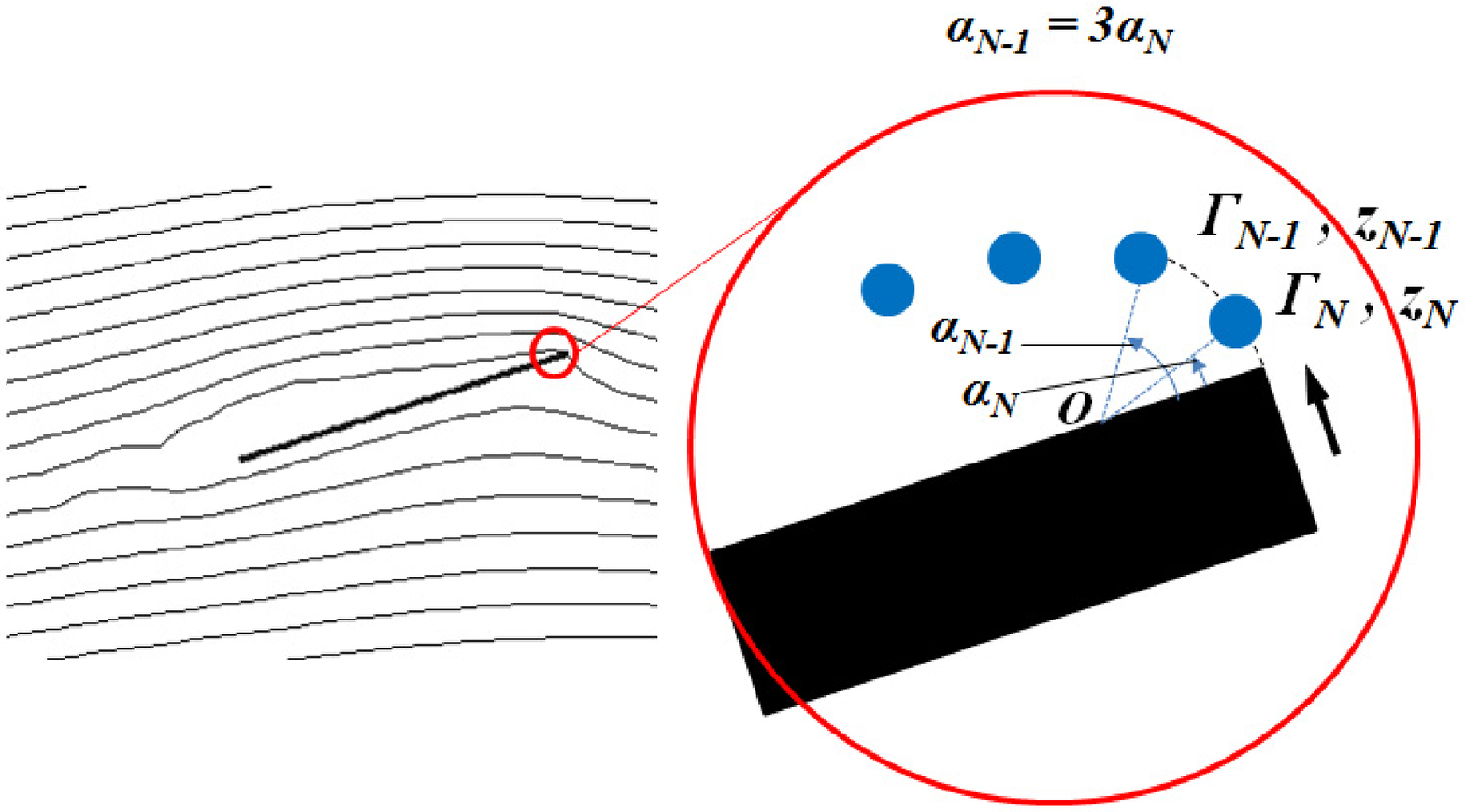}
\end{center} \end{minipage}\\

\caption{\small Kutta condition and vortex placement (Incoming flow from right to left). The black arrows represent the flow that is responsible for creating the shear layer at the leading edge. The dots represent the approximated centers of the vortex sheet segments emanated during each time step. In terms of vortex placement, both cases employ the `$1/3$ arc' approach. (a) The diagram on the left shows the streamline plot and the diagram on the right shows the vortex placement and the reverse flow at the leading edge which is similar to a Kutta-like condition of the trailing edge. (b) From the streamline plot on the left, the stagnation point is not on the bottom side of the plate but on the surface of the leading edge. Evidently, the dominant flow in this case creates a shear layer that leaves the leading edge from the top side of the plate. In this way, the leading edge would not satisfy Kutta-like condition.}
\label{fig:Kutta_plot}
\end{center}
\end{figure}

Mathematically, this condition is implemented by placing a stagnation point at the trailing edge of the cylinder in the $\zeta$-plane so that a finite velocity at the trailing edge of the flat plate in the $z$-plane is guaranteed. Thus, the velocity at the upper and lower surfaces of the plate at the trailing edge will be equal, which implies the streamline emanating from this stagnation point will be parallel to the plate, fulfilling the condition proposed in previous studies \cite{ChenSH:87a,PolingDR:87a}. While it is relatively straightforward to understand the implementation of this Kutta condition for the trailing edge, the physics around the leading edge are quite different. It is suggested by Dickinson $\&$ Gotz \cite{Dickinson:93a} that the treatment of the leading edge might depend on the size and configuration of the leading edge vortex or the separation bubble. Dickinson $\&$ Gotz recognized that for the case with a large leading edge vortex, which normally emerges for a fully separated flow at large angles of attack, the presence of the leading edge vortex eliminates the leading edge suction force by establishing a Kutta-like condition that is similar to the trailing edge Kutta condition. This can be interpreted as shown on the left image in Figure~\ref{fig:Kutta_plot}(a); a stagnation point exists on the bottom side of the plate for the case with high angle of attack which results in a reverse flow and creates a shear layer originating from the bottom of the leading edge and emanating in the tangential direction of the leading edge extension. This shear layer at the leading edge resembles that of the trailing edge and therefore it is reasonable to apply a similar Kutta condition as the one for the trailing edge. 

In the cases of large angle of attack, a classical Kutta condition are implemented at both vortex-shedding edges, which has also been applied in some previous studies \cite{MinottiFO:02a, YuY:03a}. In this potential flow model this means that stagnation should be imposed at both vortex-shedding edges in the $\zeta$-plane. Explicitly, the two new vortices near the vortex-shedding edges introduced at the $n^{\text{th}}$ time step are named as $1n$ and $2n$. Assuming the positions of the shedding vortices $\zeta_{1n}$ and $\zeta_{2n}$ are already known or can be pre-calculated, the circulations $\Gamma_{1n}$ and $\Gamma_{2n}$ are then determined by
\begin{equation}
\label{e:kutta-condition-largeA}
\frac{\partial w_{\zeta}(\zeta)}{\partial \zeta} = 0, \;\; \text{for} \;\; \zeta \rightarrow \pm a,
\end{equation}
which can be specified as
\begin{equation}
\label{e:kutta-condition2-largeA}
\frac{\partial w_{\zeta_{n-1}}(\zeta)}{\partial \zeta} - \frac{i\Gamma_{1n}}{2\pi} \frac{\partial}{\partial \zeta} \left( \ln \left( \frac{\zeta - \zeta_{1n}}{\zeta - \frac{a^2}{\bar \zeta_{1n}}} \right) \right) - \frac{i\Gamma_{2n}}{2\pi} \frac{\partial}{\partial \zeta} \left( \ln \left( \frac{\zeta - \zeta_{2n}}{\zeta - \frac{a^2}{\bar \zeta_{2n}}} \right) \right) = 0, \;\; \text{for} \;\; \zeta \rightarrow \pm a,
\end{equation}
where $w_{\zeta_{n-1}}(\zeta)$ represents the complex potential without new vortices $1n$ and $2n$. With $\zeta_{1n}$ and $\zeta_{2n}$ given, we can solve for $\Gamma_{1n}$ and $\Gamma_{2n}$ from Equation~\ref{e:kutta-condition2-largeA}.

As $\zeta_{1n}$ and $\zeta_{2n}$ are actually unknown and need to be calculated at each time step, the next focus is to find a model to determine the locations of the newly added vortices. This is of course done by first calculating the corresponding $z_{1n}$ and $z_{2n}$ in the physical plane. While a traditional way is to place the shed vortex tangential to the shedding edge and at the spot where the shedding edge was located in the previous time step, an improved approach has been used by previous researchers \cite{AnsariSA:06b, Mason:03a} and has been proven to yield decent performance. Ansari et al. \cite{AnsariSA:06b} placed the vortex at $1/3$ of the distance from the shedding edge to the previous vortex while Mason \cite{Mason:03a} placed the vortex at $1/3$ of the arc from the shedding edge to the previous vortex. We hereby adopt Mason's method because it enforces the direction of the shear layer to be tangential to the shedding edge. The use of the $1/3$ distance or the $1/3$ arc can be illustrated in Figure~\ref{fig:Kutta_plot}(a). Basically, the discrete point vortices are representative of the vortex sheet shedding from the leading or trailing edge. Therefore, each point vortex is actually a concentrated vortex sheet element with some length $\delta z_{n}$. With the assumption that the vortex sheet is continuous and the length of the vortex element does not change over single time step, it is reasonable to conclude that the shed vortex should be placed near $\delta z_{n}/2$ at the $n^{\text{th}}$ time step and the vortex center should move away about $\delta z_{n}$ at the next time step. Therefore, the new vortex is located at about $1/3$ of the distance or $1/3$ of the arc along the vortex sheet to the previous shed vortex from the shedding edge. The validity of this approach indicates that the convection of the flow near the vortex shedding edge actually determines the rate of vorticity feeding of the shear layer.
  
\subsection{Treatment for Small Angles of Attack}

The flow field for a large angle of attack case can be characterized by a large attached LEV, the thickness of which is comparable to the chord as shown in Figure~\ref{fig:Kutta_plot}(a). However, the flow pattern is dramatically different for a small angle of attack case as shown in Figure~\ref{fig:Kutta_plot}(b). Typically, the flow structure of the wake behind a flat plate with small angle of attack can be characterized by an attached flow at the leading edge or a thin separation bubble after the leading edge instead of a large leading edge vortex. The authors believe that a different leading edge treatment than the Kutta-like condition should be implemented in order to capture the physics of this flow. This is because the intensity and location of the shedding vortex are essentially determined by the feeding vorticity in the shear layer at the leading edge which is different for small and large angles of attack. As indicated previously, there is a reverse flow at the bottom of the leading edge for large angles of attack that generates a shear layer tangential to the leading edge. However, for the case of small angles of attack, this reverse flow does not exist as the lower adverse pressure gradient inside the boundary layer is unable to overcome the viscous effects. Consequently, the dominant shear flow at the leading edge follows the streamwise direction and leaves the leading edge on the top side of the flat plate almost in the tangential direction to the leading edge surface. For the case of the square leading edge in Figure~\ref{fig:Kutta_plot}(b), this is perpendicular to the plate chord; if the leading edge is rounded, the flow could remain attached and the leading edge vortex shedding location might be pushed rearward, e.g. refer to Lipinski et al. \cite{Mohseni:08l} for the vortex shedding of an airfoil at low angle of attack. With the direction of the shear layer decided, the same `$1/3$ arc' approach can be applied to calculate the placement of the vortex as shown in Figure~\ref{fig:Kutta_plot}(b). 

To determine the circulation of the newly added vortex near the leading edge, the previous used Kutta-like condition (Equation~\ref{e:kutta-condition2-largeA}) should not be used due to the small angle of attack. Therefore, a novel simple model is presented here to estimate the circulation from a 2D vortex sheet model. Basically, consider a 2D vortex sheet shed by the leading edge with elemental vorticity of $dU/d\delta$, where $d\delta$ represents the thickness of the vortex sheet. Integrating the elemental vorticity over the area of the vortex sheet gives the approximate total circulation generated during $dt$ to be $dU^2dt/2$. Here, $dU$ is determined by relating the velocity gradient to the convection velocity of the vortex sheet through a simple mass conservation of the fluid contained in the vortex sheet. The performance of this approximation of circulation will be verified in the next section.

\section{Lift Validation and Comparison}

In this section we compare our model with existing studies for two cases including a starting flat plate and a pitching flat plate problems.

\subsection{Model Validation for a Starting Plate Problem}

The objective of this section is to validate the flow field evolution as well as the lift calculations based on our multi-vortices model. The validation is done by simulating the experimental study of a flat plate start up problem done by Dickinson $\&$ Gotz \cite{Dickinson:93a}. The parameters of the physical problem are: the chord length of the flat plate is $5$ cm and the angle of attack is fixed at certain angles for each test case. The background flow accelerates at a rate of $62.5$ \text{$cms^{-2}$} from rest and reaches a steady-state velocity of $10$ \text{$cms^{-1}$} in $0.16$ s. The flat plate is brought to rest after 7.5 chord lengths of travel. Experimental runs were carried out for angle of attack ranging from $-9^o$ to $+90^o$ in increments of $4.5^o$. The Reynolds number for this experiment is $192$, which is evaluated based on the chord and the steady flow velocity. 

\begin{figure}
\begin{center}

\begin{minipage}{0.15\linewidth}\begin{center}  \end{center}
\end{minipage}
\begin{minipage}{0.35\linewidth}\begin{center} \textbf{Experiment}  \end{center}
\end{minipage}
\begin{minipage}{0.45\linewidth}\begin{center} \textbf{Simulation}  \end{center}
\end{minipage}\\

\vspace{2mm}

\begin{minipage}{0.12\linewidth}\begin{center} \textbf{(a) $s/c=1$}  \end{center}
\end{minipage}
\begin{minipage}{0.4\linewidth} \begin{center}
\includegraphics[width=.99\linewidth]{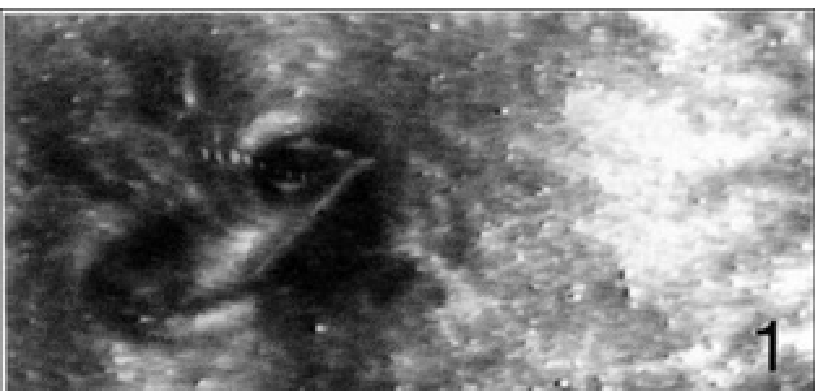}
\end{center} \end{minipage}
\begin{minipage}{0.4\linewidth} \begin{center}  
\includegraphics[width=.9\linewidth]{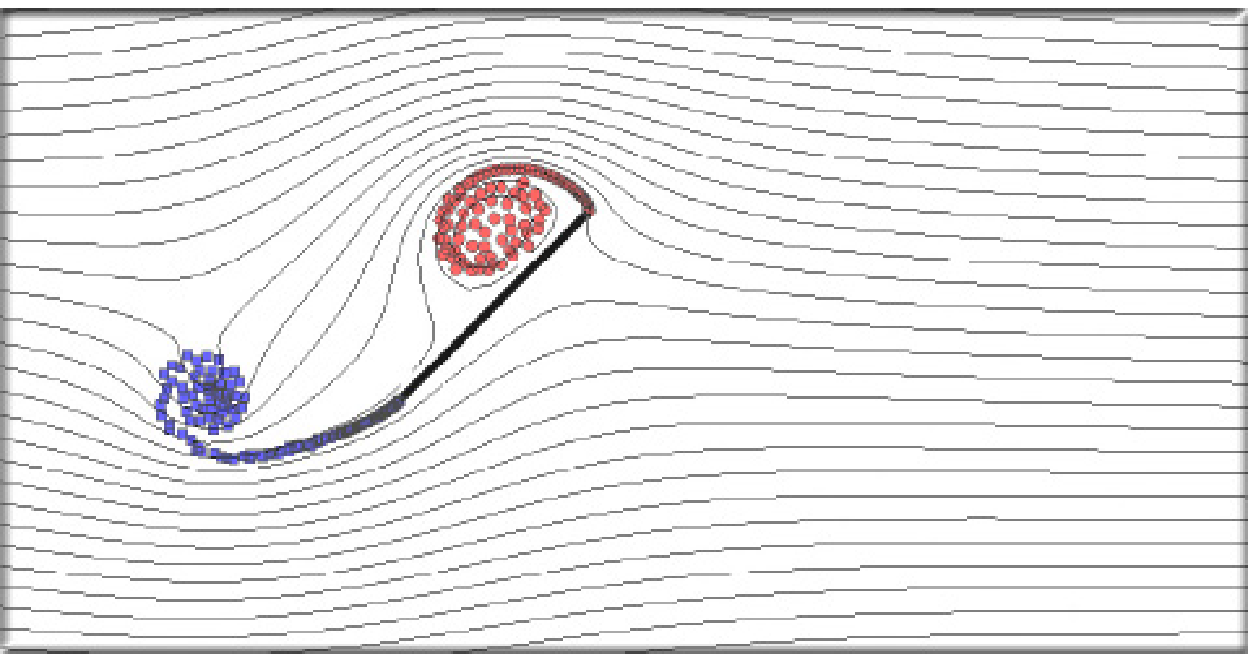}
\end{center} \end{minipage}\\

\vspace{1mm}

\begin{minipage}{0.12\linewidth}\begin{center} \textbf{(b) $s/c=2$}  \end{center}
\end{minipage}
\begin{minipage}{0.4\linewidth} \begin{center}
\includegraphics[width=.99\linewidth]{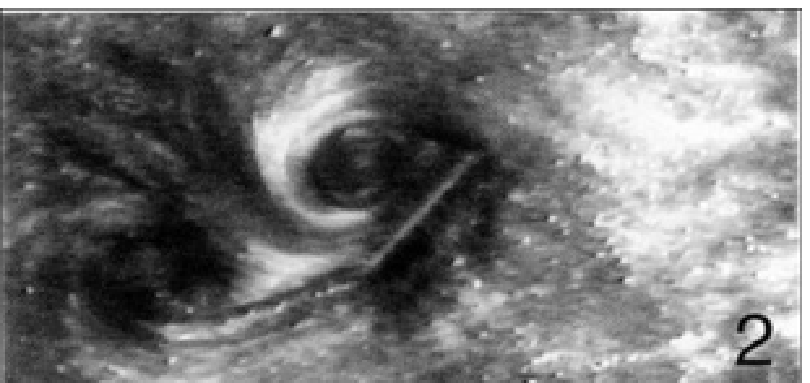}
\end{center} \end{minipage}
\begin{minipage}{0.4\linewidth} \begin{center}  
\includegraphics[width=.9\linewidth]{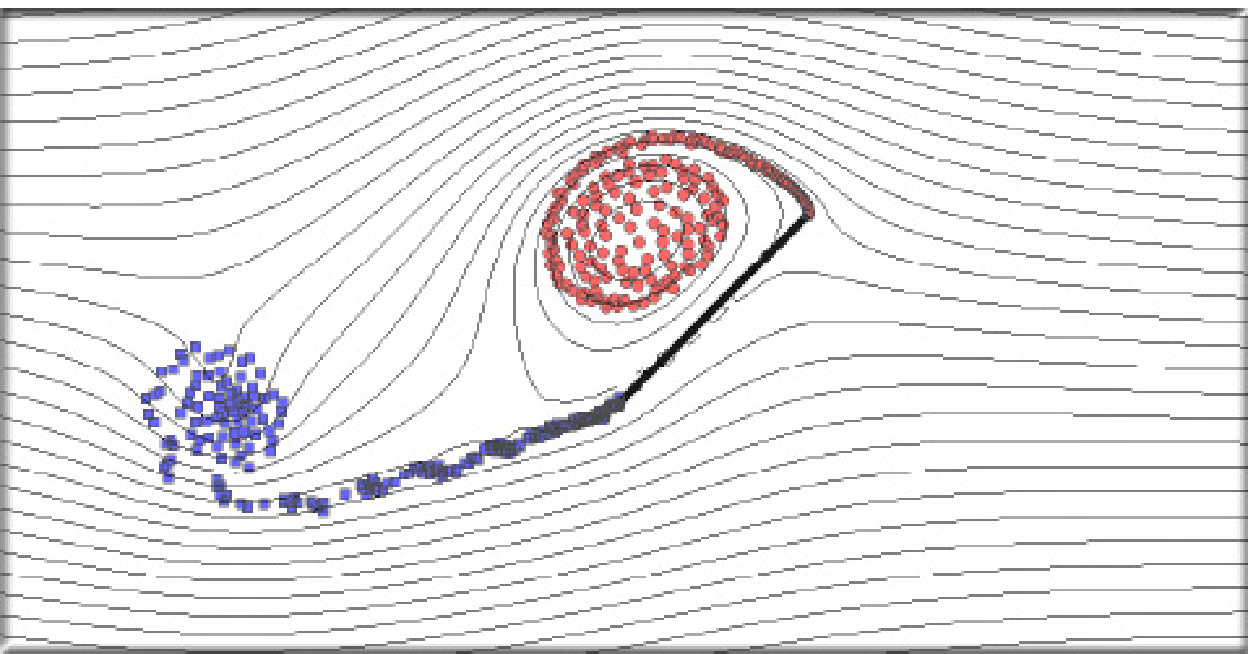}
\end{center} \end{minipage}\\

\vspace{1mm}

\begin{minipage}{0.12\linewidth}\begin{center} \textbf{(c) $s/c=3$}  \end{center}
\end{minipage}
\begin{minipage}{0.4\linewidth} \begin{center}
\includegraphics[width=.99\linewidth]{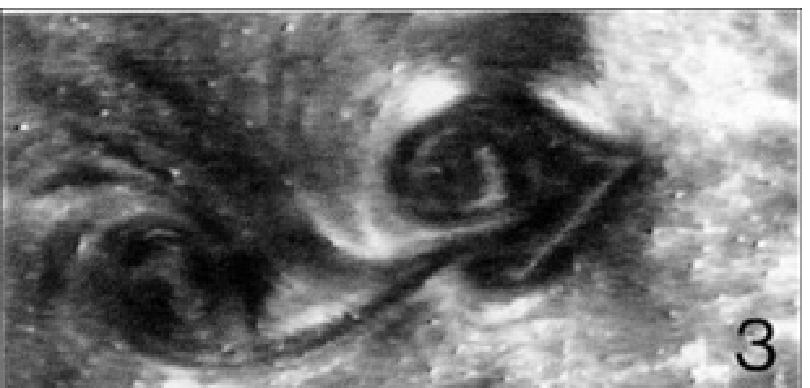}
\end{center} \end{minipage}
\begin{minipage}{0.4\linewidth} \begin{center}  
\includegraphics[width=.9\linewidth]{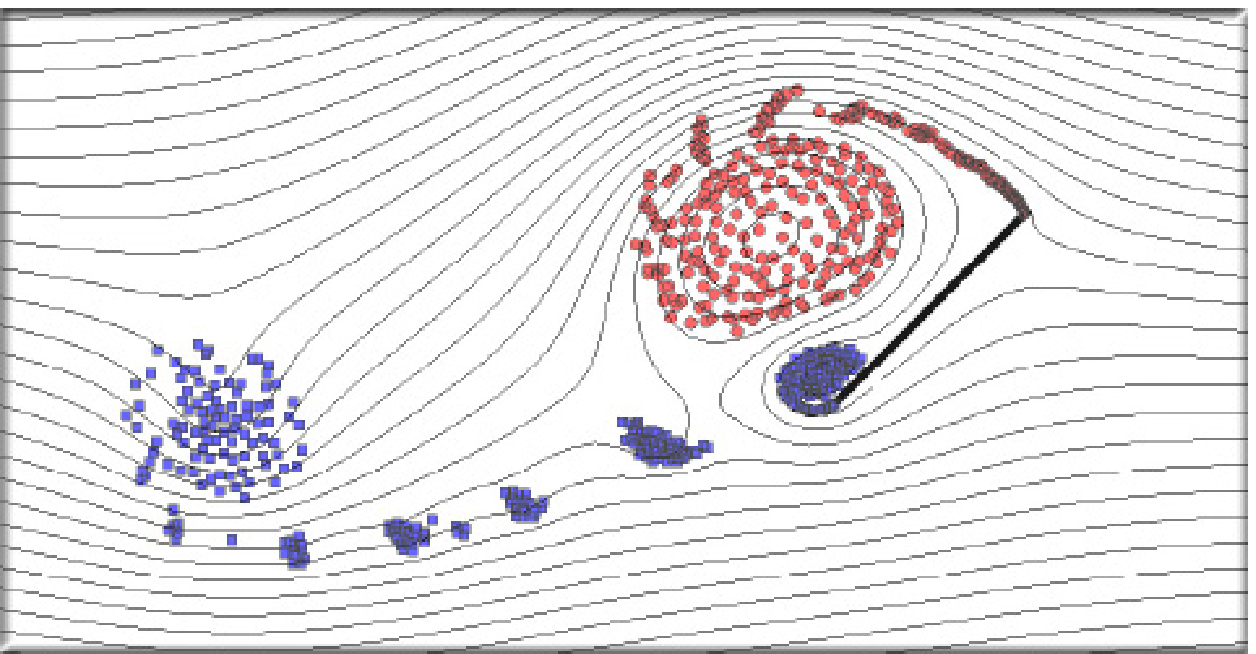}
\end{center} \end{minipage}\\

\vspace{1mm}

\begin{minipage}{0.12\linewidth}\begin{center} \textbf{(d) $s/c=4$}  \end{center}
\end{minipage}
\begin{minipage}{0.4\linewidth} \begin{center}
\includegraphics[width=.99\linewidth]{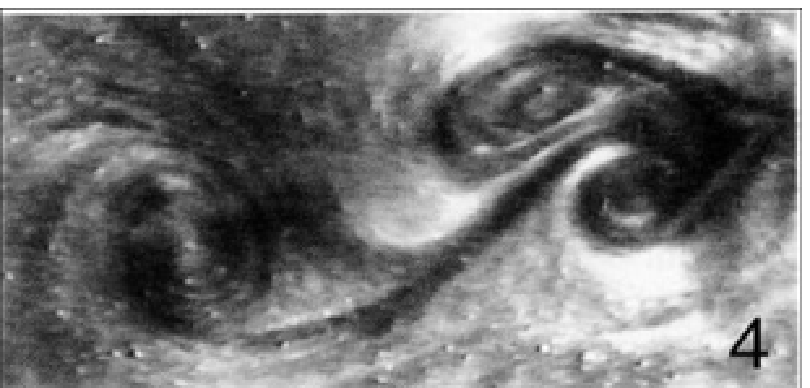}
\end{center} \end{minipage}
\begin{minipage}{0.4\linewidth} \begin{center}  
\includegraphics[width=.9\linewidth]{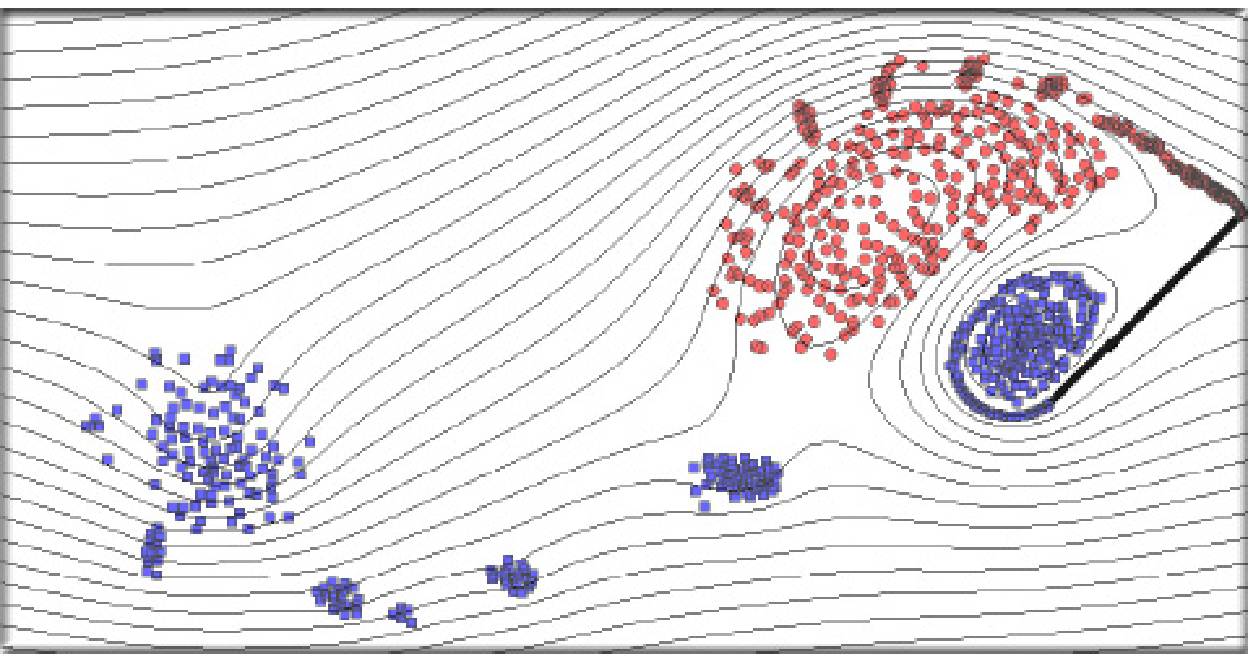}
\end{center} \end{minipage}\\

\vspace{0mm}

\caption{\small Comparison of the model in this manuscript with experimental results by Dickinson $\&$ Gotz \cite{Dickinson:93a} at $45^o$ angle of attack. (a)(b)(c)(d) correspond to different distances traveled by the flat plate. $s$ is the actual distance traveled by the flat plate.} \label{fig:Flowfield45}\vspace{-3mm}

\end{center}
\end{figure}

The flow-visualization images for the experimental case with angle of attack of $45^o$ were presented in their paper and shown in the left column of Figure~\ref{fig:Flowfield45} as a reference for comparison. The time snap-shot images correspond to the distance travelled from $1$ to $4$ measured in chord length. The right column presents the corresponding simulation snap shots predicted by the current model. As observed from the figure, vortex shedding behaviors and flow patterns match nicely between this model and Dickinson $\&$ Gotz's experiment. This qualitatively validates the accuracy of the multi-vortices model.

\begin{figure}
\begin{center}
\scalebox{0.5}{\includegraphics{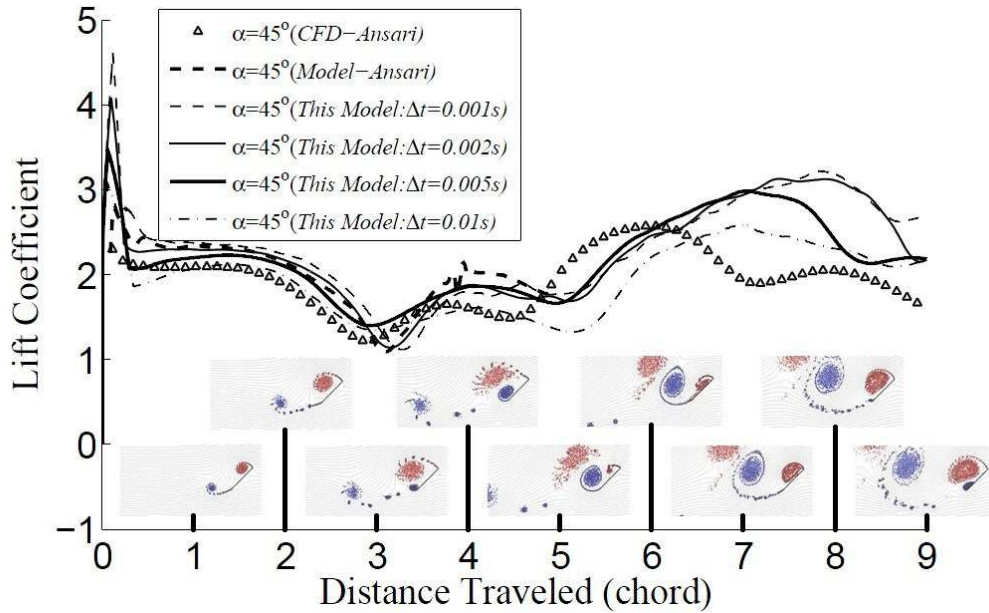}}
\vspace{0mm}
\caption{\small Lift coefficient vs. chord lengths of travel for a starting flat plate at $45^o$ angle of attack. Different results obtained from different time steps are compared to Ansari's CFD and aerodynamic model results \cite{AnsariSA:07a}. Images shown at the bottom of the figure show the evolution of flow field and vortex structures corresponding to each instant of time. As indicated from the force equation in this study, for a leading edge vortex ($\Gamma_v<0$) a motion in the positive streamwise direction ($\dot{\xi_v}>0$) will create positive lift, while for a trailing edge vortex ($\Gamma_v>0$) a motion in the positive streamwise direction ($\dot{\xi_v}>0$) will create negative lift. This indicates that LEV shedding leads to a decrease of lift while TEV shedding increases lift. This is also readily verified in this figure by relating the lift coefficient trend to the corresponding vortices structures. e.g. From $0-2$ chord lengths of travel, a constant positive lift is created due to TEV shedding and the relative stabilization of LEV on the flat plate.}
\label{fig:Lift45}
\end{center}
\end{figure}

Next, we will evaluate the proper selection of time-step by comparing the predicted lift coefficient for the case of $45^o$ angle of attack, with the results from Ansari's CFD and aerodynamic model \cite{AnsariSA:07a}. The lift coefficient is evaluated using Equation~\ref{e:Lift-Drag2}. Note that $F_y$ is the force in the $y$-direction; the actual lift should be computed from $F_x+iF_y$ based on the direction of the incoming flow. Generally, for all simulations involving time evolution, a motion with a higher Reynolds number requires a finer time resolution to guarantee accurate solutions. However, smaller time step also translates into higher computational cost. Here, to find a reasonable time step which will yield good accuracy while preserving the simulation efficiency, several time steps are attempted and the simulation results are shown for the time steps of $0.01$ s, $0.005$ s, $0.002$ s and $0.001$ s. We compared these lift calculations with those predicted by Ansari's CFD and theoretical model in Figure~\ref{fig:Lift45}. All cases match the CFD results reasonably well up to three chord lengths travelled. After $5$ chord lengths of travel, all model predictions demonstrate some `delay' behavior compared to CFD which might be caused by the lingering of the newly generated TEV in an inviscid flow model. It can be further observed that this model starts to show time convergence when time step equals $0.005$s, while the converged solution seems to have higher magnitude of lift compared to CFD. This is also one of the cons of the inviscid model, in which the flow velocity close to the flat plate is not bounded. This effect is especially profound when the time step becomes smaller. In this study, the case with time step of $0.005$ s shows both convergence and computational economy while matching with CFD results; it is therefore preferable to pick $0.005$ s as the time step for the following simulations.

Before validating this lift calculations by comparing with the experimental cases for different angles of attack, there is an important implication which needs to be pointed out regarding Equation~\ref{e:Lift-Drag2}. Basically, what this equation indicates is that the force contribution from a vortex (LEV or TEV) located at $\zeta_v = \xi_v + i\eta_v$ with a circulation $\Gamma_v$ can be expressed as $F_{vx}-iF_{vy} = -i\rho \Gamma_v (\dot{\bar{\zeta_v}} + \dot{\zeta_v}a^2/\zeta_v^2)$. Since the lift calculation is of most interest here, we would like to explicitly find the relation between the velocity of the vortex and the lift generated by the motion of the vortex. Considering the motion of a vortex in the streamwise direction, without losing generality, it is convenient to assume zero angle of attack so that the lift is purely $F_{vy}$ and $\dot{\bar{\zeta_v}} = \dot{\zeta_v} = \dot{\xi_v}$. Therefore, the lift in this case can be simplified to $F_{vy} = \rho \Gamma_v \dot{\xi_v} (1+ \text{Real}(a^2/\zeta_v^2))$. As $a < \left| \zeta_v \right|$, it yields $\left| a^2/\zeta_v^2 \right|<1$ and thus $\text{Real}(a^2/\zeta_v^2)>-1$. As a result, $(1+ \text{Real}(a^2/\zeta_v^2))>0$ which means that $F_{vy}/(\Gamma_v \dot{\xi_v})>0$. This indicates that for a leading edge vortex ($\Gamma_v<0$) a motion in the positive streamwise direction ($\dot{\xi_v}>0$) will decrease the lift, while for a trailing edge vortex ($\Gamma_v>0$) a motion in the positive streamwise direction ($\dot{\xi_v}>0$) will increase the lift; thus, LEV shedding decreases lift while TEV shedding increases lift. This might be a reason why flapping flyers try to stabilize the LEV during most of the downstroke cycle. It should also be noted that simulation by Yu et al. \cite{YuY:03a} report similar conclusions. Moreover, this interpretation can be used to explain the lift coefficient variation shown in Figure~\ref{fig:Lift45} by analyzing the flow field snap shots of the vortex evolution behavior. From $0-2$ chord lengths of travel, a constant positive lift is created due to TEV shedding and the relative stabilization of LEV on the flat plate. From $2-3$ chord lengths of travel, the LEV is still growing while a second TEV gradually forms and sticks to the trailing edge which results in the lift decrease to its first minimum. From $3-4$ chord lengths, the second TEV grows around the trailing edge and causes a stronger leading edge shear layer which potentially would become the second leading edge vortex. This addition of stronger vorticity to the LEV results in a small increase of lift and the first lift maximum. From about $4-5$ chord lengths, the first LEV is cut by the growing second TEV and starts to shed; this decreases the lift and generates the second minimum point in the lift curve. However, after about $5$ chords, the reduction of the lift is reversed due to the shedding of the second TEV, which lasts for a relatively long time and enhances the lift so significant that it reaches its greatest magnitude. From the above analysis, it can be concluded that the generation of new LEV or TEV is responsible for the first maximum or minimum of lift, while the shedding of LEV or TEV generates the second minimum or maximum of lift. 

\begin{figure}
\begin{center}

\begin{minipage}{0.05\linewidth}\begin{center} \textbf{(a) $4.5^o$}  \end{center}
\end{minipage}
\begin{minipage}{0.4\linewidth} \begin{center}
\includegraphics[width=.75\linewidth,viewport=10 0 520 440,clip]{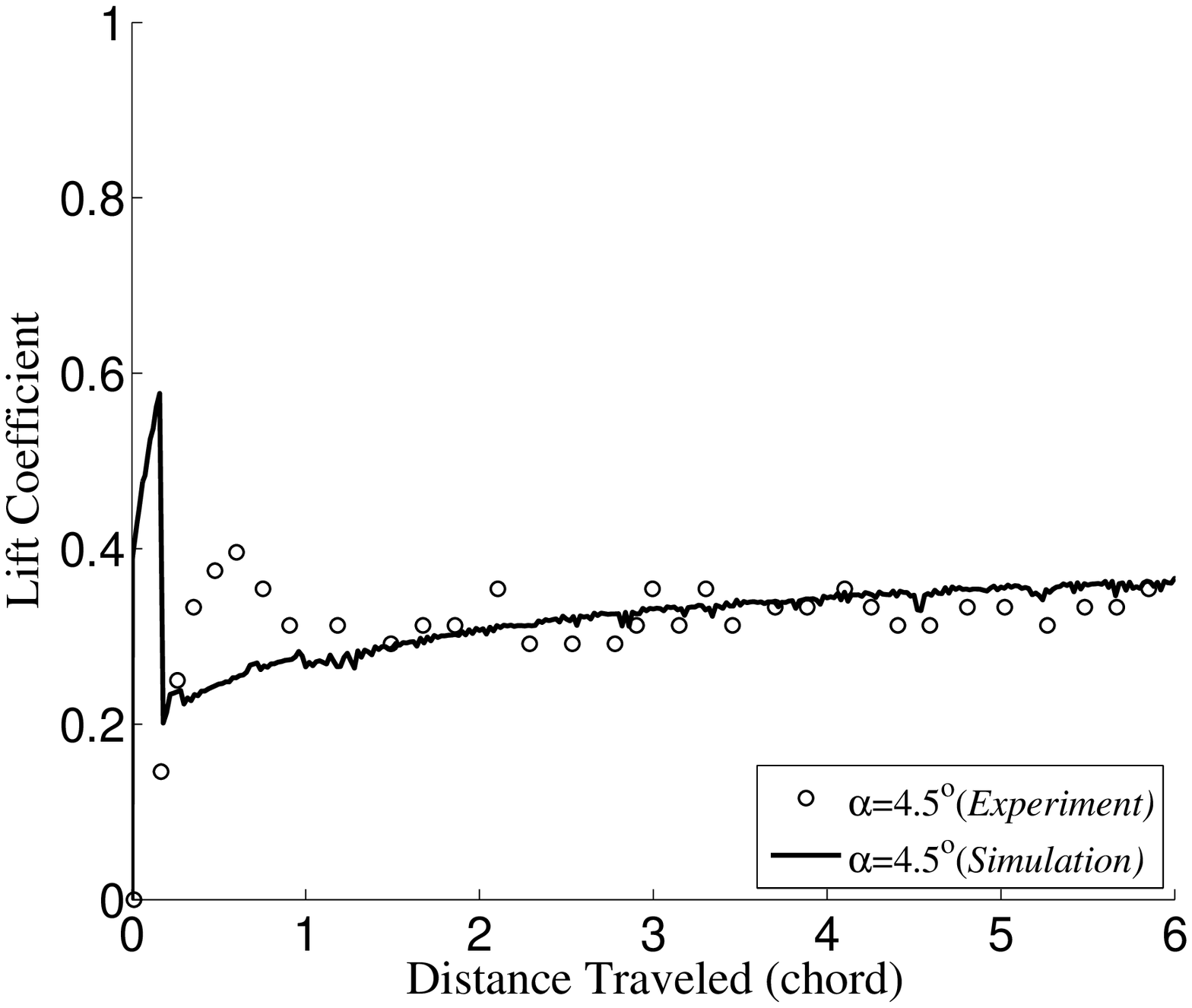}
\end{center} \end{minipage}
\begin{minipage}{0.05\linewidth}\begin{center} \textbf{(b) $9^o$}  \end{center}
\end{minipage}
\begin{minipage}{0.4\linewidth} \begin{center}  
\includegraphics[width=.75\linewidth,viewport=10 0 520 440,clip]{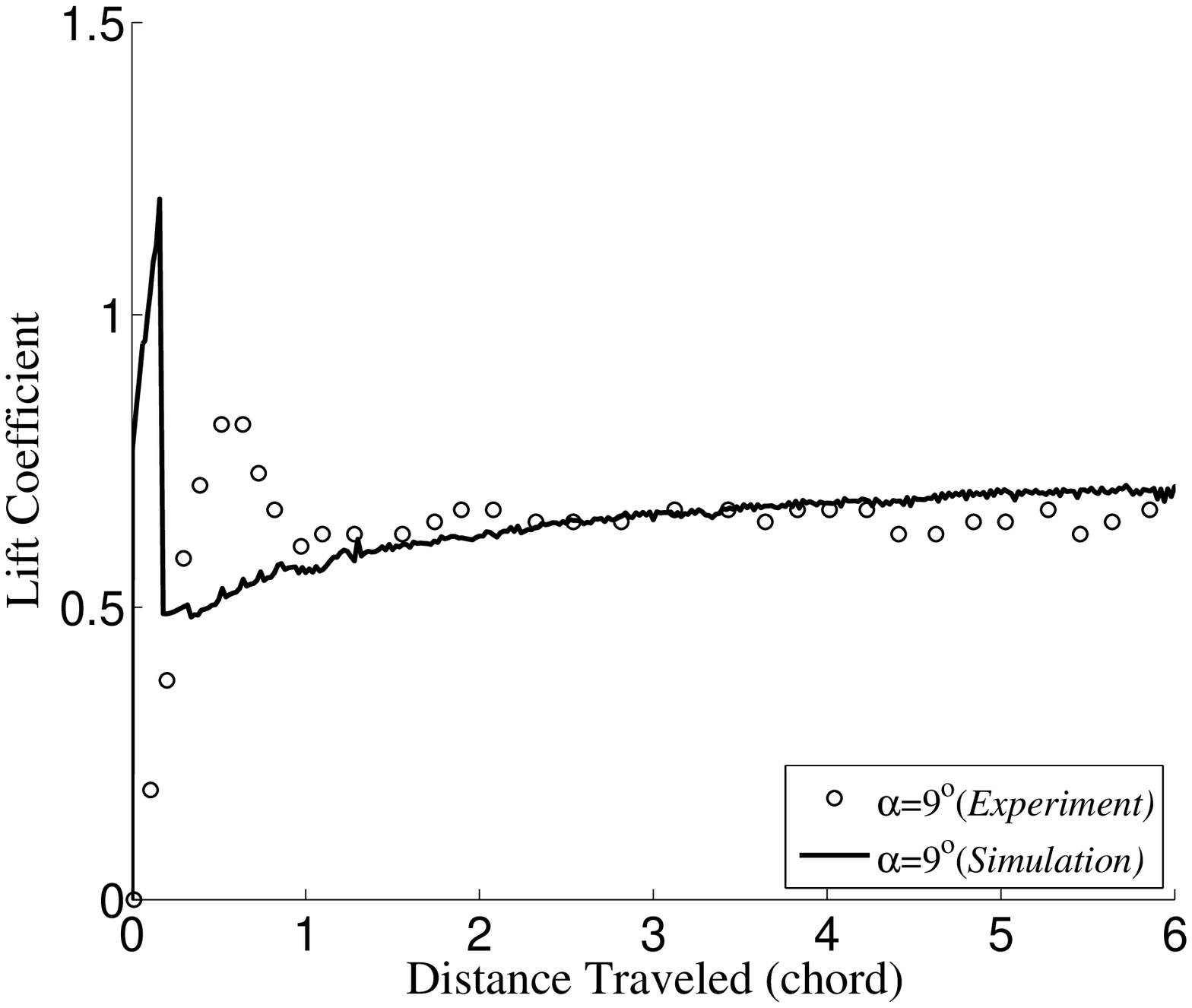}
\end{center} \end{minipage}\\
\vspace{1mm}

\begin{minipage}{0.05\linewidth}\begin{center} \textbf{(c) $13.5^o$}  \end{center}
\end{minipage}
\begin{minipage}{0.4\linewidth} \begin{center}
\includegraphics[width=.75\linewidth,viewport=10 0 520 440,clip]{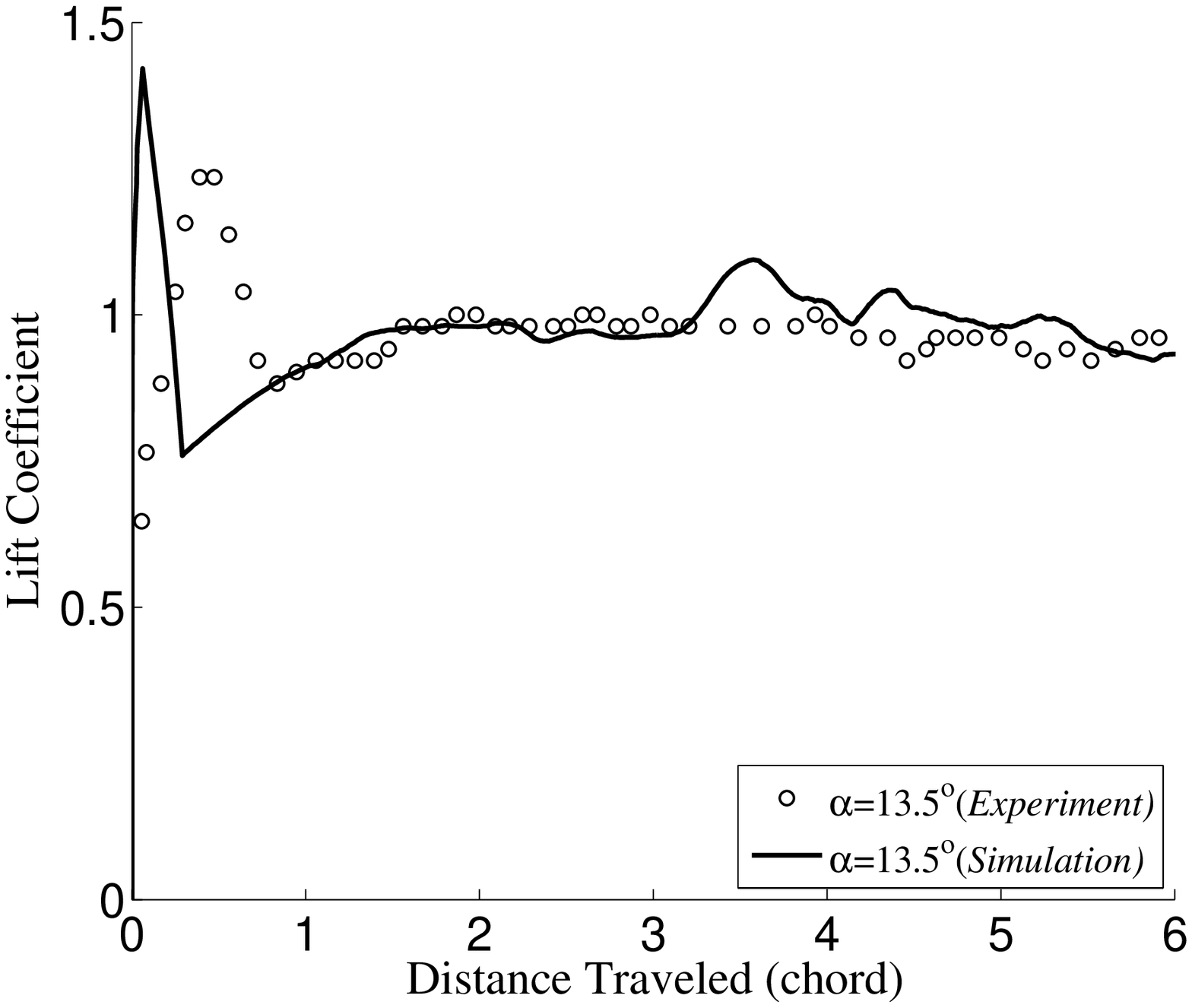}
\end{center} \end{minipage}
\begin{minipage}{0.05\linewidth}\begin{center} \textbf{(d) $18^o$}  \end{center}
\end{minipage}
\begin{minipage}{0.4\linewidth} \begin{center}  
\includegraphics[width=.75\linewidth,viewport=10 0 520 440,clip]{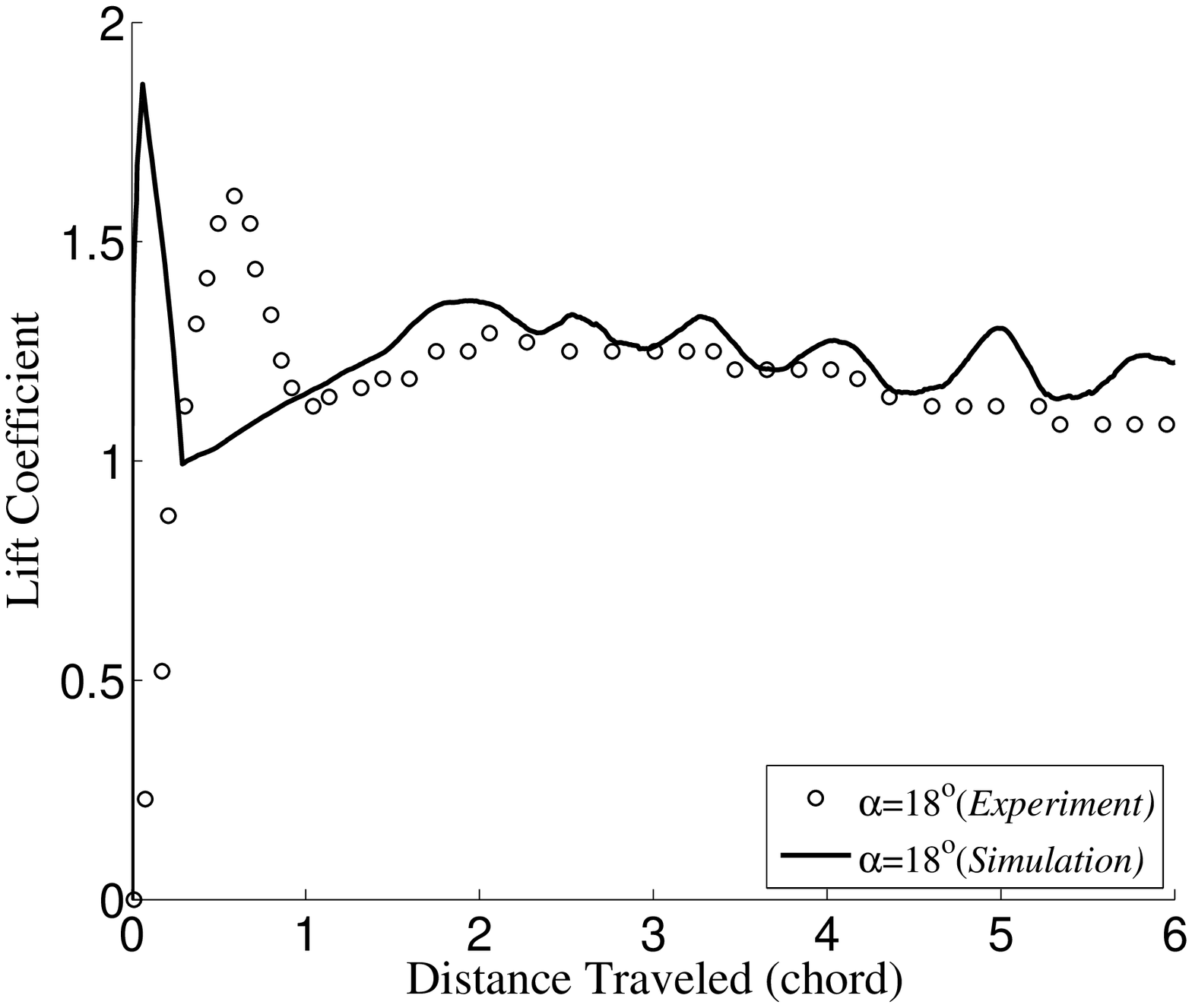}
\end{center} \end{minipage}\\
\vspace{1mm}

\begin{minipage}{0.05\linewidth}\begin{center} \textbf{(e) $22.5^o$}  \end{center}
\end{minipage}
\begin{minipage}{0.4\linewidth} \begin{center}
\includegraphics[width=.75\linewidth,viewport=10 0 520 440,clip]{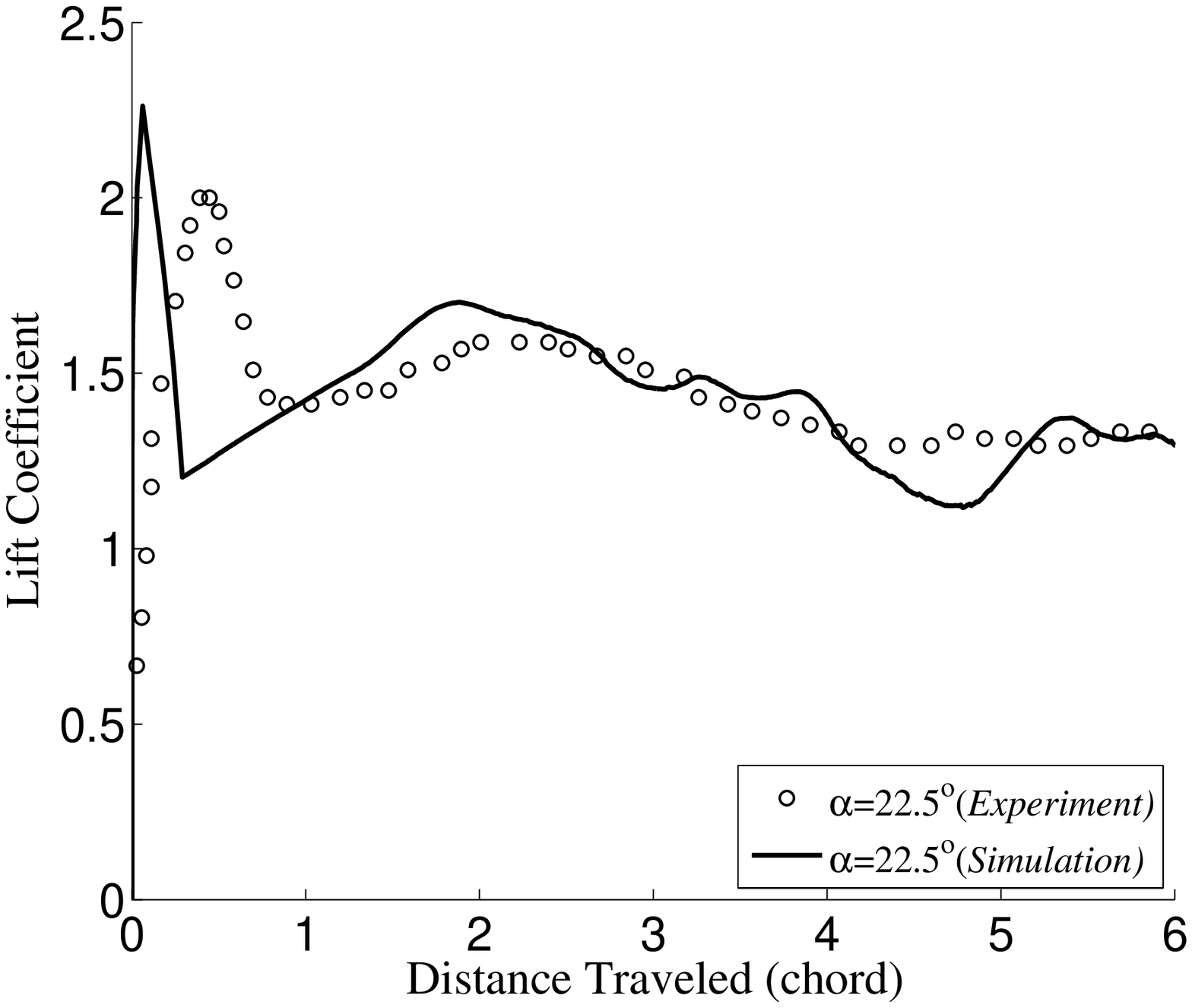}
\end{center} \end{minipage}
\begin{minipage}{0.05\linewidth}\begin{center} \textbf{(f) $27^o$}  \end{center}
\end{minipage}
\begin{minipage}{0.4\linewidth} \begin{center}  
\includegraphics[width=.75\linewidth,viewport=10 0 520 440,clip]{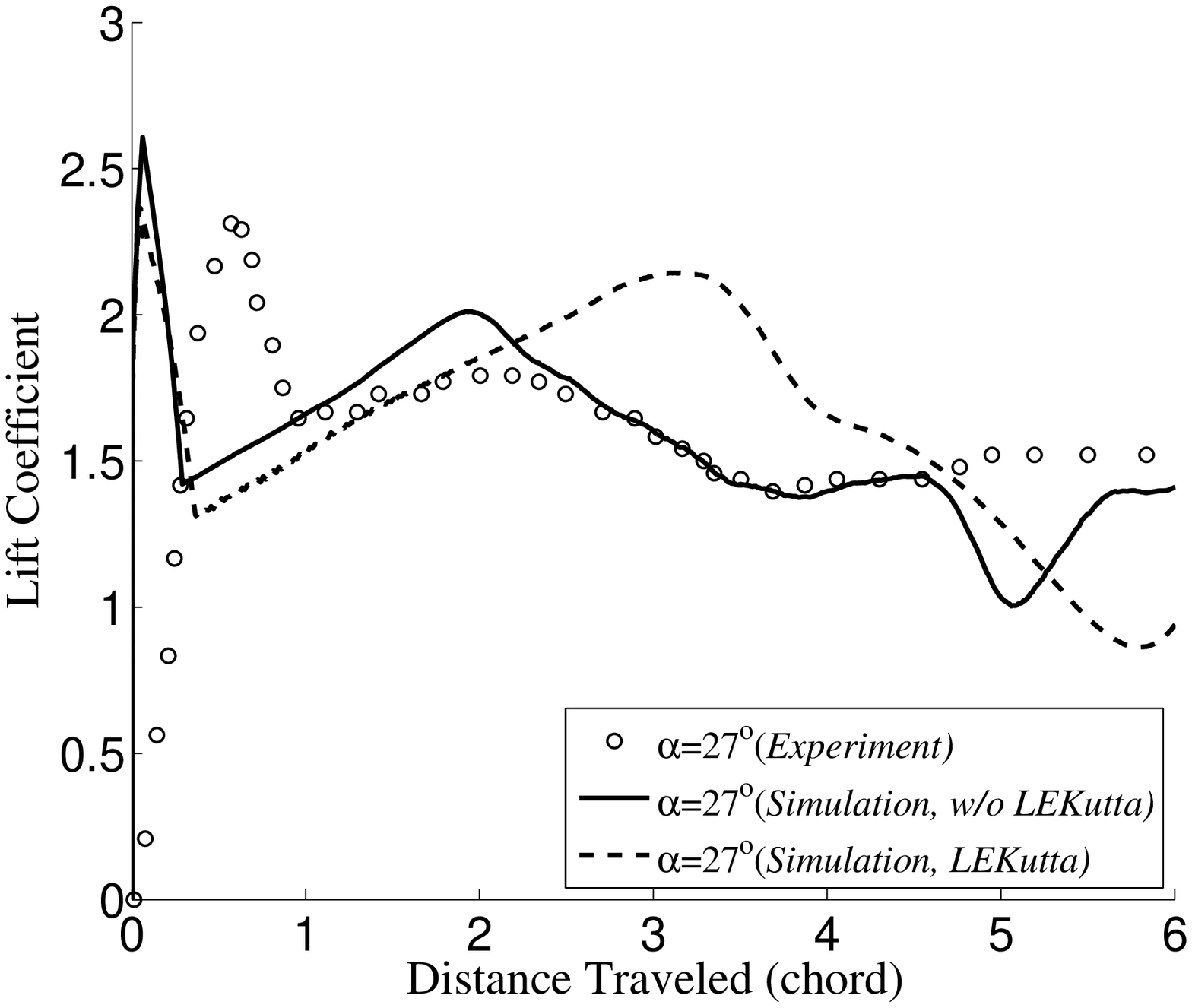}
\end{center} \end{minipage}\\
\vspace{1mm}

\begin{minipage}{0.05\linewidth}\begin{center} \textbf{(g) $31.5^o$}  \end{center}
\end{minipage}
\begin{minipage}{0.4\linewidth} \begin{center}
\includegraphics[width=.75\linewidth,viewport=10 0 520 440,clip]{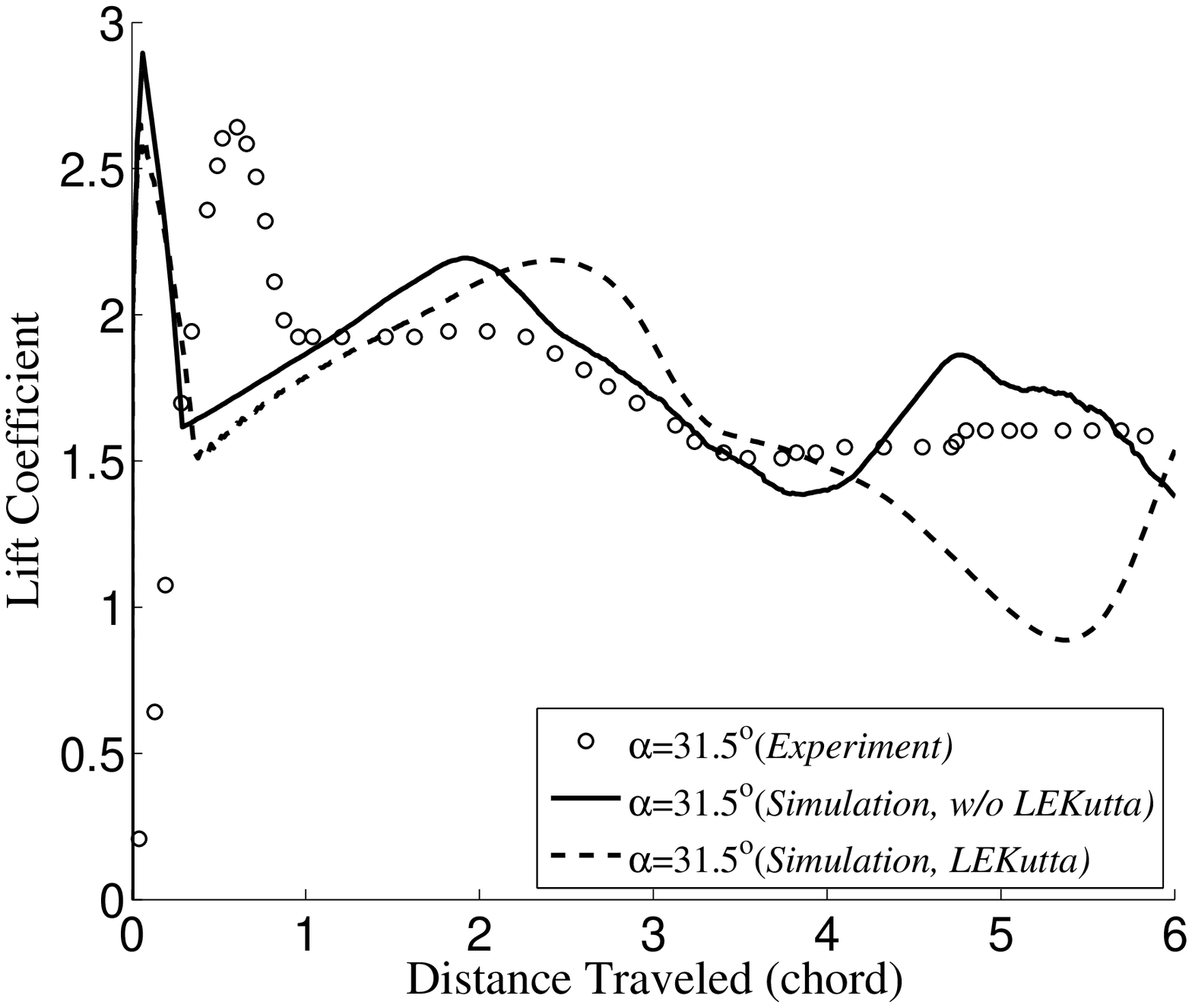}
\end{center} \end{minipage}
\begin{minipage}{0.05\linewidth}\begin{center} \textbf{(h) $36^o$}  \end{center}
\end{minipage}
\begin{minipage}{0.4\linewidth} \begin{center}  
\includegraphics[width=.75\linewidth,viewport=10 0 520 440,clip]{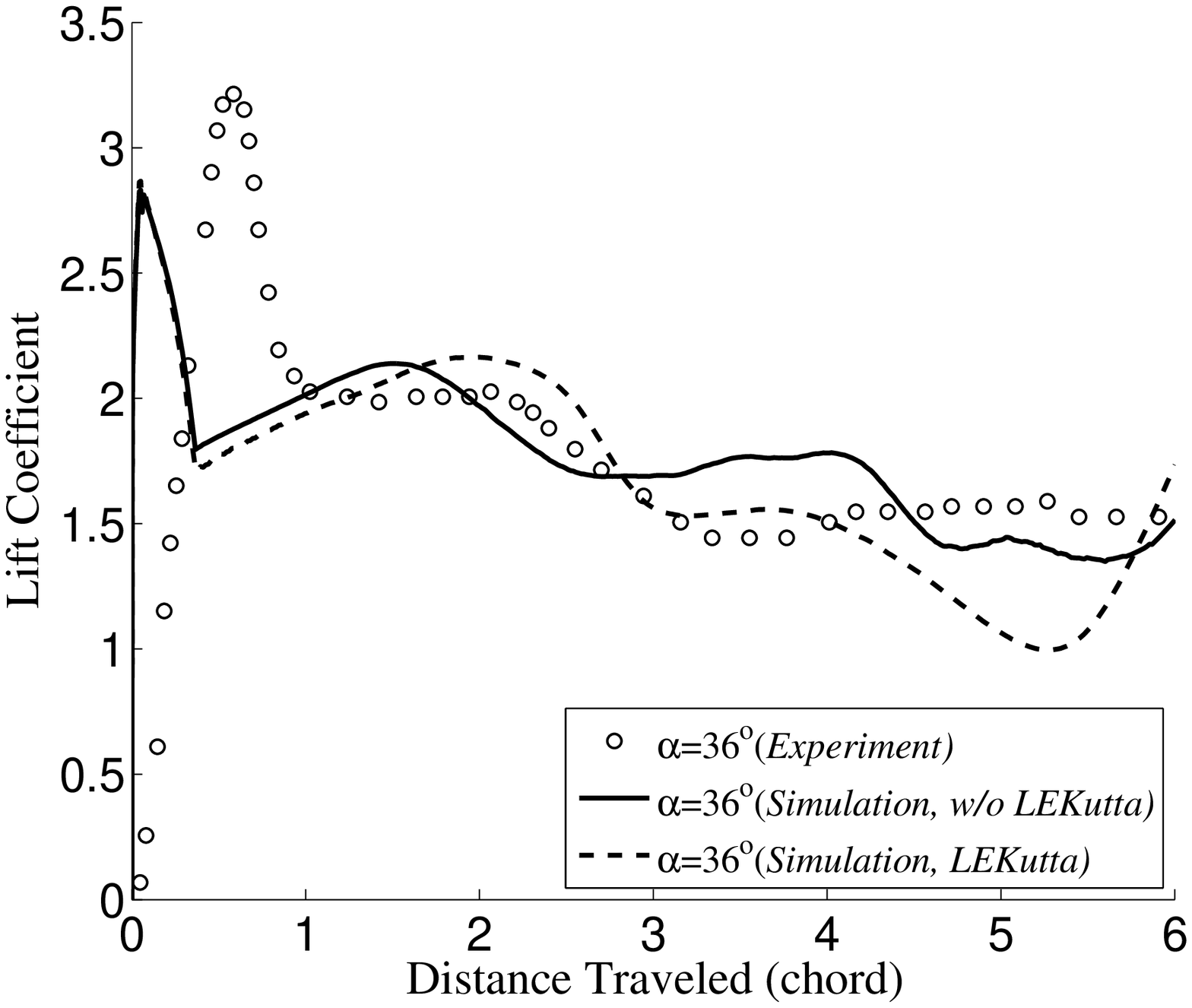}
\end{center} \end{minipage}\\
\vspace{1mm}

\begin{minipage}{0.05\linewidth}\begin{center} \textbf{(i) $40.5^o$}  \end{center}
\end{minipage}
\begin{minipage}{0.4\linewidth} \begin{center}
\includegraphics[width=.75\linewidth,viewport=10 0 520 440,clip]{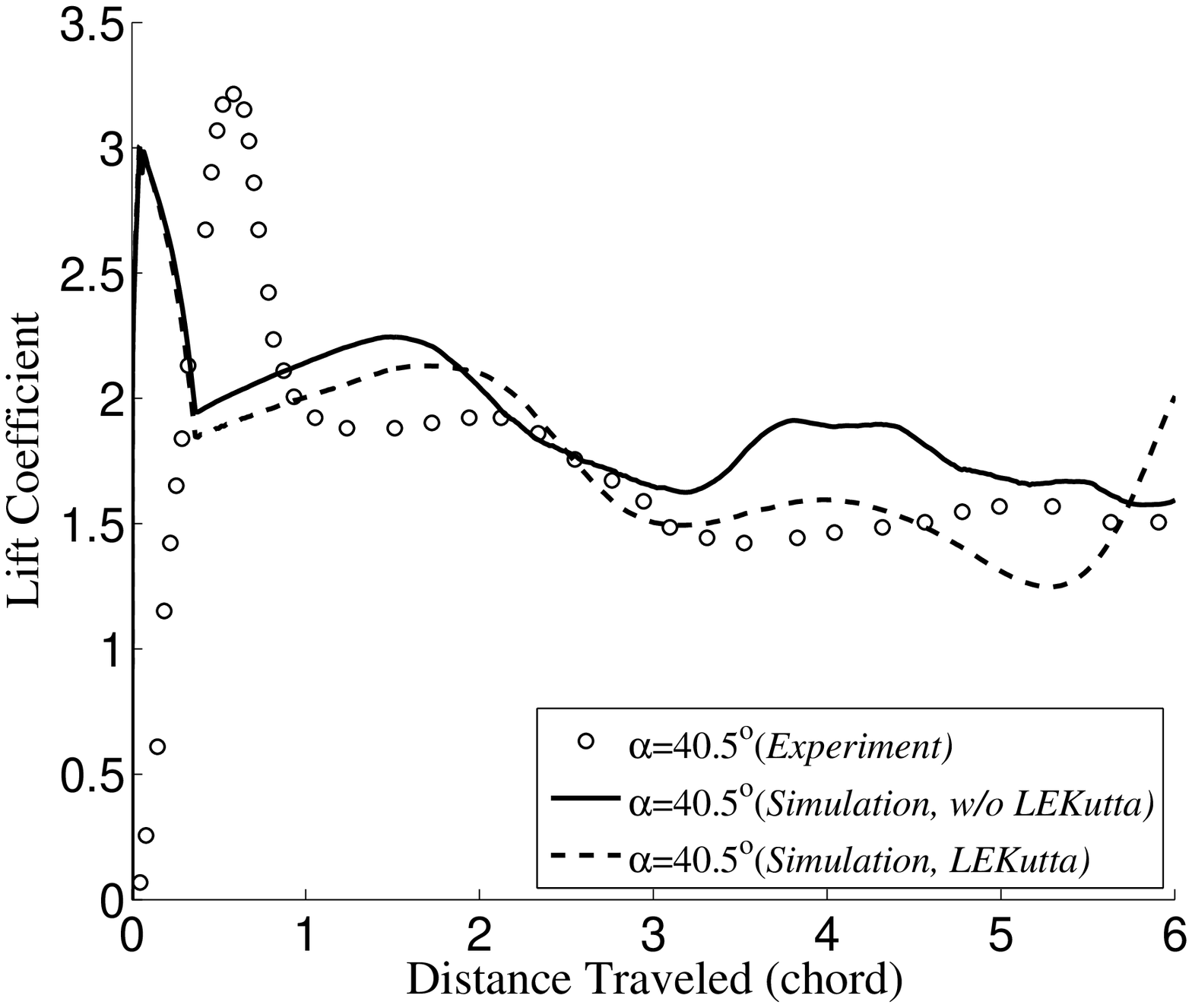}
\end{center} \end{minipage}
\begin{minipage}{0.05\linewidth}\begin{center} \textbf{(j) $45^o$}  \end{center}
\end{minipage}
\begin{minipage}{0.4\linewidth} \begin{center}  
\includegraphics[width=.75\linewidth,viewport=10 0 520 440,clip]{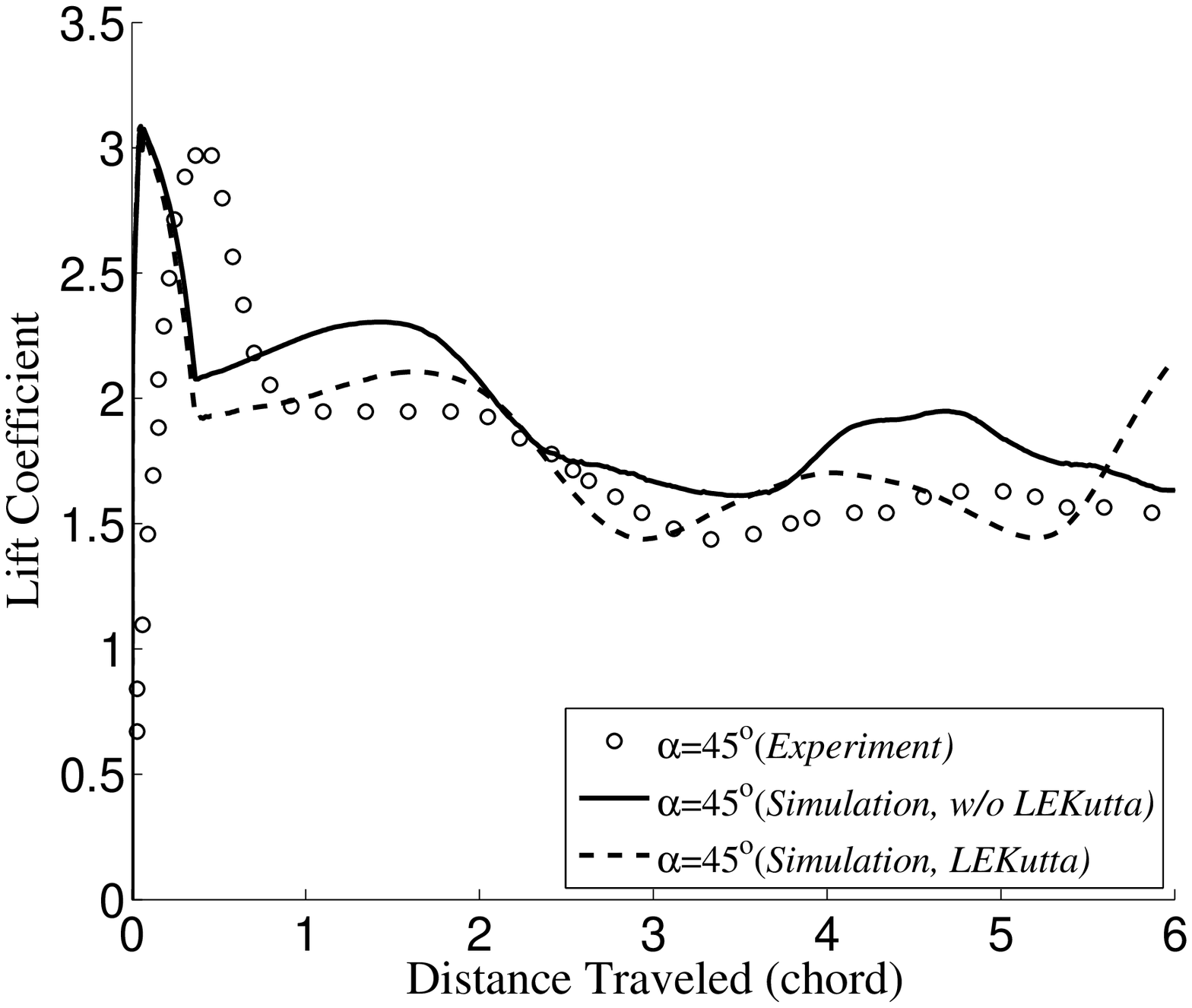}
\end{center} \end{minipage}\\
\vspace{1mm}

\caption{\small A starting flat plate. Lift coefficient compared with experiments \cite{Dickinson:93a} for angles of attack ranging from $4.5^o$ to $45^o$.} \label{fig:LiftCompareALL}\vspace{0mm}

\end{center}
\end{figure}

Next, we will validate the model and lift calculations by comparison with experimental lift coefficient data \cite{Dickinson:93a} at a variety of angles of attack. Since it is not straightforward to establish a criterion to distinguish between small and large angles of attack, we also present the results of the lift calculations using both approaches for larger angles of attack. The present criterion angle is then determined by the observation of the flow pattern and the matching with experimental lift variations. In this manuscript, $4.5^o\leq \alpha \leq 45^o$ are calculated with small angle of attack treatment while $27^o\leq \alpha \leq 45^o$ are calculated with assumption of large angle of attack. The results are compared in Figure~\ref{fig:LiftCompareALL}. Note that at $27^o\leq \alpha \leq 45^o$, both methods are tested and the results are compared with each other. We can conclude that the low angle formulation has a better performance at $\alpha=27^o$ and even at $\alpha=31.5^o$, while at $\alpha=36^o$ the advantage disappears. At $\alpha=40.5^o$ and $\alpha=45^o$, better performance from the high angle formulation can be confirmed which indicates that $\alpha=40.5^o$ is large enough to implement the Kutta condition at the leading edge. Therefore, a proper region corresponding to the transitional angles of attack here could be between $30^o \& 40^o$.

The lift calculations match nicely with experimental results for low angles of attack ranging from $4.5^o$ to $22.5^o$. There are some oscillations in the simulation results which are caused by the LE and TE shedding alternately. For high angles of attack, the magnitude of lift matches with experiment and has been shown in Figure~\ref{fig:Lift45} to match reasonably well with the CFD results presented by Ansari et al. \cite{AnsariSA:06b}.

\subsection{A Pitching Flat Plate}

This section will simulate a pitch-up, hold and pitch-down motion for a flat plate using our model and will compare the results with existing experimental, theoretical and computational results. The original experimental studies used for comparison of this case were conducted by OL \cite{OL:09a}, with the motion analytically prescribed by Eldredge et al. \cite{Eldredge:09a}. In the following discussions, four cases are simulated with the pivot about the leading edge and half-chord, and with the maximum pitch amplitudes of $25^o$ and $45^o$.

\begin{figure}
\begin{center}

\begin{minipage}{0.24\linewidth} \begin{center}
\includegraphics[width=.99\linewidth,viewport=0 0 170 120,clip]{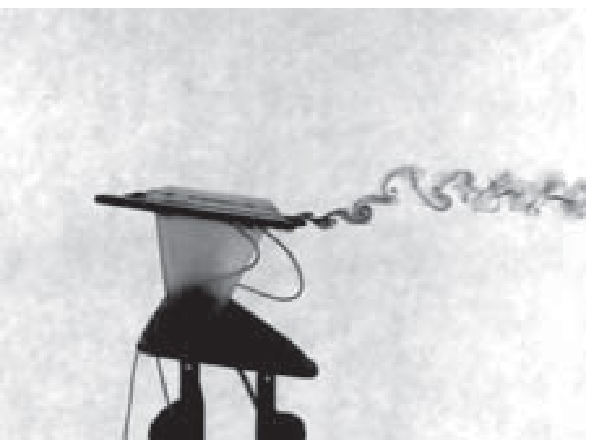}
\end{center} \end{minipage}
\begin{minipage}{0.24\linewidth} \begin{center}  
\includegraphics[width=.99\linewidth,viewport=0 0 170 120,clip]{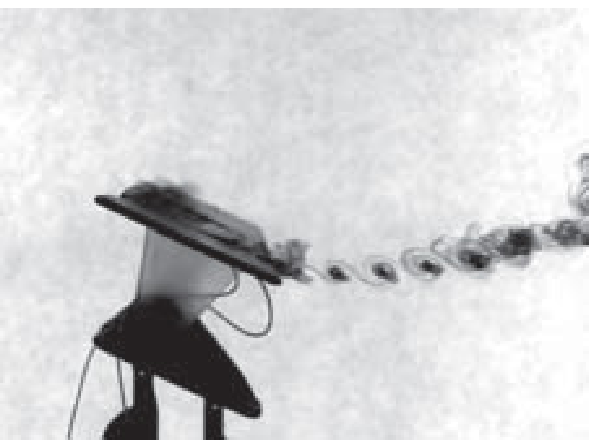}
\end{center} \end{minipage}
\begin{minipage}{0.24\linewidth} \begin{center}
\includegraphics[width=.99\linewidth,viewport=0 0 170 120,clip]{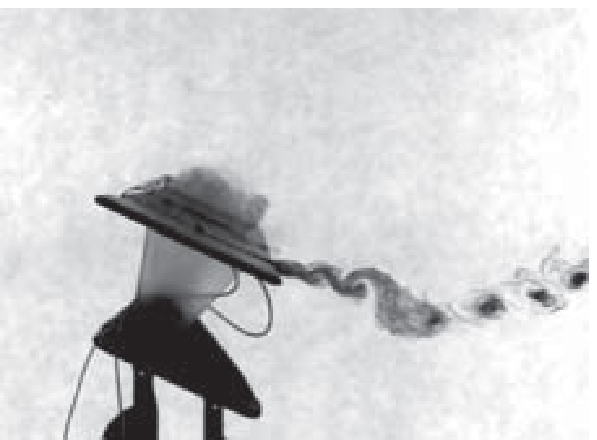}
\end{center} \end{minipage}
\begin{minipage}{0.24\linewidth} \begin{center}  
\includegraphics[width=.99\linewidth,viewport=0 0 170 120,clip]{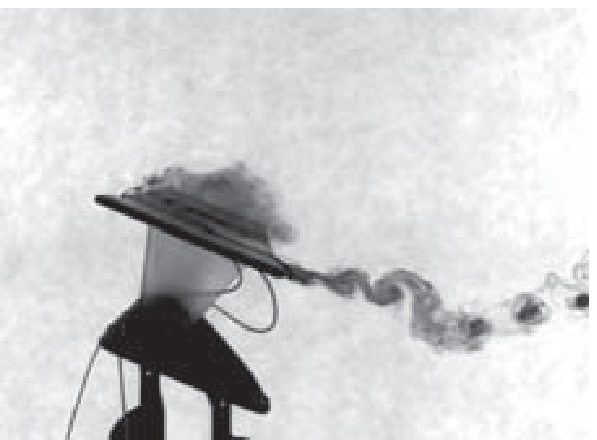}
\end{center} \end{minipage}\\

\begin{minipage}{0.24\linewidth} \begin{center}
\includegraphics[width=.99\linewidth,viewport=50 0 570 500,clip]{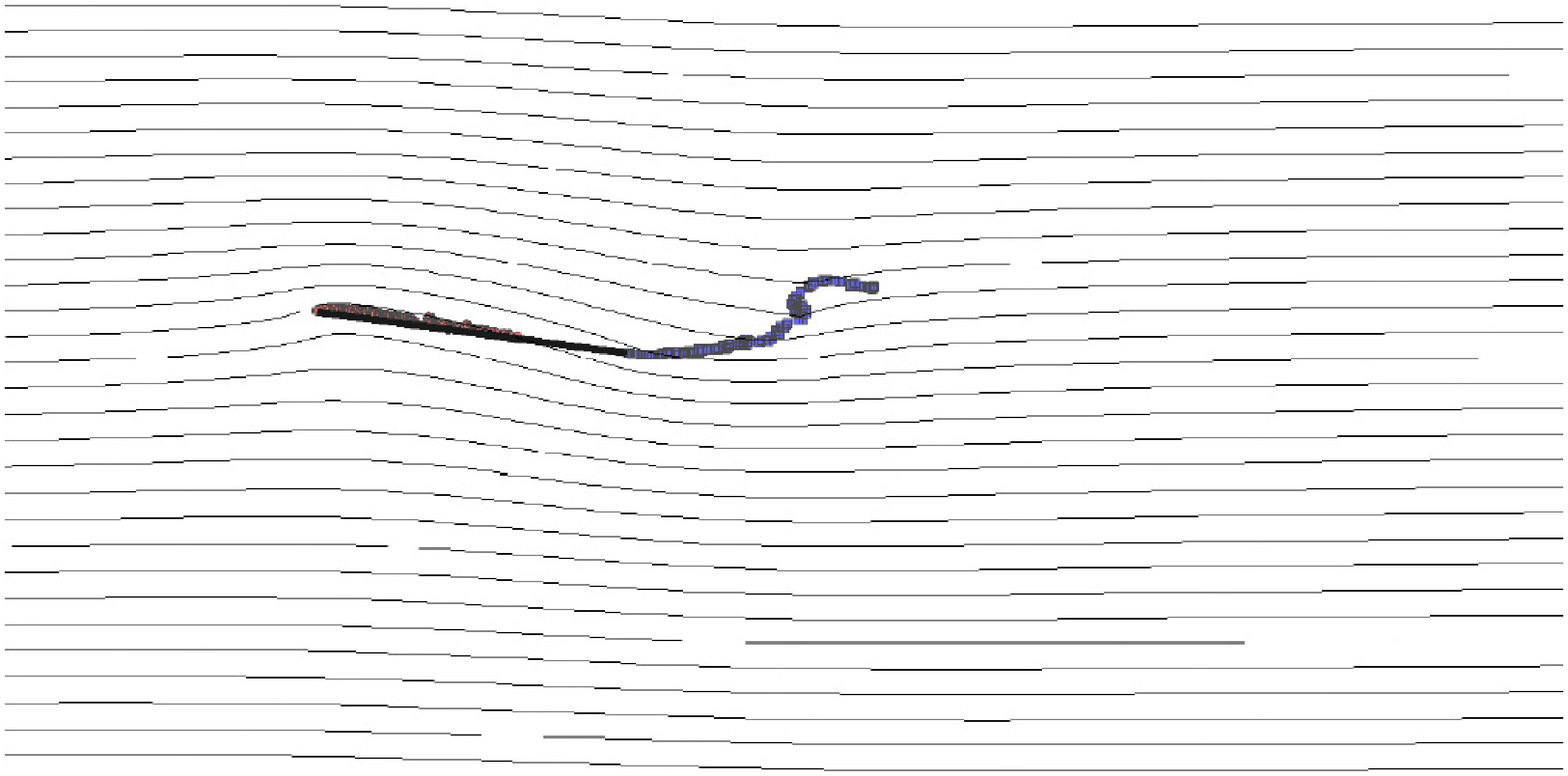}
\end{center} \end{minipage}
\begin{minipage}{0.24\linewidth} \begin{center}  
\includegraphics[width=.99\linewidth,viewport=50 0 570 500,clip]{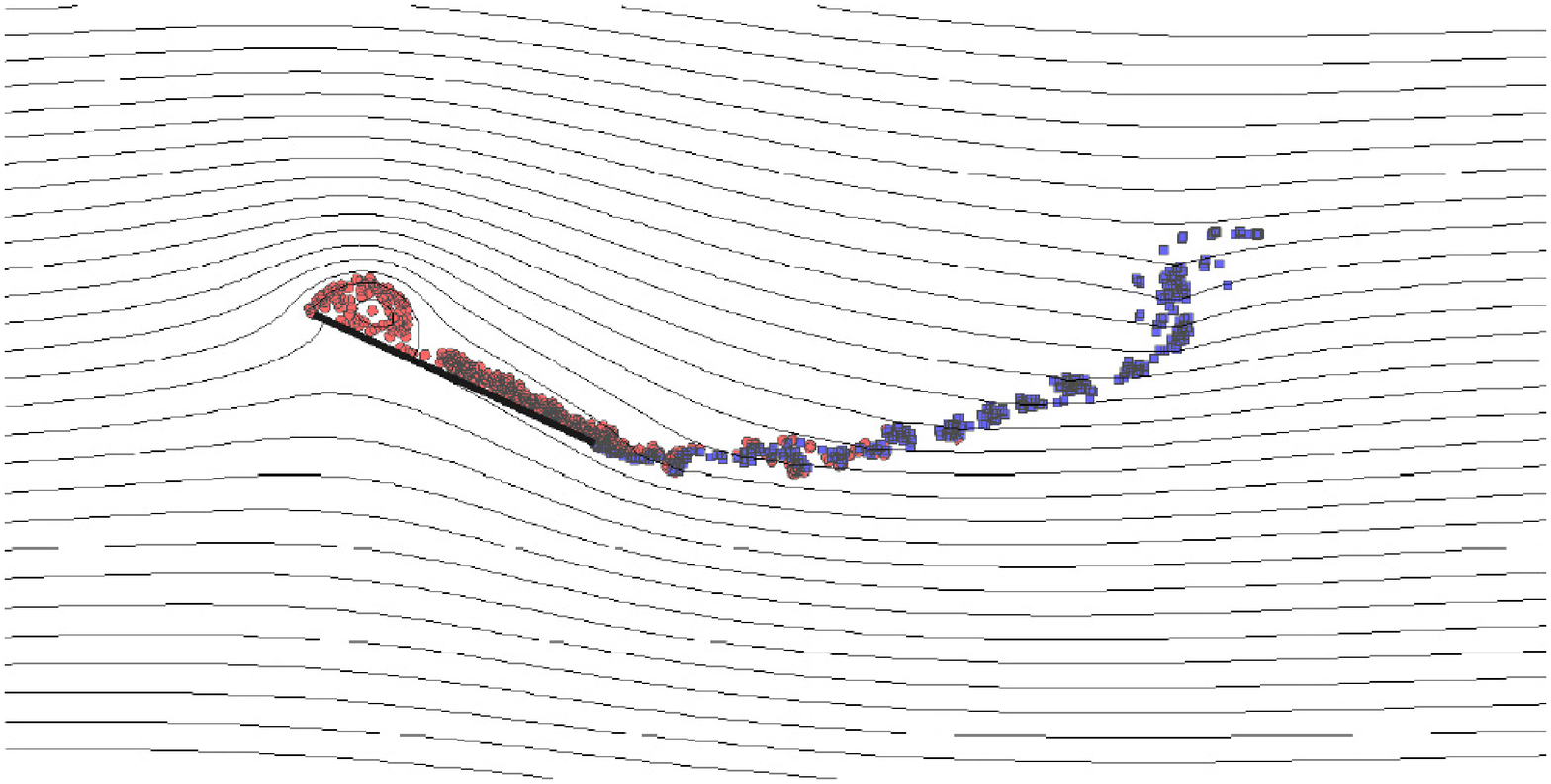}
\end{center} \end{minipage}
\begin{minipage}{0.24\linewidth} \begin{center}
\includegraphics[width=.99\linewidth,viewport=50 0 570 500,clip]{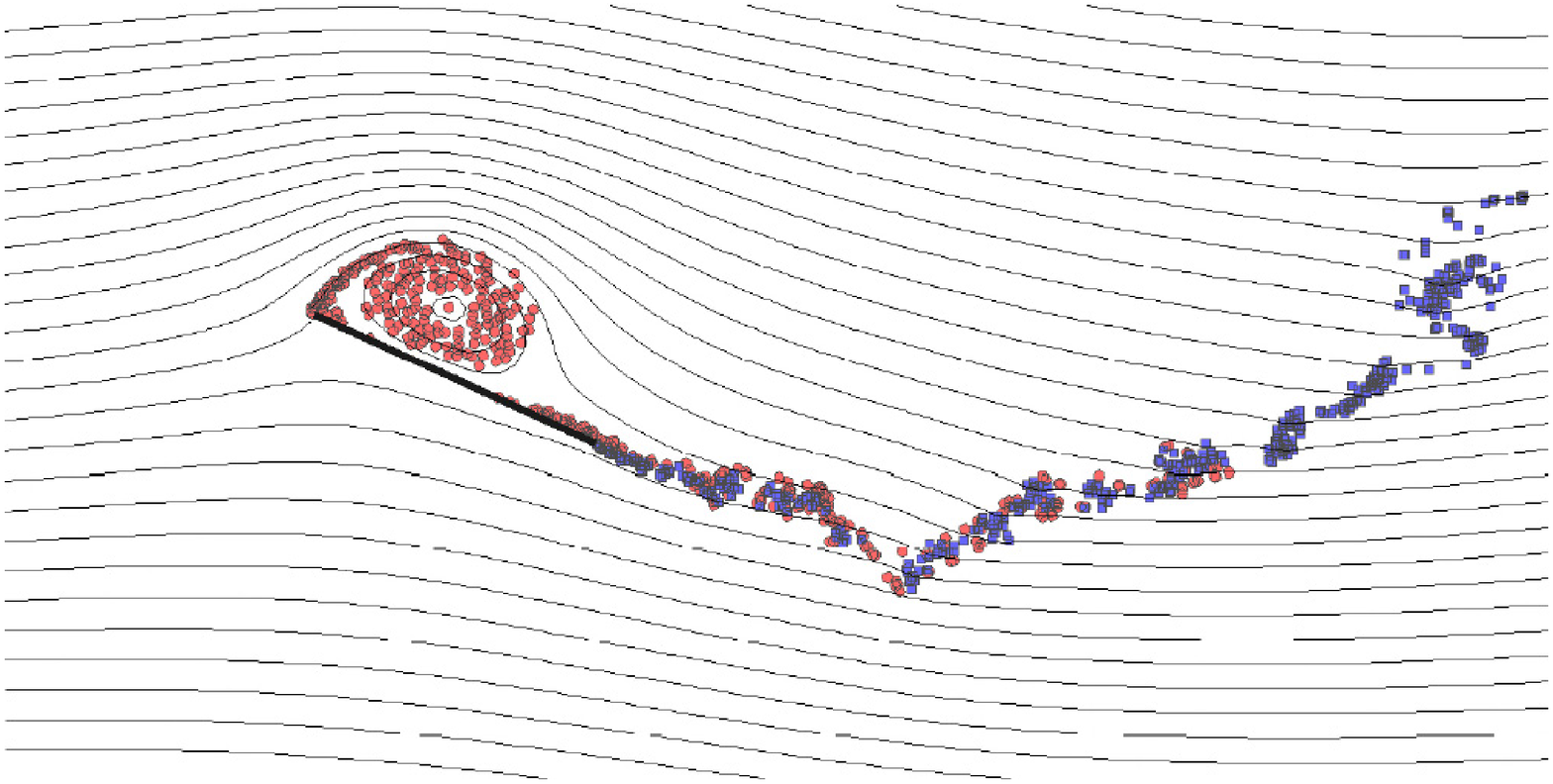}
\end{center} \end{minipage}
\begin{minipage}{0.24\linewidth} \begin{center}  
\includegraphics[width=.99\linewidth,viewport=50 0 570 500,clip]{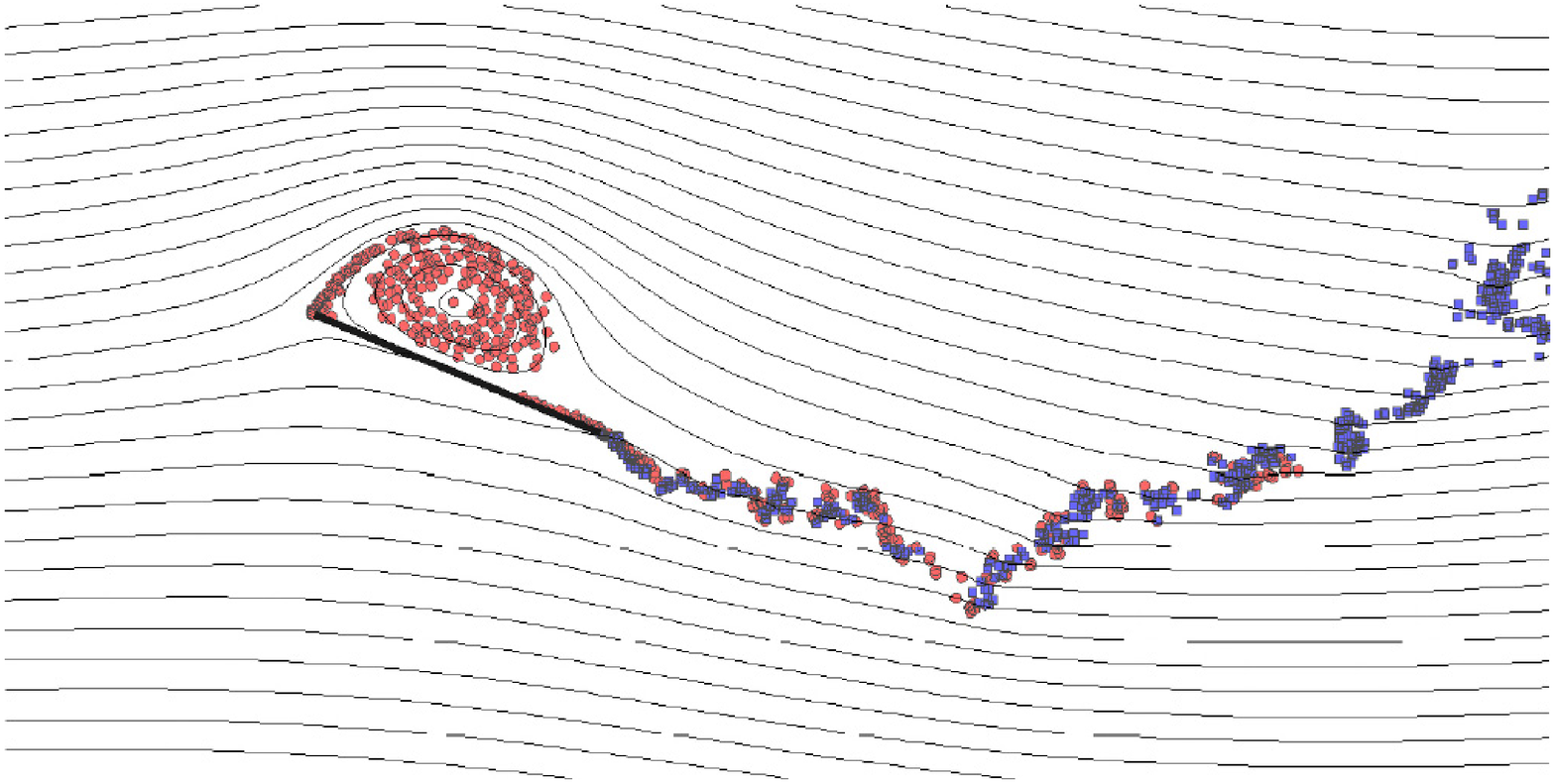}
\end{center} \end{minipage}\\

\vspace{2mm}

\begin{minipage}{0.24\linewidth}\begin{center} \textbf{$t^*=1.8$}  \end{center}
\end{minipage}
\begin{minipage}{0.24\linewidth}\begin{center} \textbf{$t^*=3.0$}  \end{center}
\end{minipage}
\begin{minipage}{0.24\linewidth}\begin{center} \textbf{$t^*=4.0$}  \end{center}
\end{minipage}
\begin{minipage}{0.24\linewidth}\begin{center} \textbf{$t^*=4.2$}  \end{center}
\end{minipage}\\

\vspace{2mm}

\begin{minipage}{0.99\linewidth}\begin{center} \textbf{(a) $25^o$ pitch amplitude rotated about the leading edge.}  \end{center}
\end{minipage}

\vspace{2mm}

\begin{minipage}{0.24\linewidth} \begin{center}
\includegraphics[width=.99\linewidth,viewport=0 0 170 120,clip]{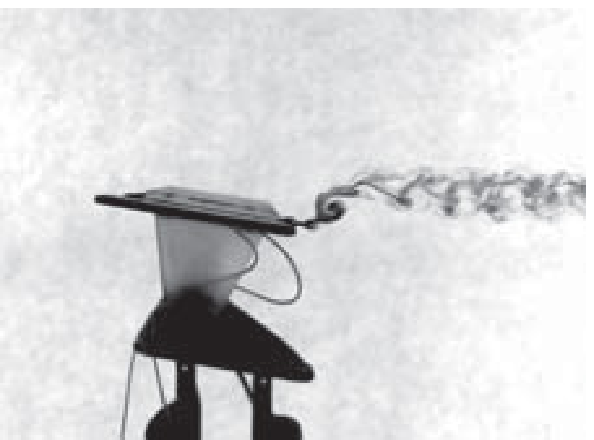}
\end{center} \end{minipage}
\begin{minipage}{0.24\linewidth} \begin{center}  
\includegraphics[width=.99\linewidth,viewport=0 0 170 120,clip]{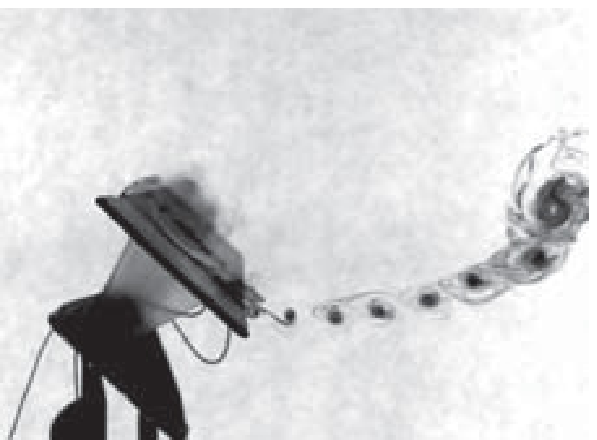}
\end{center} \end{minipage}
\begin{minipage}{0.24\linewidth} \begin{center}
\includegraphics[width=.99\linewidth,viewport=0 0 170 120,clip]{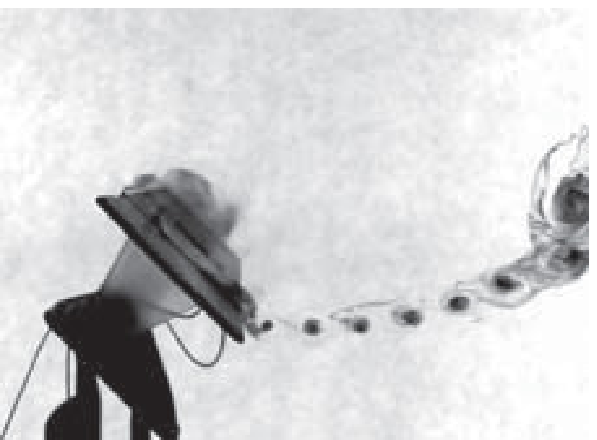}
\end{center} \end{minipage}
\begin{minipage}{0.24\linewidth} \begin{center}  
\includegraphics[width=.99\linewidth,viewport=0 0 170 120,clip]{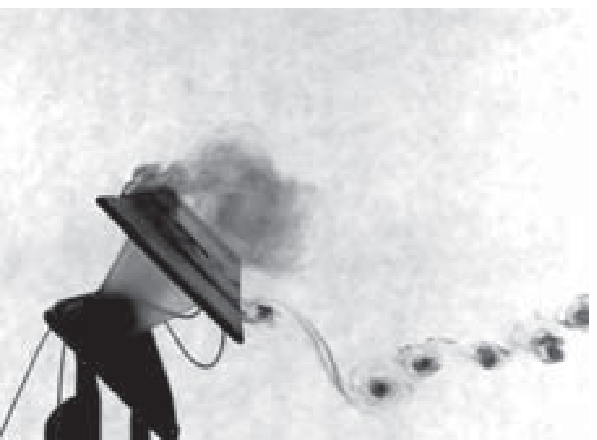}
\end{center} \end{minipage}\\

\begin{minipage}{0.24\linewidth} \begin{center}
\includegraphics[width=.99\linewidth,viewport=50 0 570 500,clip]{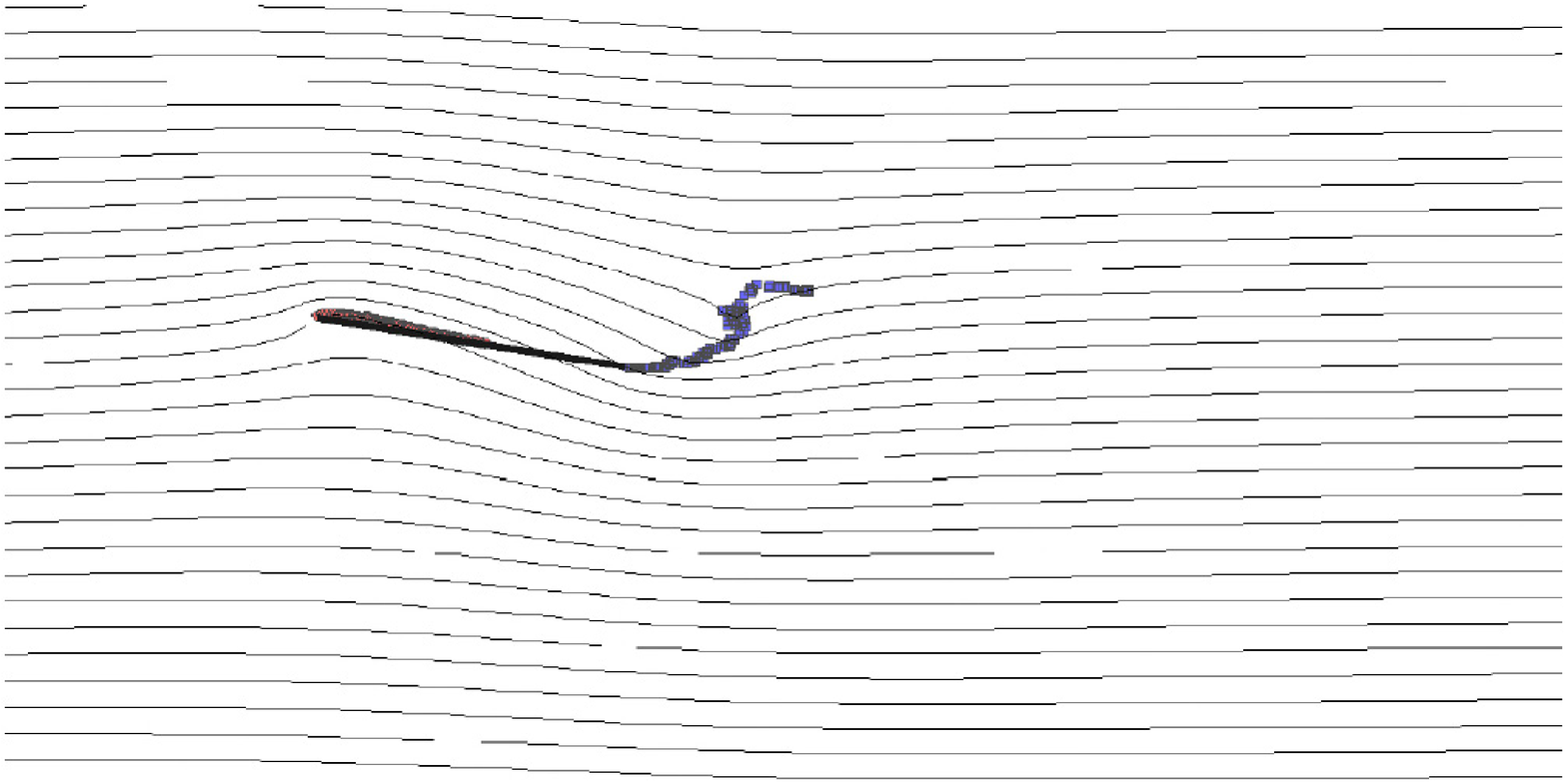}
\end{center} \end{minipage}
\begin{minipage}{0.24\linewidth} \begin{center}  
\includegraphics[width=.99\linewidth,viewport=50 0 570 500,clip]{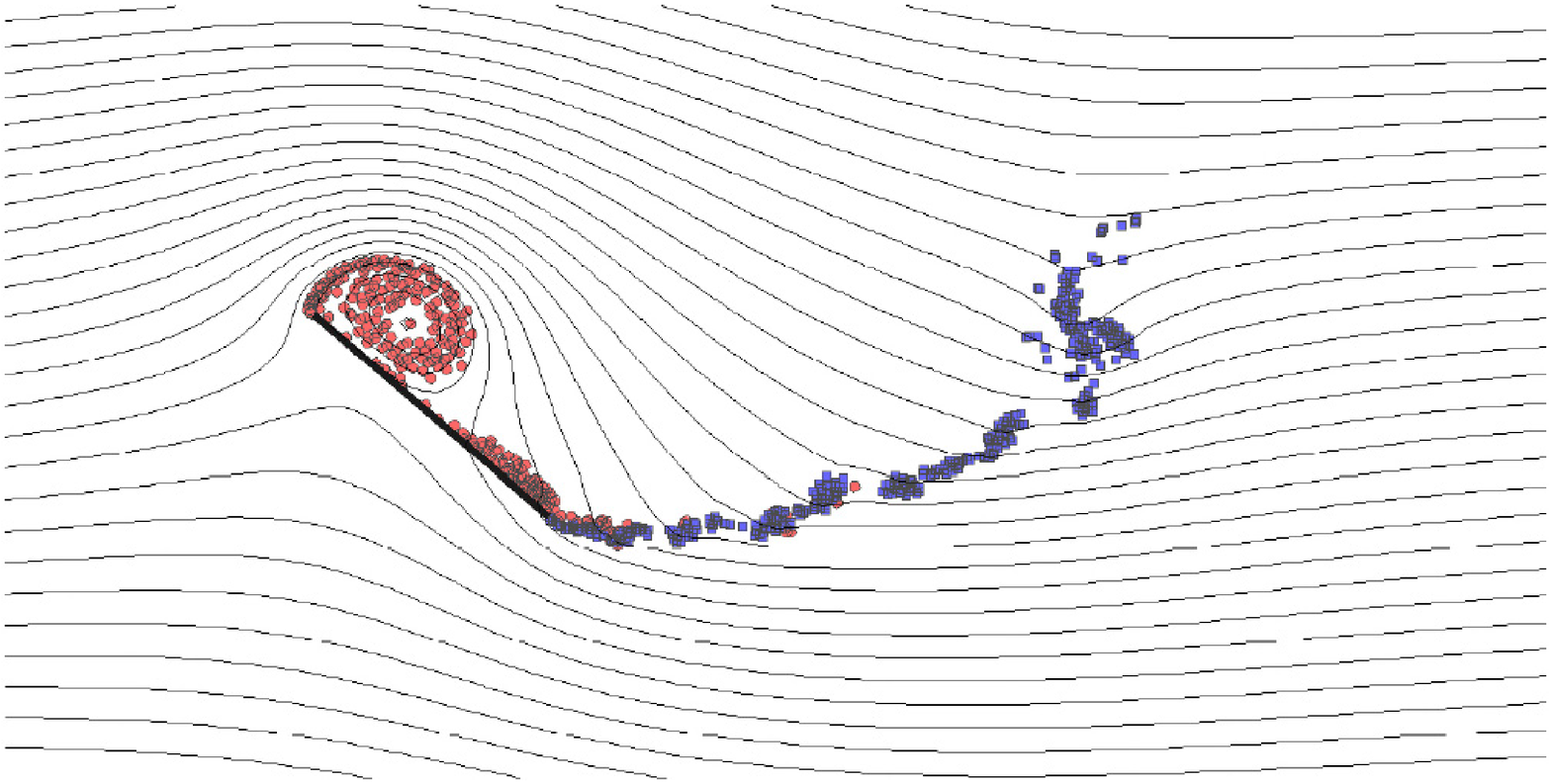}
\end{center} \end{minipage}
\begin{minipage}{0.24\linewidth} \begin{center}
\includegraphics[width=.99\linewidth,viewport=50 0 570 500,clip]{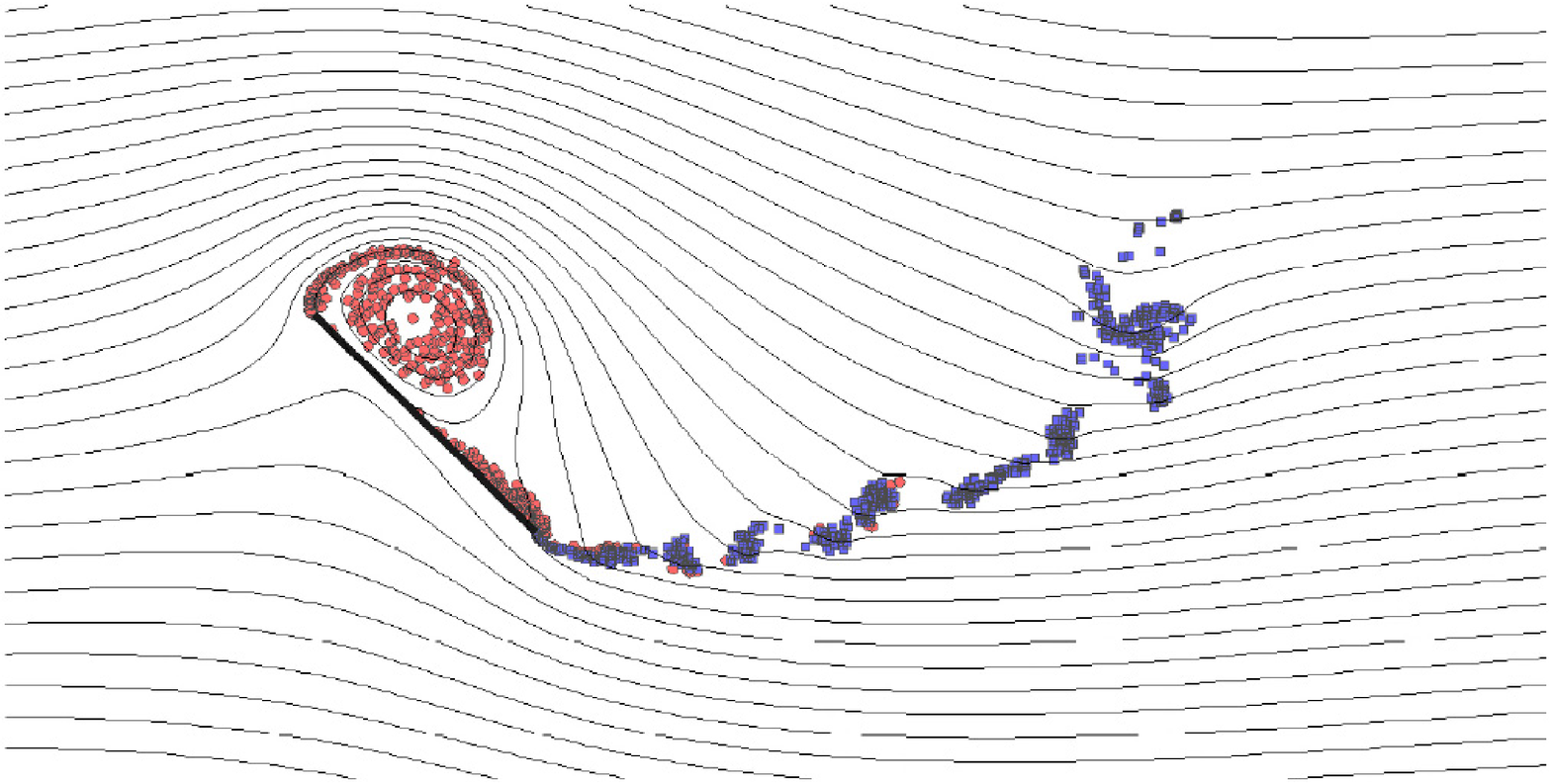}
\end{center} \end{minipage}
\begin{minipage}{0.24\linewidth} \begin{center}  
\includegraphics[width=.99\linewidth,viewport=50 0 570 500,clip]{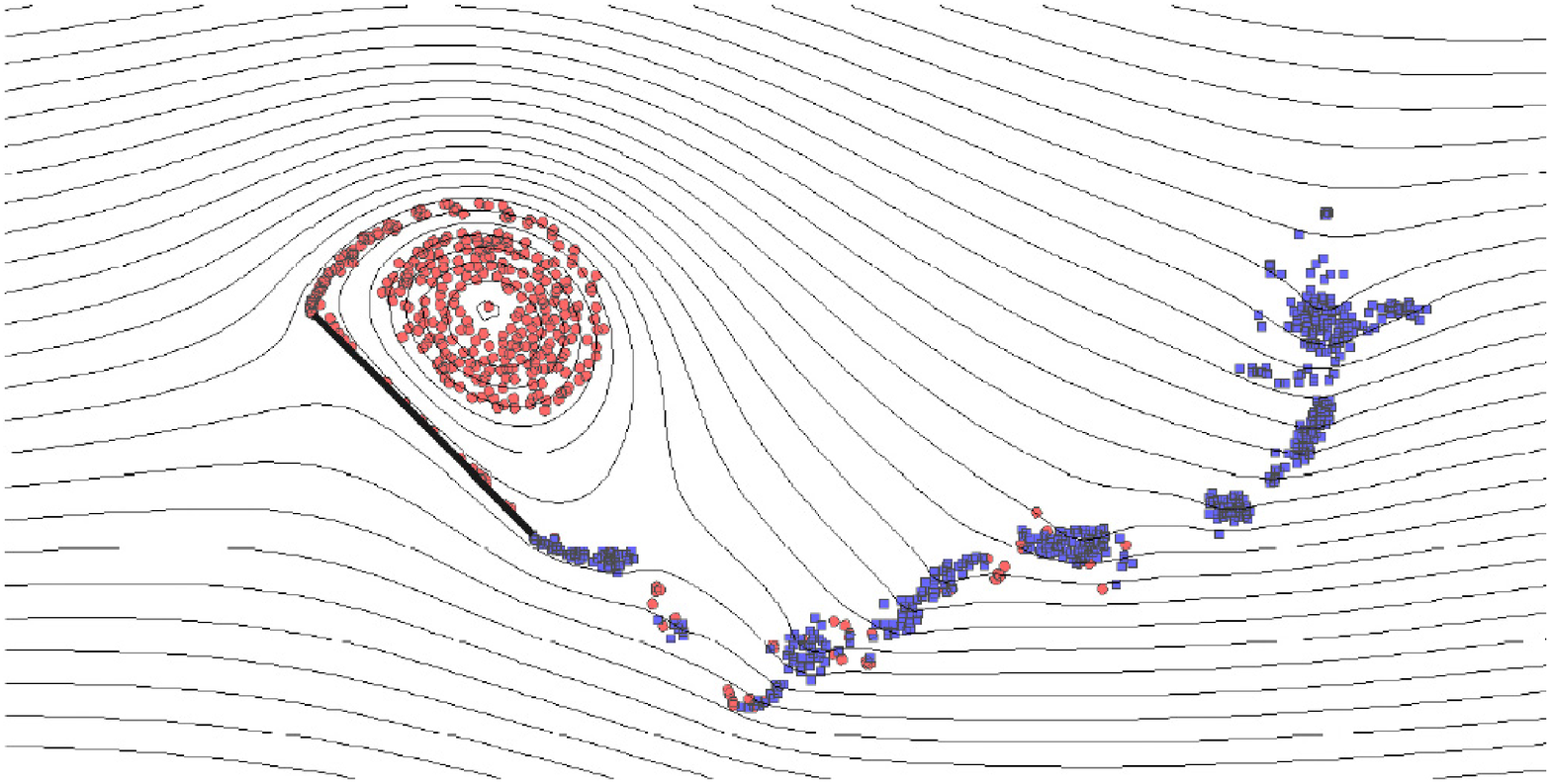}
\end{center} \end{minipage}\\

\vspace{2mm}

\begin{minipage}{0.24\linewidth}\begin{center} \textbf{$t^*=1.6$}  \end{center}
\end{minipage}
\begin{minipage}{0.24\linewidth}\begin{center} \textbf{$t^*=2.8$}  \end{center}
\end{minipage}
\begin{minipage}{0.24\linewidth}\begin{center} \textbf{$t^*=3.0$}  \end{center}
\end{minipage}
\begin{minipage}{0.24\linewidth}\begin{center} \textbf{$t^*=4.0$}  \end{center}
\end{minipage}\\

\vspace{2mm}

\begin{minipage}{0.99\linewidth}\begin{center} \textbf{(b) $45^o$ pitch amplitude rotated about the leading edge.}  \end{center}
\end{minipage}

\vspace{2mm}

\caption{\small A pitch-up, hold and pitch-down flat plate. The figure shows the comparison of flow fields and vortex structures between previous experimental work \cite{OL:11a} (top) and the potential flow model (bottom) for a pitching flat plate.} \label{fig:Flowfield_OL}\vspace{-3mm}

\end{center}
\end{figure}

The comparison between experimental flow visualization \cite{OL:11a} and potential flow simulation of two cases is shown in Figure~\ref{fig:Flowfield_OL}. Both cases pivot about the leading edge with pitch amplitudes of $25^o$ and $45^o$, respectively. The flow field and vortex structures simulated by this model match well with those from the experiment for both small and large angles of attack. This indicates that the unsteady potential flow model applied in this study well captures the flow features of unsteady motions of the flat plate. Moreover, the good matching of the vortex structures also reflects the proper implementation of vortex shedding conditions at various angles of attack. Here, it should be noted that since two different treatments are proposed for the leading edge vortex shedding, the criterion for activating the more appropriate one needs to be handled carefully, especially at mid-ranged angles of attack. Although a more delicate model is needed in the future to determine this condition, here it is justifiable that we determine this criterion by matching the configurations and size of the leading edge vortex or separation bubble with experiments. Following this rule, the critical angle is found to be about $20^o$ during the pitch-up motion for both cases.

\begin{figure}
\begin{center}

\begin{minipage}{0.49\linewidth} \begin{center}
\includegraphics[width=.99\linewidth,viewport=40 0 750 600,clip]{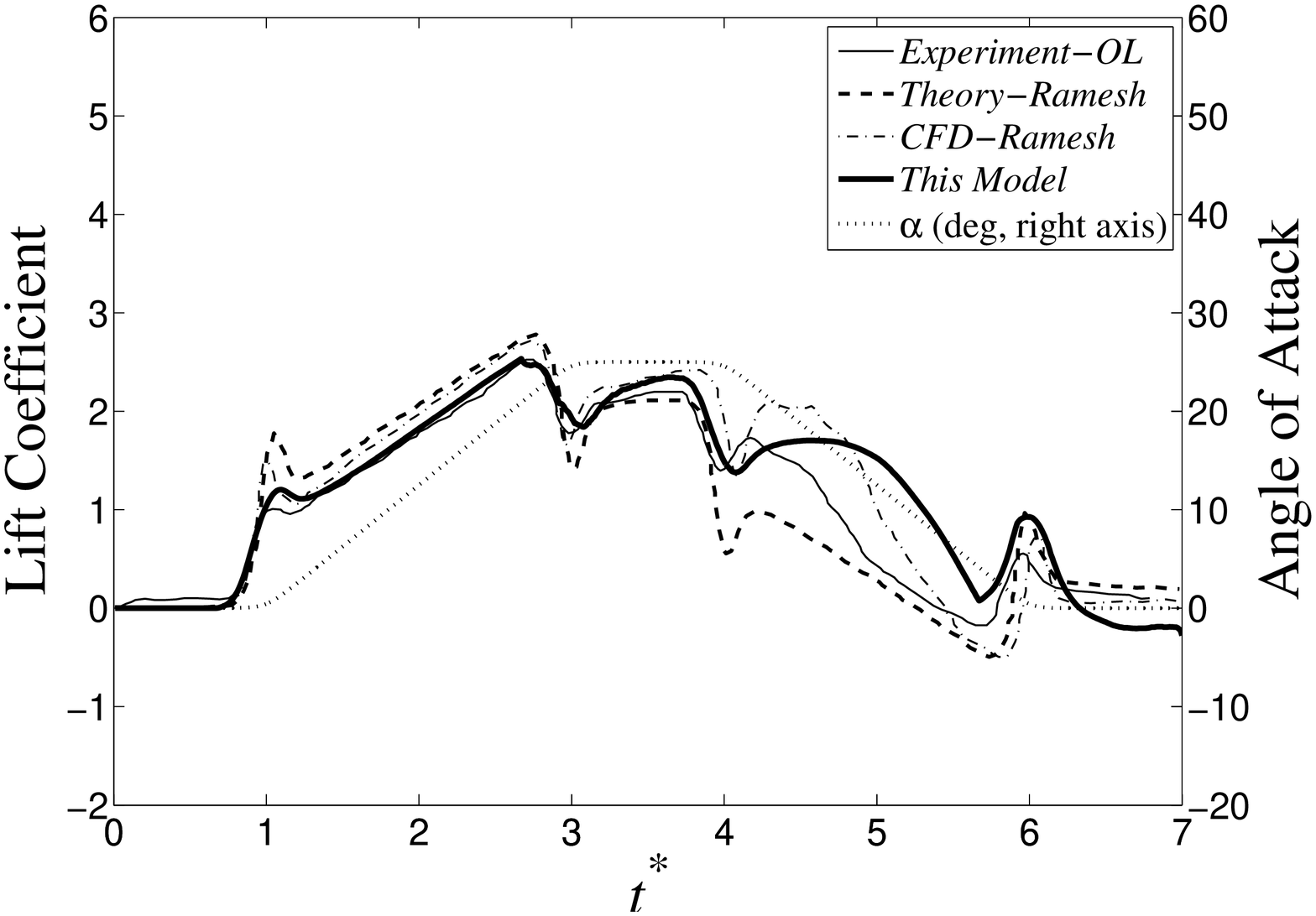}
\end{center} \end{minipage}
\begin{minipage}{0.49\linewidth} \begin{center}  
\includegraphics[width=.99\linewidth,viewport=40 0 750 600,clip]{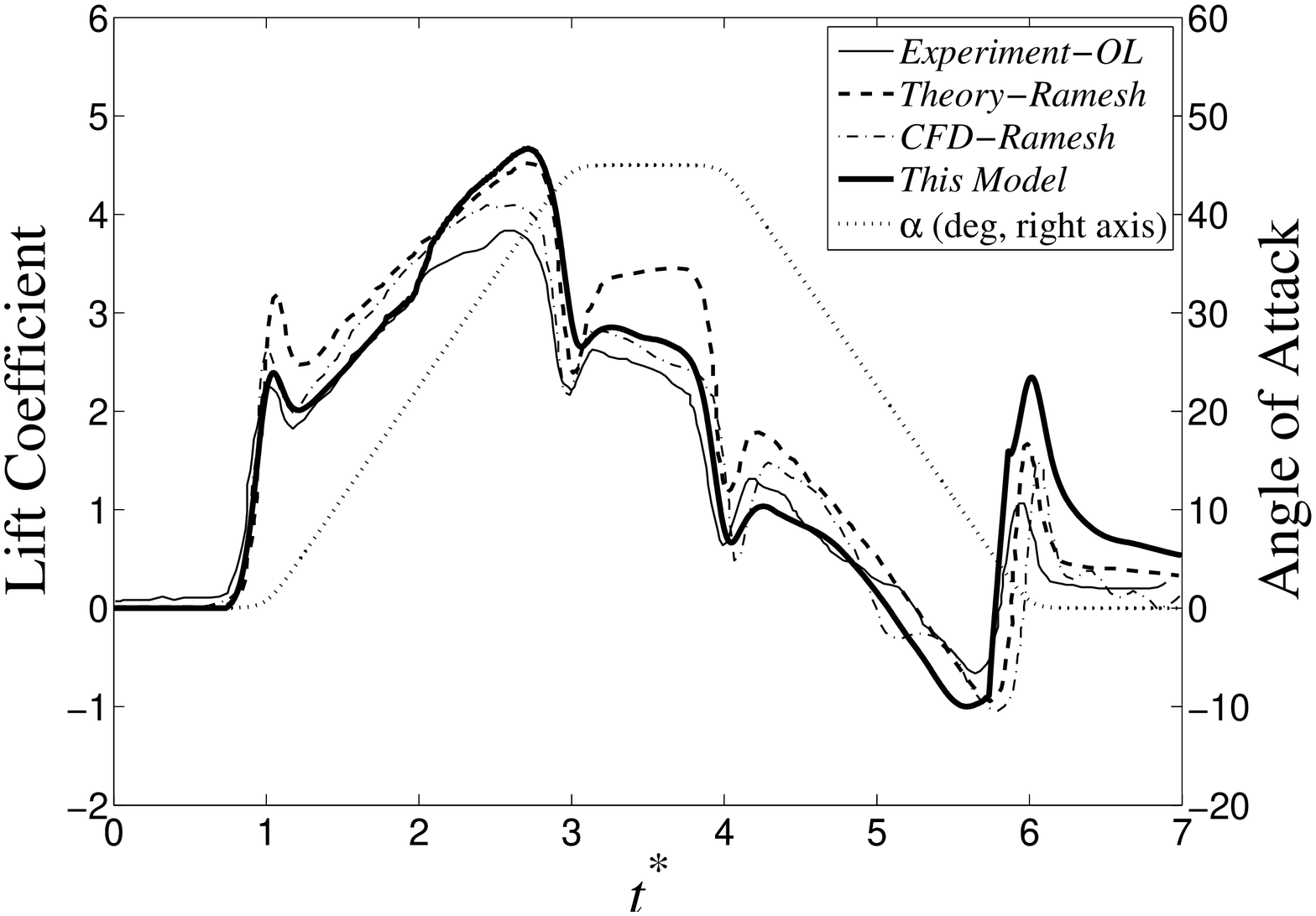}
\end{center} \end{minipage}\\
\vspace{1mm}

\begin{minipage}{0.49\linewidth}\begin{center} \textbf{(a) $25^o$ pitch, leading edge pivot}  \end{center}
\end{minipage}
\begin{minipage}{0.49\linewidth}\begin{center} \textbf{(b) $45^o$ pitch, leading edge pivot}  \end{center}
\end{minipage}\\

\begin{minipage}{0.49\linewidth} \begin{center}
\includegraphics[width=.99\linewidth,viewport=40 0 750 600,clip]{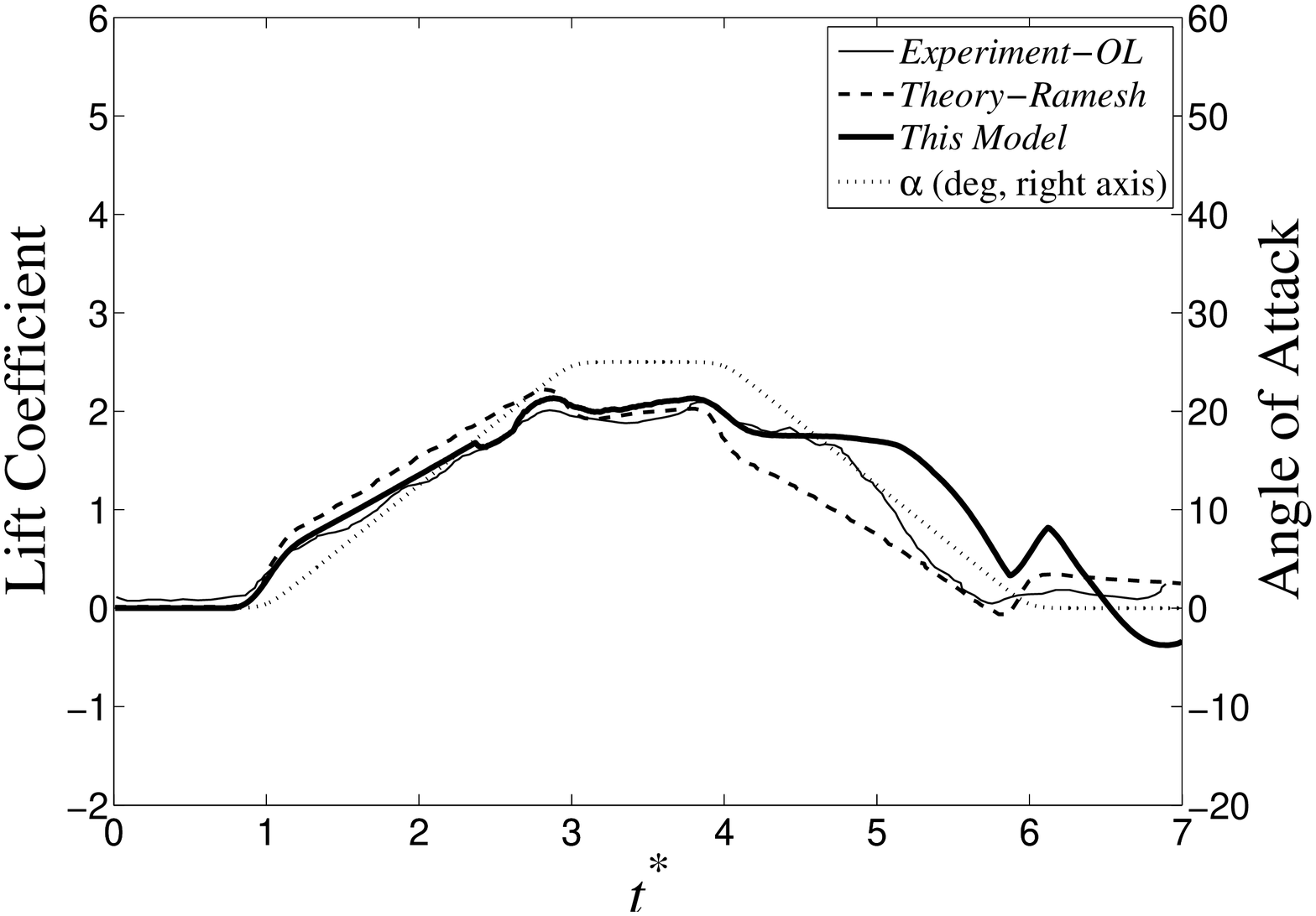}
\end{center} \end{minipage}
\begin{minipage}{0.49\linewidth} \begin{center}  
\includegraphics[width=.99\linewidth,viewport=40 0 750 600,clip]{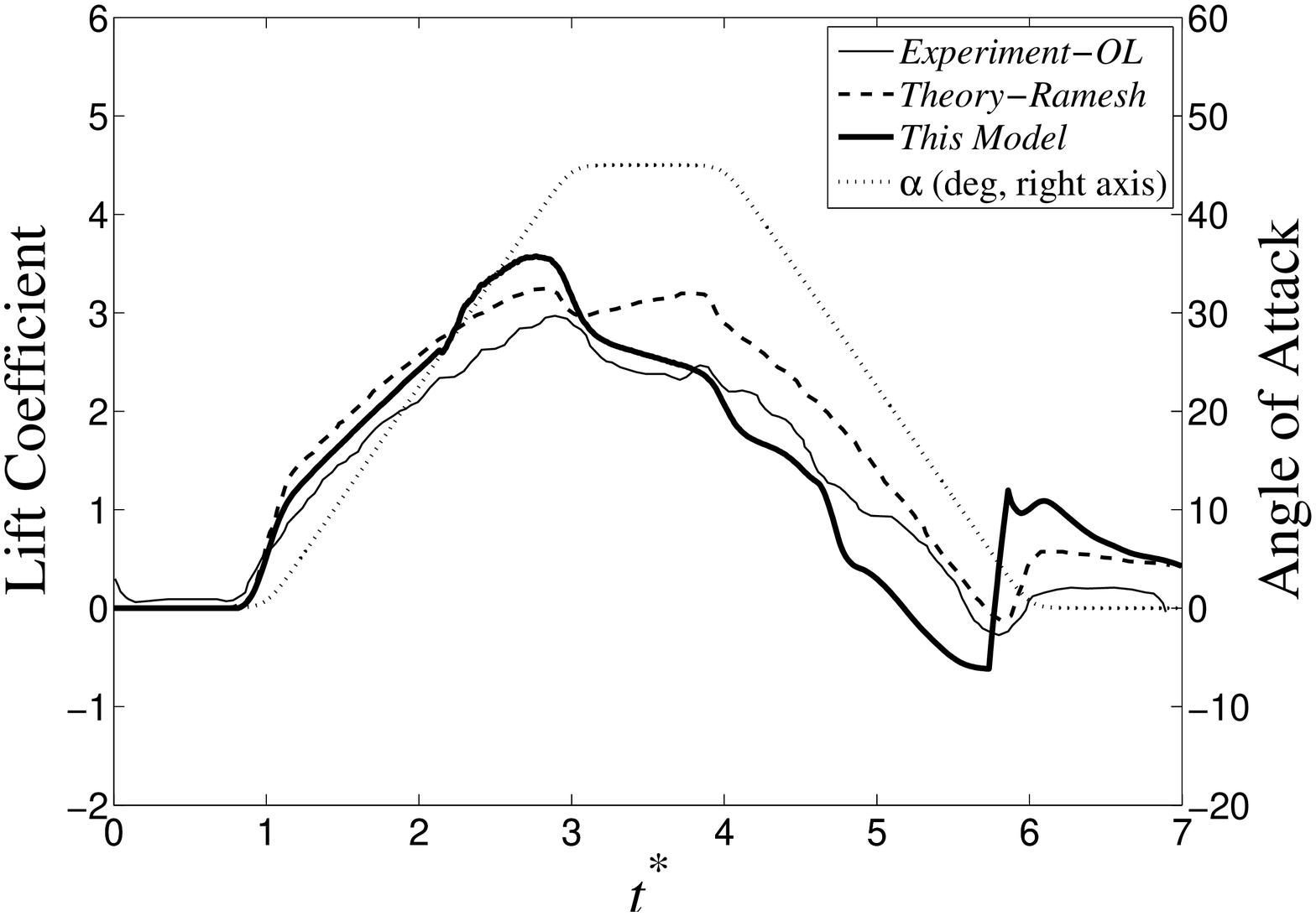}
\end{center} \end{minipage}\\
\vspace{1mm}

\begin{minipage}{0.49\linewidth}\begin{center} \textbf{(c) $25^o$ pitch, half chord pivot}  \end{center}
\end{minipage}
\begin{minipage}{0.49\linewidth}\begin{center} \textbf{(d) $45^o$ pitch, half chord pivot}  \end{center}
\end{minipage}\\

\vspace{0mm}

\caption{\small Comparison of Lift coefficient with previous results \cite{OL:11a} for a pitch up-hold-pitch down maneuver.} \label{fig:Lift_OL_compare}\vspace{0mm}

\end{center}
\end{figure}

Next, we compare the lift calculations of four test cases with OL's experimental, Ramesh et al.'s computational and theoretical approaches \cite{OL:11a} as shown in Figure~\ref{fig:Lift_OL_compare}. Ramesh et al.'s theoretical model adopted Theodorsen's method, based on thin airfoil theory, which does not resolve explicit vortices in the wake. It can be observed that the overall time evolution trends of our lift calculations resemble experimental data in most of the periods for all cases. Actually, the lift calculations from this model match well with experiments during the pitch-up and hold phase of the flapping motion ($t^*<4$, $t^*$ denotes the non-dimensional time.) in all cases. When further compared with others' work (including CFD simulations), this model seems to display a better performance at small angles of attack during the pitch-up motion and comparable performance with CFD during the hold phase. This indicates that the implementation of the vortex shedding condition for small angles of attack accurately interprets the physics that occur at the leading edge shear layer. However, at higher angles of attack during the pitch-up motion (above $30^o$), this model tends to overestimate the lift to some extent. The reason for this might be the abrupt transition between the two vortex shedding schemes, which creates an initial offset that increases the lift when the high angle of attack shedding condition is activated. Additional work need to be done to smooth the transition between the vortex shedding methods. 

In Figure~\ref{fig:Lift_OL_compare}, it is also notable that the gap between model prediction and experimental data are significant for most cases (except for case (b)) during the pitch-down motion ($4<t^*<6$). It seems that the decreasing lift trends are either delayed (case (a) and (c)) or advanced (case (b)) relative to the experimental values. The results from CFD and other models apparently suffer from the same issue as neither of their predictions show convincing matching of lift in this region. Besides the influence of the 3D effect near the tip region, there are two potential reasons for the difficulty of lift prediction here. First of all, the leading edge vortex becomes extremely large and moves rearward before the pitch-down motion as separated and reverse flows develop on top of the flat plate, which would result in the failure of the traditional Kutta condition at the trailing edge. Second, during the pitch-down motion the trailing edge has a possibility of interfering with the growing leading edge vortex which would add further complexity to the flow near the trailing edge. With the awareness of these two circumstances, the authors believe that traditional Kutta condition should be relaxed at the trailing edge during the pitch-down motion to better reflect the reality (refer to Ansari et al. \cite{AnsariSA:06b} for similar discussions about releasing Kutta condition  at the shedding edges). As a result, a similar implementation to that of the leading edge condition at small angle of attack is adopted with the circulation determined by a simple equation: $dU^2dt/2$. However, the question still remains as to when this condition should take an effect. It has been implied from the results that an arbitrary activation of this condition only results in the delay or advance of the lift variation trends.

In the starting flat plate problem, it is apparent that lift enhancement is achieved when the leading edge vortex grows and the trailing edge vortex sheds. This can also be verified by the pitching problem during the hold phase ($3<t^*<4$). In cases (a) and (c) in Figure~\ref{fig:Lift_OL_compare} where the pitch amplitude is small, the LEV is still growing and the trailing edge is shedding vortices without interference from the LEV during the hold phase (see Figure~\ref{fig:Flowfield_OL}(a)). As a result, a conservation or even slight increase of lift is observed. In cases (b) and (d) during the same period with larger pitch amplitude, the leading edge vortex is much larger and affects the trailing edge vortex shedding significantly (see Figure~\ref{fig:Flowfield_OL}(b)). In this case, even as the leading edge vortex is growing, the trailing edge vortex shedding is altered which results in a declining trend in the lift coefficient. To this end, it should be reasonable to conclude that maintaining the growth of the LEV and simultaneously preventing the trailing edge flow from being interrupted by the LEV or other unsteady flows should be a main method to enhance lift (or prevent lift loss) in high angle of attack situations. This might be the reason why stabilizing the LEV will enhance the lift. For future active flow control applications on MAVs, control actuations should be aimed at creating and maintaining the LEV, while eliminating the effect of LEV on the trailing edge and maintaining the normal shedding condition at the trailing edge.

\section{Conclusions}

The problem of a flapping flat plate is investigated in this paper. First, an unsteady potential flow model is presented with a single vortex-sink singularity attached to a flat plate. Then, a multi-vortices model is extended from the single singularity model to discretely simulate the shedding vortices from the leading and trailing edges. The lift calculation is performed by integrating the unsteady Blasius equation. This is an improvement based on Minotti's unsteady lift calculations \cite{MinottiFO:02a} in the sense of identifying the velocity contributions from the LEV and TEV. It is shown that the shedding of TEV and stabilization of LEV would increase the lift. 

Shedding conditions required to determine the intensities and placement of the shedding vortices at the shedding edges are discussed. It is suggested that a Kutta-like condition should be satisfied at the leading edge for cases with large angles of attack, while a vortex sheet based model is proposed to calculate the circulation for cases with small angles of attack. The shedding conditions, together with the dynamic model and lift estimation, are verified by comparing the simulation of a starting flat plate problem with previous experimental, CFD and model approaches. The results show good performance of lift coefficient at various angles of attack. Furthermore, several cases of the canonical pitch up and pitch down motion are simulated using this model and the results are compared with experiments, numerical simulations and other previous models. The results show promising performance of lift prediction during the pitch-up and hold phases of the flapping motion. Excellent matching behaviors are observed at pitching cycles with small angles of attack which verifies the new shedding model for the leading edge. Moreover, since this model is based on the flow configuration with wake and vortex movement being resolved, it can be expected to provide more accuracy than the thin airfoil model during the pitch-down phase of flapping motion as well. However, to better improve this model, future work should be focused on modeling the shedding condition with complex flow configurations such as the presence of strong edge-wake interactions. In addition, the time variations of the lift coefficient in both the starting plate problem and, the pitching up and down problem provide insights to future flow control applications that stabilizing the LEV and at the same time maintaining the normal shedding of trailing edge vortices are required to enhance the lift of MAVs. 

\begin{acknowledgments}
The authors would like to thank Christopher Davis and Shervin Shajiee for their preliminary work in developing the basics of the potential flow model as discussed in Reference~\cite{Mohseni:09s}. The authors would like to thank Matthiew Shields for his comments in improving the presentation of this manuscript.
\end{acknowledgments}


\begin{thebibliography}{10}
\newcommand{\enquote}[1]{``#1''}

\bibitem{EllingtonCP:84a}
Ellington, C., \enquote{The aerodynamics of hovering insect flight. IV.
  Aerodynamic mechanisms,} {\em Phil. Trans. R. Soc. Lond. B\/}, Vol.~305,
  1984, pp.~79--113.

\bibitem{Dickinson:93a}
Dickinson, M. and Gotz, K., \enquote{Unsteady aerodynamic performance of model
  wings at low Reynolds numbers,} {\em \JEB\/}, Vol.~174, 1993, pp.~45--64.

\bibitem{Dickinson:04a}
Wang, Z., Birch, J., and Dickinson, M., \enquote{Unsteady forces and flows in a
  low Reynolds number hovering flight: Two-dimensional computations vs. robotic
  wing experiments,} {\em \JEB\/}, Vol.~207, 2004, pp.~449--460.

\bibitem{EllingtonCP:98a}
Lia, H., Ellington, C.~P., Kawachi, K., Berg, C. V.~D., and Willmott, A.~P.,
  \enquote{A computational fluid dynamic study of hawkmoth hovering,} {\em
  \JEB\/}, Vol.~201, No. 461-477, 1998.

\bibitem{SunM:02a}
Sun, M. and Tang, J., \enquote{Unsteady aerodynamic force generation by a model
  fruit fly wing in flapping motion,} {\em The Journal of Experimental
  Biology\/}, Vol.~205, 2002, pp.~55--70.

\bibitem{Peskin:04a}
Miller, L. and Peskin, C., \enquote{When vortices stick: An aerodynamic
  transition in tiny insect flight,} {\em The Journal of Experimental
  Biology\/}, Vol.~207, 2004, pp.~3073--3088.

\bibitem{Saffman:77a}
Saffman, P. and Sheffield, J., \enquote{Flow over a wing with an attached free
  vortex,} {\em Studies in Applied Mathematics\/}, Vol.~57, 1977, pp.~107--117.

\bibitem{ChowCY:81a}
Huang, M.-K. and Chow, C.-Y., \enquote{Trapping of a free vortex by Joukowski
  airfoils,} {\em AIAA\/}, Vol.~20, No.~3, 1981.

\bibitem{RossowVJ:78a}
Rossow, V., \enquote{Lift enhancement by an externally trapped vortex,} {\em
  Journal of Aircraft\/}, Vol.~15, No.~9, 1978, pp.~618--625.

\bibitem{MourtosNJ:96a}
Mourtos, N. and Brooks, M., \enquote{Flow past a flat plate with a vortex/sink
  combination,} {\em Journal of Applied Mechanics\/}, Vol.~63, 1996,
  pp.~543--550.

\bibitem{EllingtonCP:97c}
Willmott, A., Ellington, C., and Thomas, A., \enquote{Flow visualization and
  unsteady aerodynamics in the flight of the hawkmoth, {Manduca} sexta,} {\em
  Phil. Trans. R. Soc. Lond. B\/}, Vol.~352, 1997, pp.~303--316.

\bibitem{Dickinson:01a}
Birch, J. and Dickinson, M., \enquote{Spanwise flow and the attachment of the
  leading edge vortex on insect wings,} {\em Nature\/}, Vol.~412, 2001,
  pp.~729--733.

\bibitem{MinottiFO:02a}
Minotti, F.~O., \enquote{Unsteady two-dimensional theory of a flapping wing,}
  {\em Phys. Rev. E\/}, Vol.~66, 2002, pp.~(051907) 1--10.

\bibitem{YuY:03a}
Yu, Y., Tong, B., and Ma, H., \enquote{An analytic approach to theoretical
  modeling of highly unsteady viscous flow excited by wing flapping in small
  insects,} {\em Acta Mechanica Sinica\/}, Vol.~19, 2003.

\bibitem{AnsariSA:06b}
Ansari, S., Zbikowski, R., and Knowles, K., \enquote{Non-linear unsteady
  aerodynamic model for insect-like flapping wings in the hover. Part 2:
  Implementation and validation,} {\em Proceedings IMechE Part G: Journal of
  Aerospace Engineering\/}, Vol.~220, No.~G3, 2006, pp.~169--186.

\bibitem{AnsariSA:06a}
Ansari, S., Zbikowski, R., and Knowles, K., \enquote{A nonlinear unsteady
  aerodynamic model for insect-like flapping wings in the hover: Part I.
  Methodology and analysis,} {\em Proceedings IMechE Part G: Journal of
  Aerospace Engineering\/}, Vol.~220, No.~G2, 2006, pp.~61--83.

\bibitem{Dickinson:99a}
Dickinson, M., Lehmann, F., and Sane, S., \enquote{Wing rotation and the
  aerodynamic basis of insect flight,} {\em Science\/}, Vol.~284, 1999,
  pp.~1954--1960.

\bibitem{Pullin:04b}
Pullin, D. and Wang, Z., \enquote{Unsteady forces on an accelerating plate and
  application to hovering insect flight,} {\em J. Fluid Mech\/}, Vol.~509,
  No.~07, 2004, pp.~1--21.

\bibitem{LlewellynSmith:09a}
Michelin, S. and Smith, S.~L., \enquote{An unsteady point vortex method for
  coupled fluid-solid problems,} {\em Theor. Comput. Fluid Dyn.\/}, Vol.~23,
  2009, pp.~127--153.

\bibitem{BrownCE:54a}
Brown, C. and Michael, W., \enquote{Effect of leading edge separation on the
  lift of a delta wing,} {\em J. Aero. Sci\/}, Vol.~21, 1954, pp.~690--694.

\bibitem{Mason:03a}
Mason, R.~J., {\em Fluid Locomotion and Trajectory Planning for Shape-changing
  Robots\/}, Ph.D. thesis, California Institute of Technology, Pasadena, CA,
  June 2002.

\bibitem{Kelly:09a}
Cochran, J., Kelly, S., Xiong, H., and Krstic, M., \enquote{Source seeking for
  a joukowski foil model of fish locomotion,} Tech. rep., 2009 American Control
  Conference, St. Louis, MO, June 10-12 2009.

\bibitem{Eldredge:09a}
Eldredge, J., Wang, C., and OL, M., \enquote{A computational study of a
  canonical pitch-Up, pitch-down wing maneuver,} AIAA paper 2009-3687, San
  Antonio, TX, June 2009, 39$^{th}$ AIAA Fluid Dynamics Conference.

\bibitem{Wu:81a}
Wu, J., \enquote{Theory for aerodynamic force and moment in viscous flows,}
  {\em AIAA Journal\/}, Vol.~19, 1981, pp.~432--441.

\bibitem{Dudley:02a}
Dudley, R., {\em The Biomechanics of Insect Flight: Form, Function,
  Evolution\/}, Princeton University Press, 2002.

\bibitem{MilneThomsonLM:58a}
Milne-Thomson, L.~M., {\em Theoretical Hydrodynamics\/}.

\bibitem{LinCC:41a}
Lin, C.~C., \enquote{On the motion of vortices in two dimensions-I. Existence
  of the Kirchhoff-Routh function,} {\em Physics\/}, Vol.~27, 1941,
  pp.~570--575.

\bibitem{ClementsRR:73a}
Clements, R.~R., \enquote{An inviscid model of two-dimensional vortex
  shedding,} {\em J. Fluid Mech.\/}, Vol.~57, 1973, pp.~321--336.

\bibitem{CrightonDG:85a}
Crighton, D., \enquote{The Kutta condition in unsteady flow,} {\em \ARFM\/},
  Vol.~17, 1985, pp.~411--445.

\bibitem{ChenSH:87a}
Chen, S. and Ho, C.~M., \enquote{Near wake of an unsteady symmetric airfoil,}
  {\em Journal of Fluids and Structures\/}, Vol.~1, 1987, pp.~151--164.

\bibitem{PolingDR:87a}
Poling, D. and Telionis, D., \enquote{The trailing edge of a pitching airfoil
  at high reduced frequency,} {\em Journal of Fluid Engineering\/}, Vol.~109,
  1987, pp.~410--414.

\bibitem{Mohseni:08l}
Lipinski, D., Cardwell, B., and Mohseni, K., \enquote{A {L}agrangian analysis
  of a two-dimensional airfoil with vortex shedding,} {\em J. Phys. A\/},
  Vol.~41, 2008, pp.~344011.

\bibitem{AnsariSA:07a}
Knowsles, K., Wilkins, P., Ansari, S., and Zbikowski, R., \enquote{Integrated
  computational and experimental studies of flapping-wing micro air vehicle
  aerodynamics,} Tech. rep., U.S. Air Force Academy, CO, 20-21 June 2007,
  $3^{rd}$ International Symposium on Integrating CFD and Experiments in
  Aerodynamics.

\bibitem{OL:09a}
OL, M.~V., \enquote{The high-frequency, high-amplitude pitch problem: Airfoils,
  plates and wings,} AIAA paper 2009-3686, San Antonio, TX, June 2009,
  39$^{th}$ AIAA Fluid Dynamics Conference.

\bibitem{OL:11a}
Ramesh, K., Gopalarathnam, A., Edwards, J.~R., OL, M.~V., and Granlund, K.,
  \enquote{Theoretical, computational and experimental studies of a flat plate
  undergoing high-amplitude pitching motion,} AIAA paper 2011-217, Orlando, FL,
  January 2011, 49$^{th}$ AIAA Aerospace Sciences Meeting including the New
  Horizons Forum and Aerospace Exposition.

\bibitem{Mohseni:09s}
Davis, C., Shajiee, S., and Mohseni, K., \enquote{Lift enhancement in a
  flapping airfoil by an attached free vortex and sink,} AIAA paper 2009-0390,
  47$^\textrm{th}$ AIAA Aerospace Sciences Meeting and Exhibit, Orlando, FL,
  January 5-8 2009.

\end{thebibliography}

\newcommand{\AIAAJ}{AIAA J.} \newcommand{\AIAAP}{AIAA Paper}
  \newcommand{\ARMA}{Archive for Rational Mechanics and Analysis}
  \newcommand{\ASMEJFE}{J. Fluids Eng., Trans. ASME} \newcommand{\ASR}{Applied
  Scientific Research} \newcommand{\CF}{Computers Fluids}
  \newcommand{\CJFAS}{Can. J. Fish. Aquat. Sci.}
  \newcommand{\ETFS}{Experimental Thermal and Fluid Science}
  \newcommand{\EF}{Experiments in Fluids} \newcommand{\FDR}{Fluid Dynamics
  Research} \newcommand{\IJHMT}{Int. J. Heat Mass Transfer}
  \newcommand{\JASA}{J. Acoust. Soc. Am.} \newcommand{\JCP}{J. Comp. Physics}
  \newcommand{\JEB}{J. Exp. Biol.} \newcommand{\JFM}{J. Fluid Mech.}
  \newcommand{\JMP}{J. Math. Phys.} \newcommand{\JSC}{J. Scientific Computing}
  \newcommand{\JSP}{J. Stat. Phys.} \newcommand{\JSV}{J. of Sound and
  Vibration} \newcommand{\MC}{Mathematics of Computation}
  \newcommand{\MWR}{Monthly Weather Review} \newcommand{\PAS}{Prog. in
  Aerospace. Sci.} \newcommand{\PCPS}{Proc. Camb. Phil. Soc.}
  \newcommand{\PD}{Physica D} \newcommand{\PRA}{Physical Rev. A}
  \newcommand{\PRE}{Physical Rev. E} \newcommand{\PRL}{Phys. Rev. Lett.}
  \newcommand{\PF}{Phys. Fluids} \newcommand{\PFA}{Phys. Fluids A.}
  \newcommand{\PL}{Phys. Lett.} \newcommand{\PRSLA}{Proc. R. Soc. Lond. A}
  \newcommand{\SIAMJMA}{SIAM J. Math. Anal.} \newcommand{\SIAMJNA}{SIAM J.
  Numer. Anal.} \newcommand{\SIAMJSC}{SIAM J. Sci. Comput.}
  \newcommand{\SIAMJSSC}{SIAM J. Sci. Stat. Comput.}
  \newcommand{\TCFD}{Theoret. Comput. Fluid Dynamics} \newcommand{\ZAMM}{ZAMM}
  \newcommand{\ZAMP}{ZAMP} \newcommand{\ICASER}{ICASE Rep. No.}
  \newcommand{\NASACR}{NASA CR} \newcommand{\NASATM}{NASA TM}
  \newcommand{\NASATP}{NASA TP} \newcommand{\ARFM}{Ann. Rev. Fluid Mech.}
  \newcommand{\WWW}{from {\tt www}.} \newcommand{\CTR}{Center for Turbulence
  Research, Annual Research Briefs} \newcommand{\vonKarman}{von Karman
  Institute for Fluid Dynamics Lecture Series}

\end{document}